\documentclass[pdflatex,sn-mathphys-num]{sn-jnl}

\usepackage[utf8]{inputenc}
\usepackage[T1]{fontenc}
\usepackage{url}
\usepackage{siunitx}
\usepackage{booktabs}
\usepackage{amsfonts}
\usepackage{nicefrac}
\usepackage{microtype}
\usepackage{csvsimple}
\usepackage{graphicx}
\usepackage{amsmath}
\usepackage{mathtools}
\usepackage{float}
\usepackage{subfig}
\usepackage{adjustbox}
\usepackage{tikz}
\usepackage{algorithm}
\usepackage{subcaption}
\usepackage{float}
\usepackage[algo2e, ruled,vlined, boxed, linesnumbered]{algorithm2e}
\SetArgSty{textnormal}

\makeatletter
\renewcommand*{\@fnsymbol}[1]{\ensuremath{\ifcase#1\or \dagger\or *\or \mathsection\or
   \ddagger\or \mathparagraph\or \|\or **\or \dagger\dagger
   \or \ddagger\ddagger \else\@ctrerr\fi}}
\makeatother
   
\definecolor{darkblue}{HTML}{1A254B}
\definecolor{lightblue}{HTML}{A7BED3}
\definecolor{blue}{HTML}{114083}
\definecolor{green}{HTML}{81B5AE}
\definecolor{pink}{HTML}{F2545B}
\definecolor{red}{HTML}{A4243B}

\DeclarePairedDelimiter\norm{\lVert}{\rVert}

\renewcommand{\epsilon}{\ensuremath\varepsilon}

\renewcommand{\phi}{\ensuremath{\varphi}}

 \newcommand{\hb}{\mathbf{h}}

 \newcommand{\mb}{\mathbf{m}}

 \newcommand{\xb}{\mathbf{x}}

\newcommand{\Ecal}{\mathcal{E}}

\newcommand{\Ncal}{\mathcal{N}}

\newcommand{\angstrom}{\mbox{\normalfont\AA}\xspace}

\usepackage{arydshln}

\usepackage[capitalise]{cleveref}
\usepackage{xcolor}
\definecolor{cite_color}{HTML}{114083}
\definecolor{link_color}{RGB}{153, 0,0}  
\definecolor{url_color}{RGB}{153, 102,  0}
\definecolor{emp_color}{RGB}{0,0,255}
\hypersetup{
 colorlinks,
 citecolor=cite_color,
 linkcolor=link_color,
 urlcolor=url_color}
 
\usepackage{etoc}
\etocdepthtag.toc{mtchapter}
\etocsettagdepth{mtchapter}{subsection}
\etocsettagdepth{mtappendix}{none}

\usepackage{microtype}
\usepackage{graphicx}
\usepackage{booktabs} 
\usepackage{amsmath}
\usepackage{amsfonts}
\usepackage{nicefrac}
\usepackage{multirow}
\usepackage{multicol}
\usepackage{lineno}
\usepackage{enumitem}
\usepackage{mathtools}
\usepackage[font=small,labelfont=bf,tableposition=top]{caption}

\DeclareCaptionLabelFormat{andtable}{#1~#2  \&  \tablename~\thetable}

\newcommand{\rmm}{\mathbf{m}}
\newcommand{\rmh}{\mathbf{h}}
\newcommand{\rmx}{\mathbf{x}}

\usepackage{hyperref}

\begin{document}

\title[SurGBSA: Learning Representations From Molecular Dynamics Simulations]{SurGBSA: Learning Representations From Molecular Dynamics Simulations}

\author*[1,2]{\fnm{Derek} \sur{Jones}}\email{wdjones@ucsd.edu}

\author[3]{\fnm{Yue}\sur{Yang}}

\author[3]{\fnm{Felice C.}\sur{Lightstone}}

\author[1]{\fnm{Niema} \sur{Moshiri}}
\author*[2]{\fnm{Jonathan E.} \sur{Allen}}
\author*[1]{\fnm{Tajana S.} \sur{Rosing}}

\affil*[1]{\orgdiv{Department of Computer Science and Engineering}, \orgname{University of California, San Diego}, \orgaddress{\city{La Jolla}, \postcode{92093}, \state{CA}, \country{USA}}}

\affil[2]{\orgdiv{Global Security Computing Applications Division}, \orgname{Lawrence Livermore National Laboratory}, \orgaddress{\city{Livermore}, \postcode{94550}, \state{CA}, \country{USA}}}

\affil[3]{\orgdiv{Biosciences and Biotechnology Division}, \orgname{Lawrence Livermore National Laboratory}, \orgaddress{\city{Livermore}, \postcode{94550}, \state{CA}, \country{USA}}}

\abstract{
Self-supervised pretraining from static structures of drug-like compounds and proteins enable powerful learned feature representations. Learned features demonstrate state of the art performance on a range of predictive tasks including molecular properties, structure generation, and protein-ligand interactions. The majority of approaches are limited by their use of static structures and it remains an open question, how best to use atomistic molecular dynamics (MD) simulations to develop more generalized models to improve prediction accuracy for novel molecular structures. We present SURrogate mmGBSA (SurGBSA) as a new modeling approach for MD-based representation learning, which learns a surrogate function of the Molecular Mechanics Generalized Born Surface Area (MMGBSA). We show for the first time the benefits of physics-informed pre-training to train a surrogate MMGBSA model on a collection of over 1.4 million 3D trajectories collected from MD simulations of the CASF-2016 benchmark. SurGBSA demonstrates a dramatic 27,927$\times$ speedup versus a traditional physics-based single-point MMGBSA calculation while nearly matching single-point MMGBSA accuracy on the challenging pose ranking problem for identification of the correct top pose ($-0.4\%$ difference). Our work advances the development of molecular foundation models by showing model improvements when training on MD simulations. Models, code and training data are made publicly available. 
}


\maketitle

\section{Introduction}\label{sec:into}

Artificial Intelligence (AI) has made a significant impact on computer based drug-design~\cite{Jumper2021-xh, Baek2021-wq, Abramson2024-xu,Xu2025-ih, Zhang2025-ma}. New tools enable the prediction of structure from sequence~\cite{Jumper2021-xh, Baek2021-wq, Abramson2024-xu, Wohlwend2024-ag,Passaro2025-ci}, binding poses~\cite{Corso2022-hd, Lu2024-xh}, and binding affinities~\cite{Stepniewska-Dziubinska2018-fo,Jimenez2018-ei,Jones2021-ar,Volkov2022-vc,Wang2024-lx}. 
These advances are driven in part by the development of sample-efficient inductive biases~\cite{Fuchs2020-cx, Satorras2021-tw}, allowing invariance to arbitrary rotation and translation in Euclidean space, coupled with momentum towards the development of increasingly large \textit{foundation} models~\cite{Bommasani2021-uo}. In recent years, foundation models, which are often trained using self-supervision, have enabled significant advances across nearly every application domain of AI and machine learning~\cite{Bommasani2021-uo}. Methods that process text and graphs have been further adapted to develop biological foundation models trained using publicly available sequence data for ligands and proteins~\cite{Chithrananda2020-el, Ross2022-cm, Wang2022-hg, Singh2025}.
In contrast to NLP and Computer Vision, datasets for drug discovery are comparatively limited in the number of training samples, posing unique challenges to foundation model development. Acquiring additional training data remains cost prohibitive on the scales typically observed for foundation model development~\cite{Bommasani2021-uo}. 

Atomistic Molecular Dynamics~\cite{Allen1989-nk, Dror2012-di} (MD) has traditionally served as a compliment to physical experiments that use physics theory to sample the dynamics of the protein-ligand co-complex which can then be used to estimate binding free energies in order to prioritize compounds for experimental validation. The \textit{trajectories} generated by MD contain dynamical aspects of the binding interaction that static representations do not fully capture.
However, despite extensive research on MD acceleration, MD simulations continue to require significant computational resources~\cite{Jones2022-xt}, especially when considering the approximately 20,000 protein-ligand co-complex structures currently in the PDB~\cite{Berman2000-kn}, and with generative AI poised to play a significant role in increasing the amount of co-complex data~\cite{Abramson2024-xu}. Consequently, computational effort should be allocated based on the confidence of predicted binding interactions, typically prioritizing the best-scoring poses from molecular docking calculations. Furthermore, the required simulation time varies considerably across different free-energy methods, with some approaches requiring multiple MD trajectories under specific parameter sets~\cite{Wang2019-ja}. Molecular Mechanics Generalized Borne Surface Area (MMGBSA) and Molecular Mechanics Poission Boltz Surface Area (MMPBSA) serve as alternatives to the more expensive free-energy estimation methods and are commonly used in conjunction with molecular docking in virtual screening to prioritize co-complex structures for more expensive methods. Typically, the MMGBSA and MMPBSA re-scoring methods only require a single MD simulation of the co-complex structure to make their free-energy estimation as the ensemble average over all of the trajectory frames. These methods require generally less simulation time compared to the more expensive Alchemical Free Energy methods. The MMGBSA method makes approximations that increase the efficiency of the calculation considerably~\cite{Wang2019-ja} and so we based our study on this method specifically. As an additional step for efficiency, ``single-point'' MMGBSA (and MMPBSA) calculations can be done using a single 3D structure as input, providing a method to re-rank docking poses prior to running more expensive methods~\cite{Wang2019-ja}. 

Recent research in molecular representation learning suggests promising results when integrating MD simulations for model development~\cite{Wang2022-hg,Ross2022-cm,Wu2022-bw, Jing2024-wj, Siebenmorgen2024-pn,Prat2024-vp, Liu2025-ef, Passaro2025-ci, Libouban2025-aw}. \citet{Libouban2025-aw} investigated training convolution and recurrent architectures on MD simulation frames to augment the PDBBind dataset. The authors show an improvement for predicting binding affinities for the PDBbind v.2016 core set, their MDbind test set, and the FEP dataset. This study did not consider equivariant architectures so it is not clear how these architectures would perform in this context. The earliest work to develop a model trained on MMPBSA energetics was demonstrated by \citet{Korlepara2022-ff, Korlepara2024-wb} based on 4ns MD simulations for 14,500 structures from the PDB~\cite{Berman2000-kn}. However, the authors of PLAS20k only provide the ensemble-average of the MMPBSA calculation for each PDB co-complex and the models are trained using crystal structures. While the study provides insight into the general problem of learning the MMPBSA energetics, it also doesn't explicitly model the MD simulation data itself, making it unclear how a model trained using a re-scoring method for label generation, would perform when tasked with re-scoring novel structures in practice.
The Misato dataset~\cite{Siebenmorgen2024-pn} aggregates MD simulations and quantum mechanics (QM) properties for 19,443 protein-ligand structures from PDBbind~\cite{Su2019-ni} while also providing MMPBSA energies for each trajectory. A total of 100 snapshots are collected from each trajectory over the 8ns production MD runs~\cite{Siebenmorgen2024-pn}. For training models on MD trajectory samples, the authors predict the atomic adaptability, which aims to quantify local fluctuations over the trajectory. The authors don't explore the development of an MMGBSA surrogate model, examine the ranking performance of such a model, or compare the scoring cost versus physics-based methods. The authors do demonstrate an improvement on predicting experimental binding affinities using their MD and QM-derived features~\cite{Siebenmorgen2024-pn}. 
However, this work is limited by the architecture choice of the GNN based on 3D coordinates, likely decreasing sample efficiency and representation robustness~\cite{Satorras2021-tw}. 
Recently, \citet{Liu2025-ef} address the gap of MMPBSA model development using MISATO~\cite{Siebenmorgen2024-pn}. The authors frame predicting MMPBSA energetics as a pretraining task, in which the training objective encourages the model to make predictions of the trajectory ensemble-averaged MMPBSA score using ANY frame from that simulation. This objective used by \citet{Xu2025-ih} effectively encourages the model to make predictions that are invariant to the subtle changes in the atomic coordinates that vary from snapshots or the docking pose. The authors then consider finetuning their model on the LBA and LEP tasks from ATOM3D~\cite{Townshend2020-eh}, where an improvement in performance is achieved by their model on both tasks. While this work represents the first example of ``energy-guided pretraining'' in the literature, this step is not studied in detail, and it is unclear how well their method generalizes across changes in the 3D structures such as distinguishing multiple poses of the same co-complex structure.

In this Article, we present Surrogate mmGBSA (SurGBSA) for binding 3D structure ranking. Our SurGBSA method comprises (1) the first study of surrogate MMGBSA (or MMPBSA) models trained on snapshot level energetics with an order of magnitude higher snapshot resolution than previously studied and (2) demonstrate a significant improvement in learning the MMGBSA energetics due to physics-informed pretraining. We give the first practical evaluation of MMGBSA surrogate models based on docking pose ranking performance, compared to both the ensemble-averaged MMGBSA and single-point MMGBSA baseline methods. Additionally, we (3) investigate a range of molecular graph encoder modules (GNN, EGNN, EGMN) and their learned representations in detail. 
Our best-performing SurGBSA model achieves an approximate 27,927$\times$ speedup in pose re-ranking while preserving comparable ranking performance to the most efficient MMGBSA baseline, with the difference in correct pose identification being a single sample ($-0.4\%$). This dramatic efficiency gain enables the high-throughput generation of physics-informed pose ranking, providing the computational foundation necessary for training large-scale MD models that could accelerate virtual screening and drug discovery pipelines. Additionally, in contrast to ensemble-averaged MMGBSA based upon MD, SurGBSA can perform inference without the need to perform additional simulation, greatly reducing the computational resources required for virtual screening. SurGBSA can thus allow for rapid pose ranking comparable to single-point MMGBSA with a dramatic increase in computational efficiency while learning a robust molecular representation. Lastly, we make our implementation and data publicly accessible to support future work in MD foundation model development.

\section{Results}\label{sec:results}

\begin{figure}[ht]
    \centering
    \includegraphics[width=\linewidth]{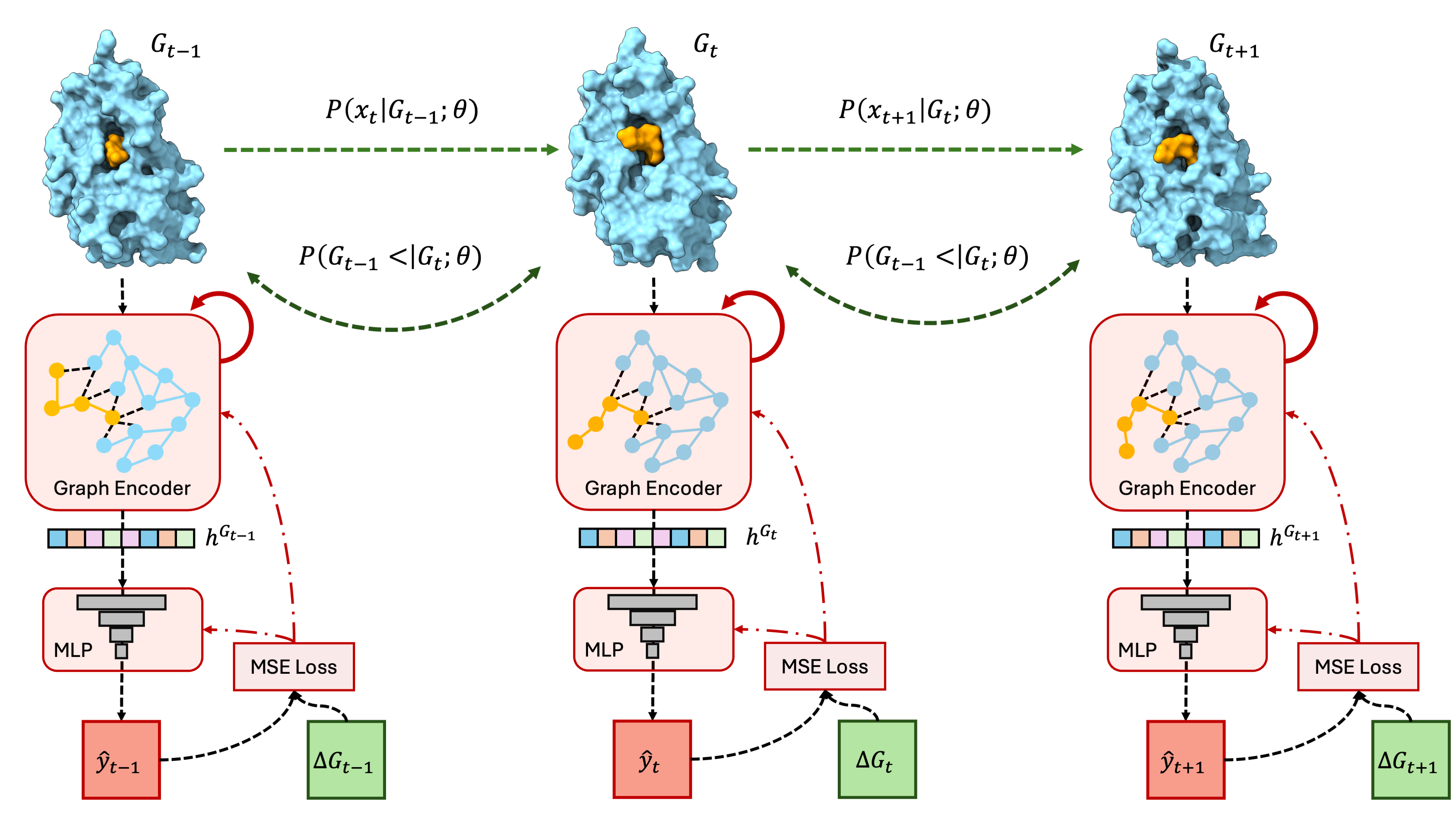}
    \caption{Schematic of the SurGBSA training pipeline. Input MD simulations (a) are given as input to the Graph Encoder network (GNN, EGNN, or EGMN) and to extract the graph embedding $h^{G_t}$ (eq. \ref{eq:avg_pool}). The $h^{G_t}$ vector is then given as input to an MLP network which produces a scalar $\hat{y_t}$ as output. The MSE loss is then computed using the corresponding MMGBSA value $\Delta G_t$. All weights in the encoder network and MLP receive gradient updates during training.}
    \label{fig:surgbsa_schem}
\end{figure}

\subsection{Learning a free-energy surrogate model from MD}\label{subsec:surgbsa_result}

We give the schematic of our SurGBSA approach in Figure~\ref{fig:surgbsa_schem} and give model training details in Section~\ref{subsec:model_train_optim}. We consider each frame of simulation as a unique graph instance $\mathcal{G}=\{\rmh^0, \rmx^0\}$ paired with a corresponding MMGBSA energy. We train a molecular encoder network with $\mathcal{L}$ layers to map the 3D atomic structure graph $\mathcal{G}$, using atom types and coordinates as inputs, to a set of atomic embeddings $\rmh^{\mathcal{L}}$ which are then average-pooled on the atomic dimension producing a graph summary vector, $\rmh^{\mathcal{G}}$. The vector $\rmh^{\mathcal{G}}$ is then given as input to a multilayer perceptron (MLP) network which projects $\rmh^{\mathcal{G}}$ to a scalar used to predict the MMGBSA energy. 

\begin{table}[ht]
    \centering
    \begin{tabular}{c|c|c}
        \textbf{Method} & 
        \textbf{Pearson} $r$ & \textbf{Spearman} $\rho$  \\ \hline

         GNN & .266 (.065) & .253 (.062) \\

         EGNN & .311 (.056) & .293 (.049) \\

         EGMN (Scratch) & .436 (.048) & .406 (.044) \\

         EGMN (Pretrain) & \textbf{.573} (.082) & \textbf{.547} (.078) \\

         \hline
         
                  EGMN (Best) & \textbf{.702} (.045) & \textbf{.680} (.043) \\
    \end{tabular}
        \caption{Test set metrics for each graph encoder network type used for SurGBSA model training. Each model is evaluated using the MD trajectories for the top 5 docking poses and their corresponding single-point MMGBSA (SP-GBSA) scores. Each metric is calculated separately for each PDB in the test set using all corresponding MD frames and is then averaged across all PDBs. The metrics are then averaged across the fold and seed dimensions. All models above the horizontal line are trained for 100 epochs using the MD simulations corresponding to the top 5 docking poses and the crystal structure, i.e. 6 simulations in total per PDB structure. Our results suggest that the EGMN model using the ProtMD~\cite{Wu2022-bw} pre-trained checkpoint achieves the best performance in replicating the single-point MMGBSA score. We then chose to run the pre-trained EGMN model for an additional 500 epochs of training which further improves the Pearson $r$ and Spearman $\rho$ by 22.5\% and 24.5\% respectively.}
    \label{tab:spgbsa_pred}
\end{table}

\begin{figure}[ht]
    \centering
    \includegraphics[width=\linewidth]{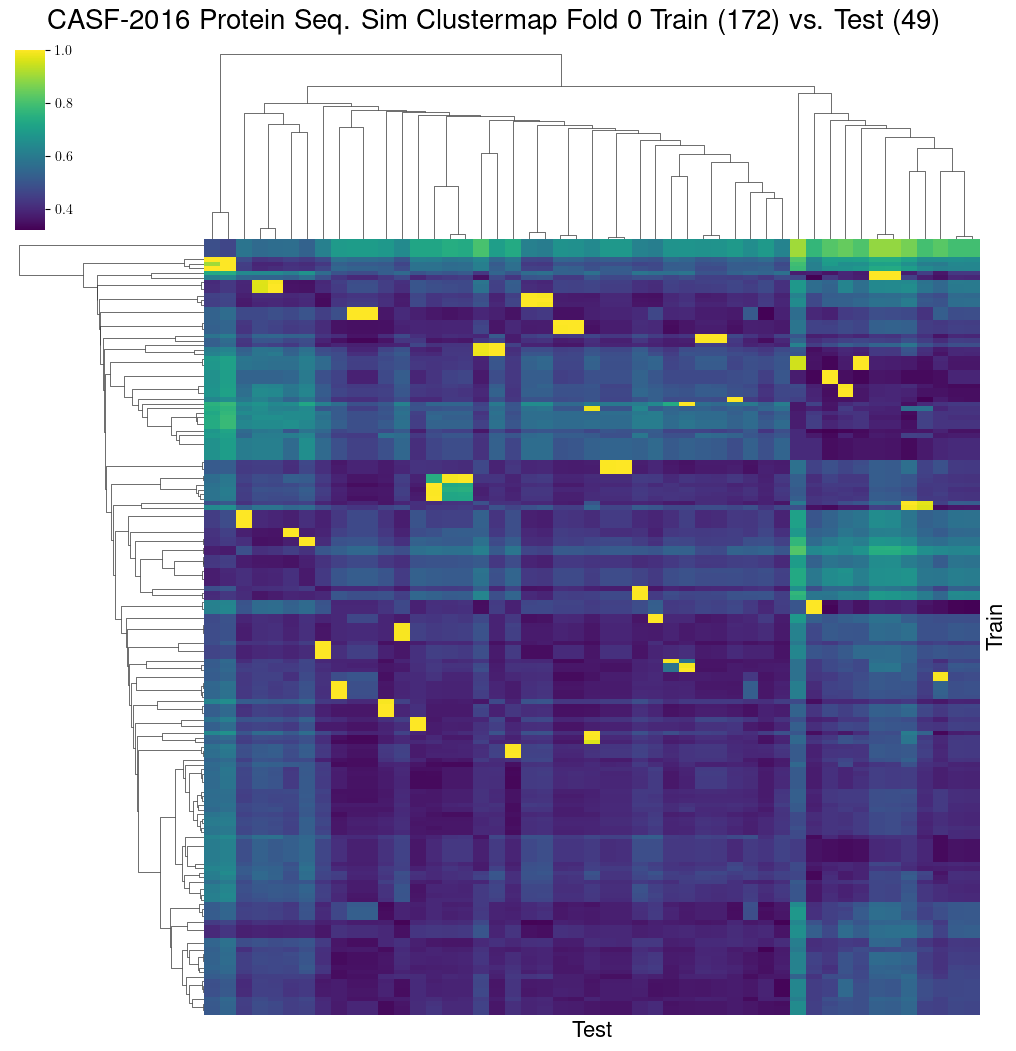}
    \caption{Clustered heatmap of the PDB structures contained within the training set (columns) versus the PDB structures contained within the test set (rows). Rows and columns are clustered according to protein sequence similarity. Fold 0.}
    \label{fig:heatmap_coremd_train_vs_coremd_test_seqsim_0}
\end{figure}

\begin{figure}[ht]
    \centering
    \includegraphics[width=\linewidth]{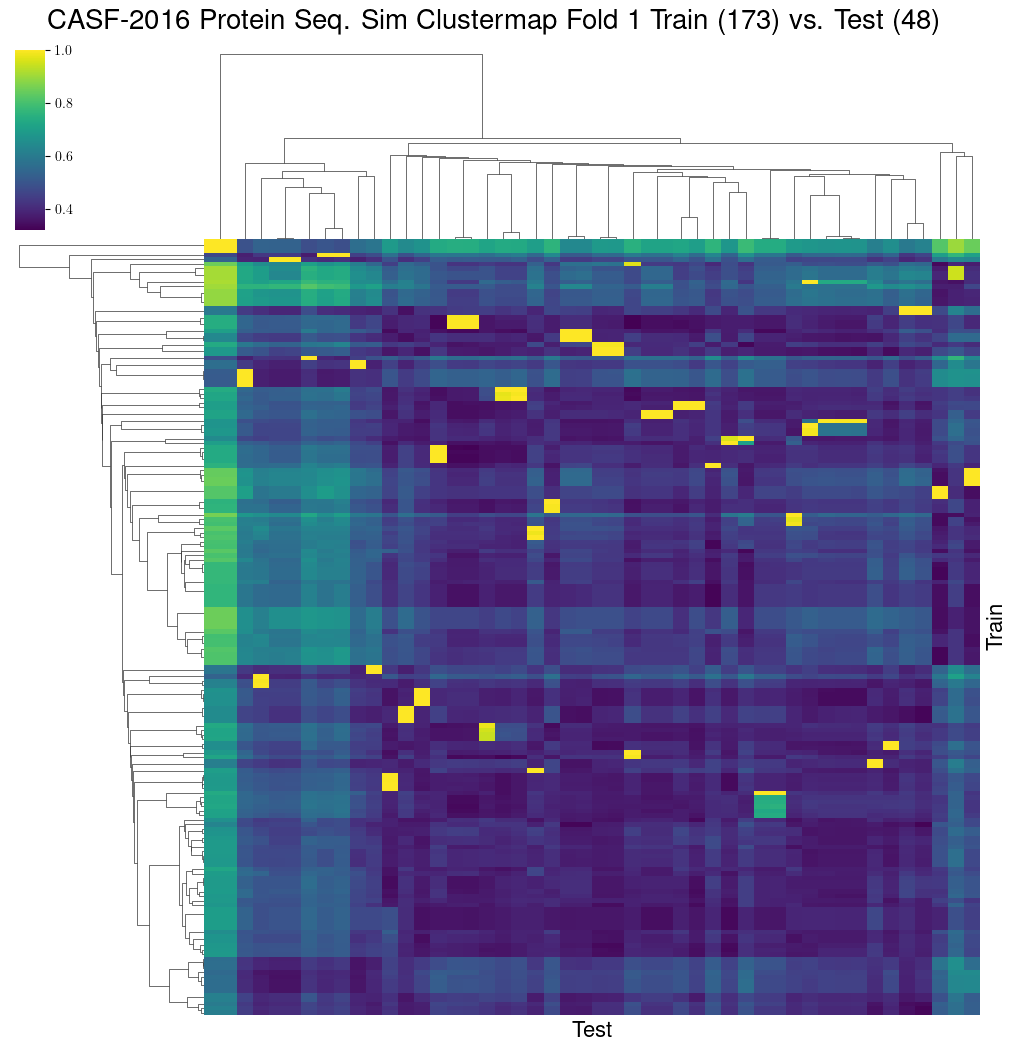}
    \caption{Clustered heatmap of the PDB structures contained within the training set (columns) versus the PDB structures contained within the test set (rows). Rows and columns are clustered according to protein sequence similarity. Fold 1.}
    \label{fig:heatmap_coremd_train_vs_coremd_test_seqsim_1}
\end{figure}

\begin{figure}[ht]
    \centering
    \includegraphics[width=\linewidth]{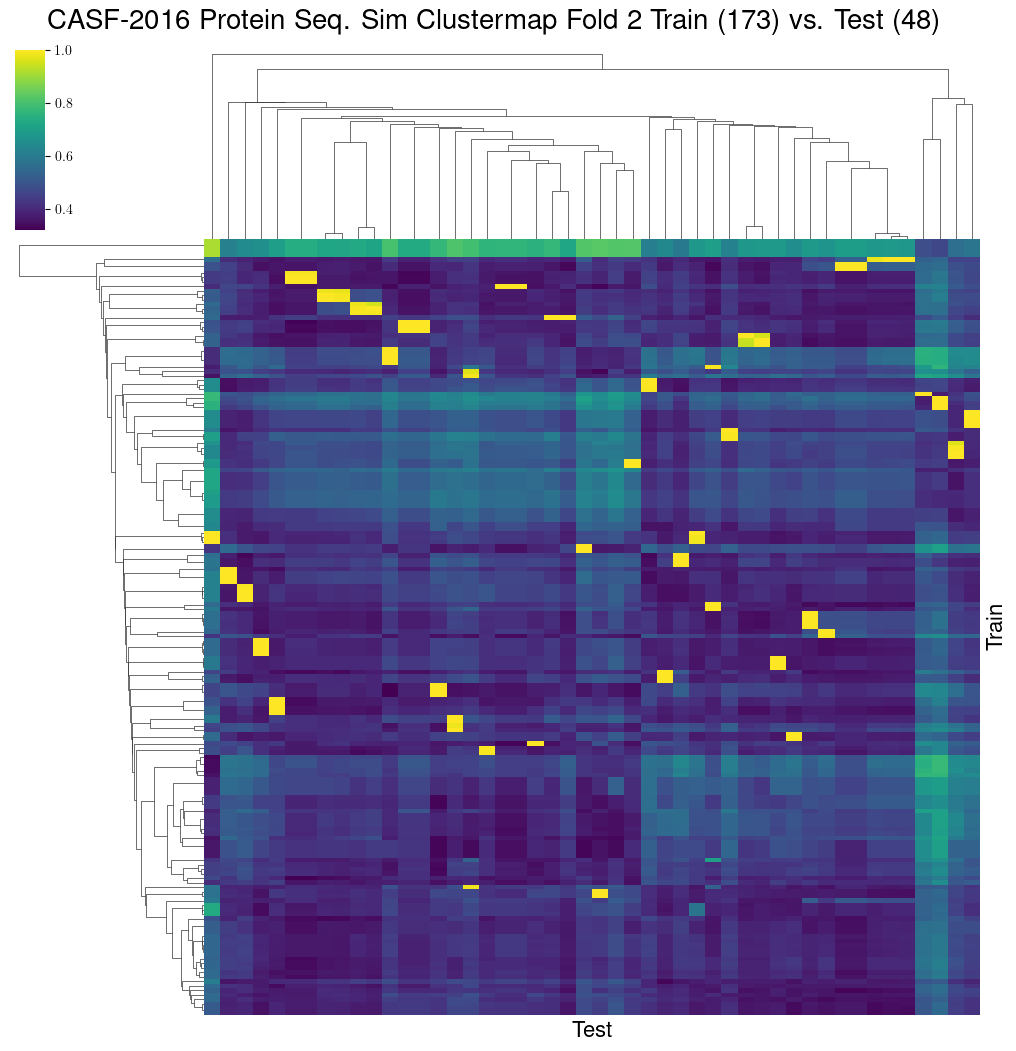}
    \caption{Clustered heatmap of the PDB structures contained within the training set (columns) versus the PDB structures contained within the test set (rows). Rows and columns are clustered according to protein sequence similarity. Fold 2.}
    \label{fig:heatmap_coremd_train_vs_coremd_test_seqsim_2}
\end{figure}

\begin{figure}[ht]
    \centering
    \includegraphics[width=\linewidth]{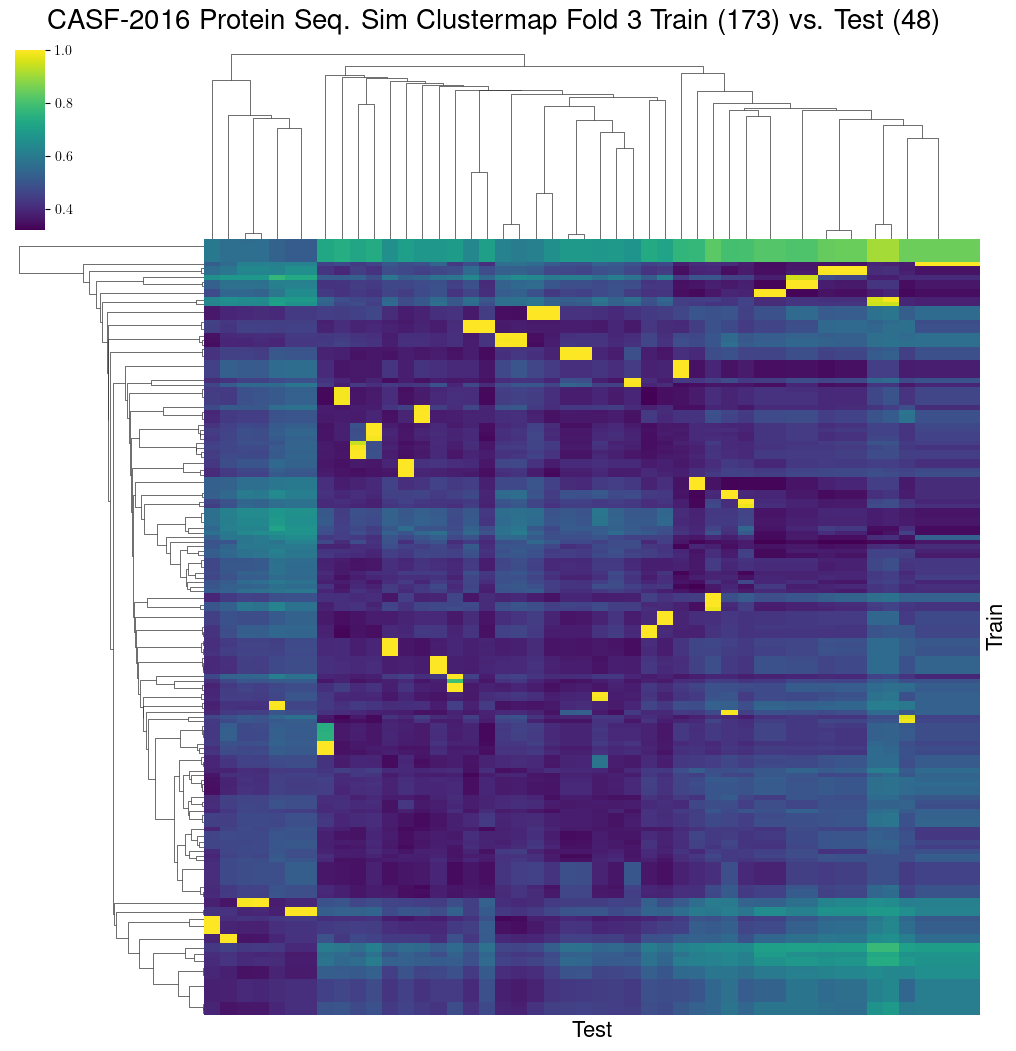}
    \caption{Clustered heatmap of the PDB structures contained within the training set (columns) versus the PDB structures contained within the test set (rows). Rows and columns are clustered according to protein sequence similarity. Fold 3.}
    \label{fig:heatmap_coremd_train_vs_coremd_test_seqsim_3}
\end{figure}

\begin{figure}[ht]
    \centering
    \includegraphics[width=\linewidth]{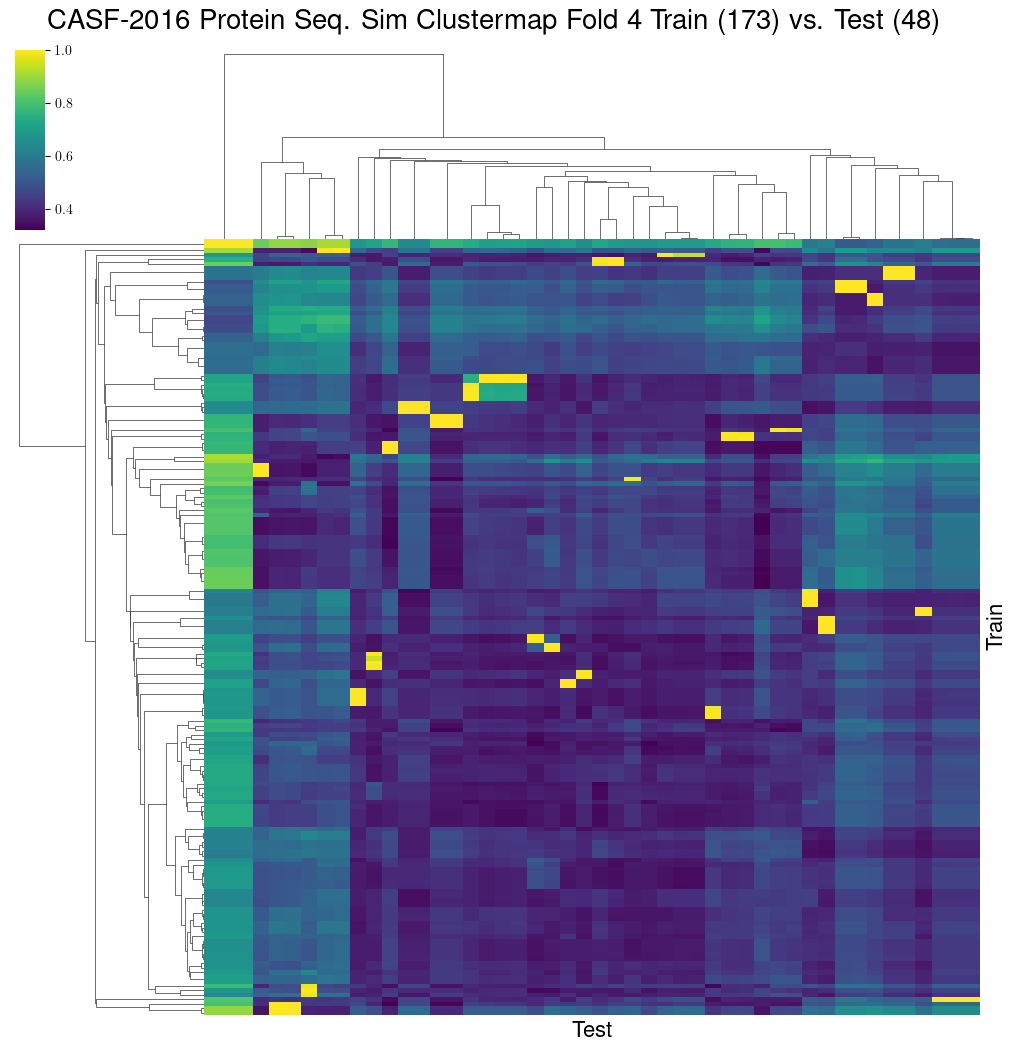}
    \caption{Clustered heatmap of the PDB structures contained within the training set (columns) versus the PDB structures contained within the test set (rows). Rows and columns are clustered according to protein sequence similarity. Fold 4.}
    \label{fig:heatmap_coremd_train_vs_coremd_test_seqsim_4}
\end{figure}

\begin{figure}[ht]
    \centering
    \includegraphics[width=\linewidth]{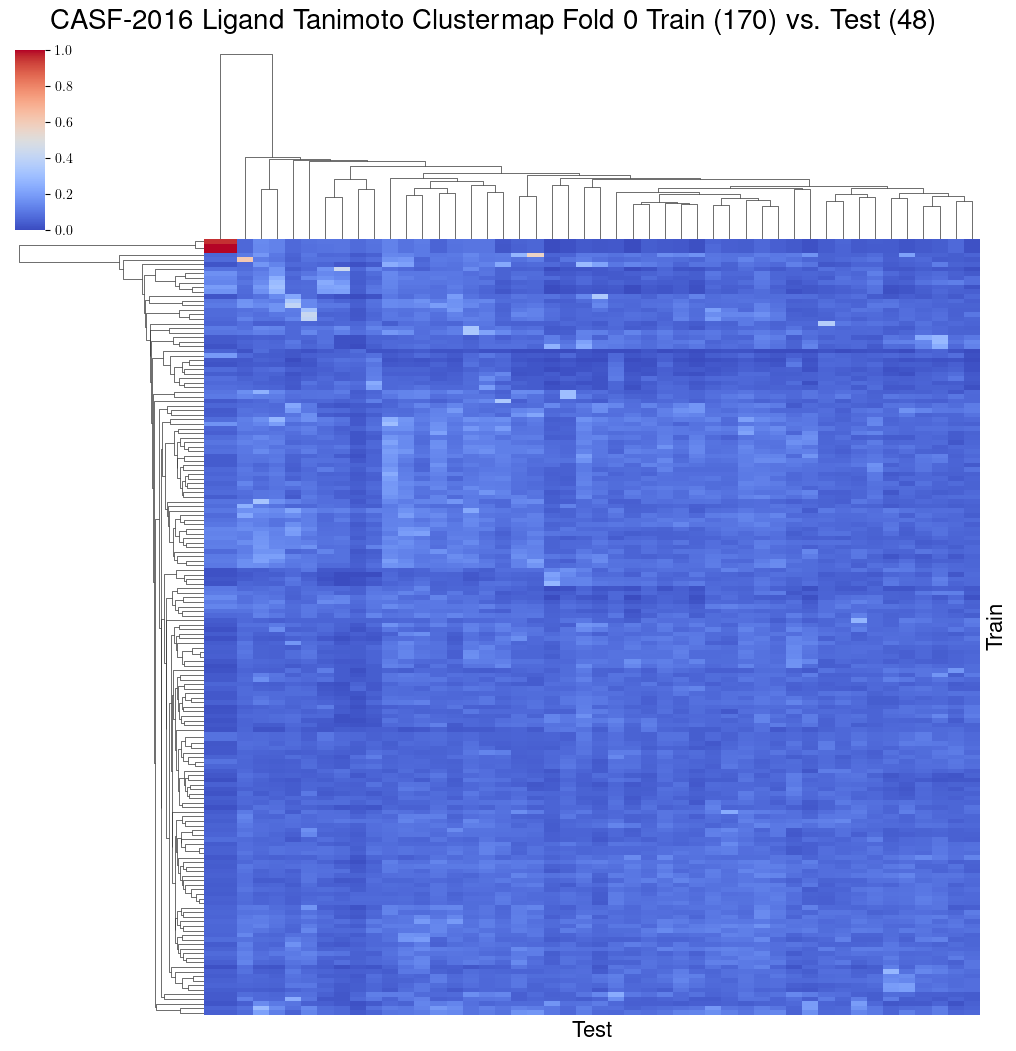}
    \caption{Clustered heatmap of the ligands from each PDB structure in the training set (rows) versus the ligands from the test set (columns). ECFP fingerprints are computed using RDKit~\cite{RDKit} with length 2048 and radius of 2. Fold 0.}
    \label{fig:heatmap_coremd_train_vs_coremd_test_tani_0}
\end{figure}

\begin{figure}[ht]
    \centering
    \includegraphics[width=\linewidth]{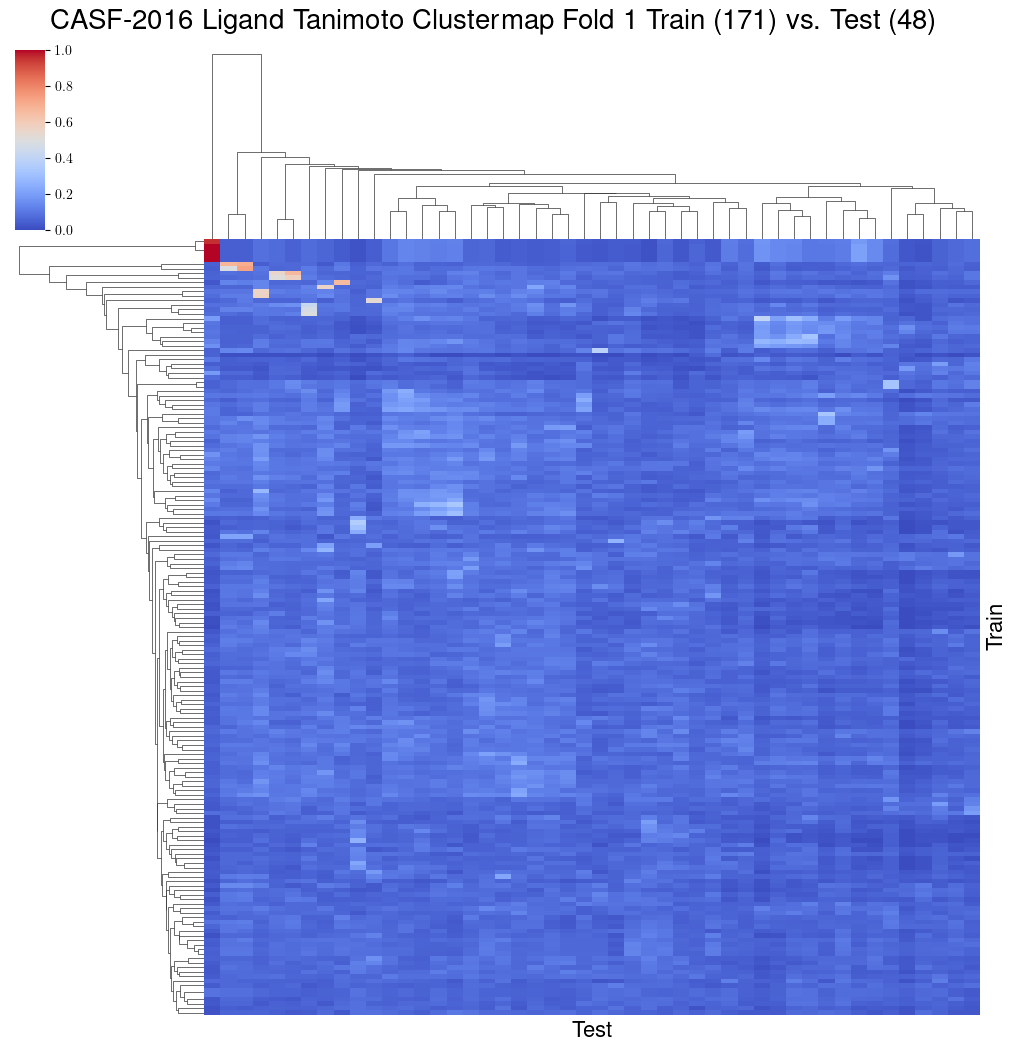}
    \caption{Clustered heatmap of the ligands from each PDB structure in the training set (rows) versus the ligands from the test set (columns). ECFP fingerprints are computed using RDKit~\cite{RDKit} with length 2048 and radius of 2. Fold 1.}
    \label{fig:heatmap_coremd_train_vs_coremd_test_tani_1}
\end{figure}

\begin{figure}[ht]
    \centering
    \includegraphics[width=\linewidth]{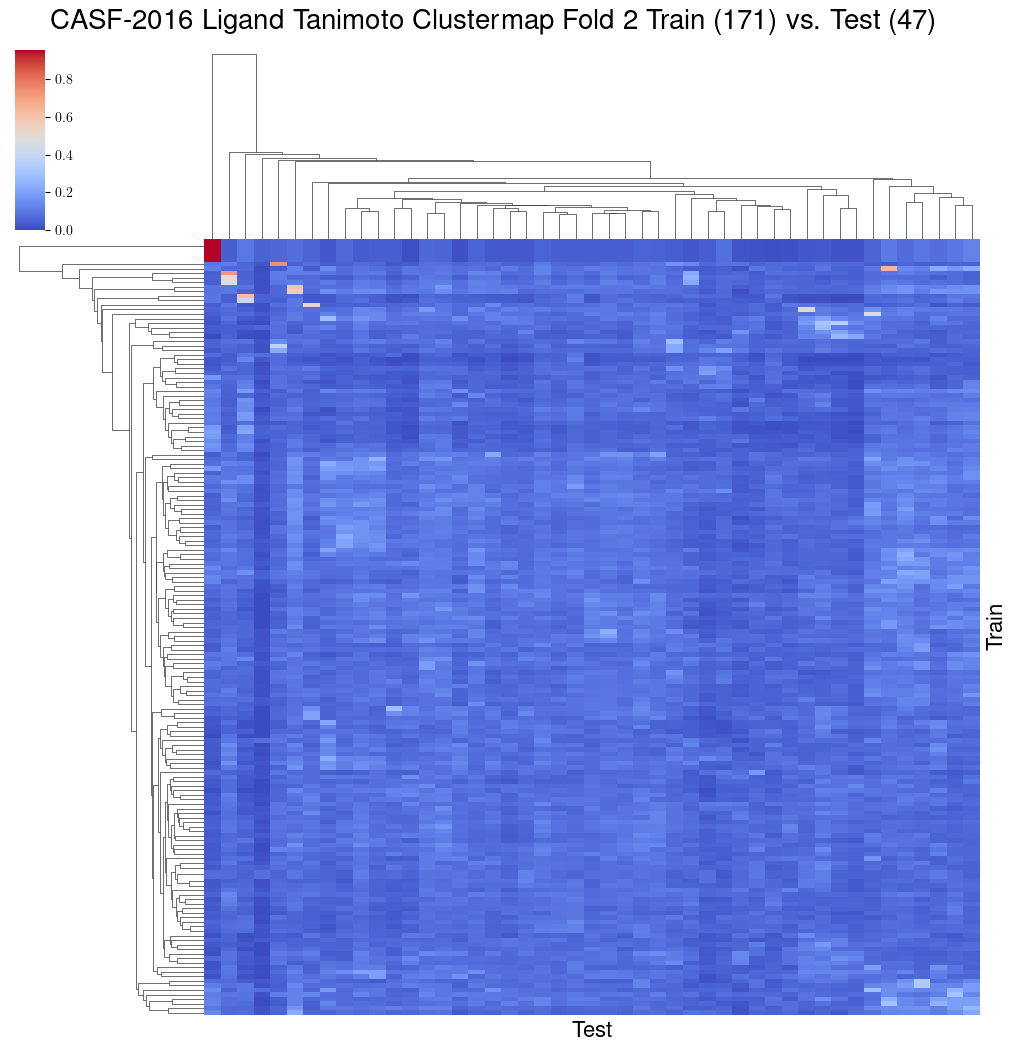}
    \caption{Clustered heatmap of the ligands from each PDB structure in the training set (rows) versus the ligands from the test set (columns). ECFP fingerprints are computed using RDKit~\cite{RDKit} with length 2048 and radius of 2. Fold 2.}
    \label{fig:heatmap_coremd_train_vs_coremd_test_tani_2}
\end{figure}

\begin{figure}[ht]
    \centering
    \includegraphics[width=\linewidth]{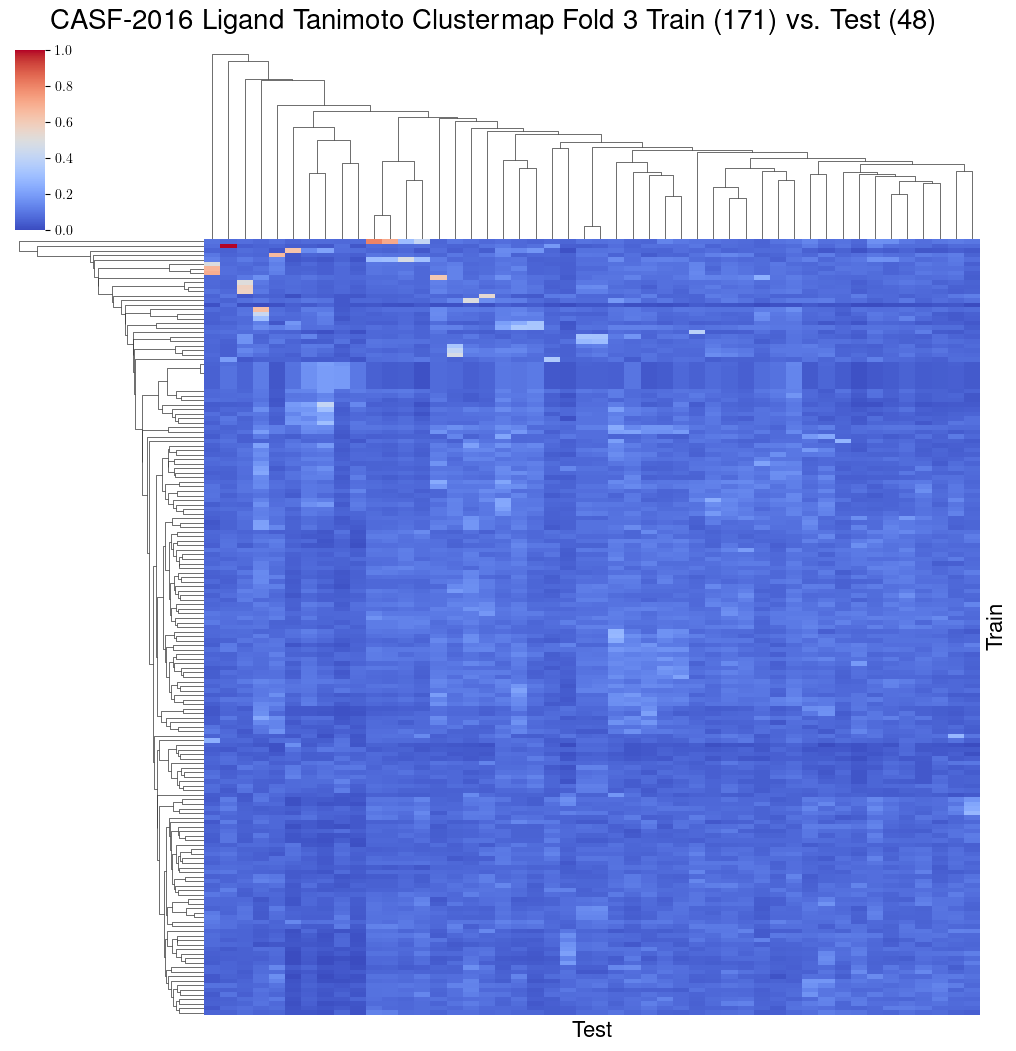}
    \caption{Clustered heatmap of the ligands from each PDB structure in the training set (rows) versus the ligands from the test set (columns). ECFP fingerprints are computed using RDKit~\cite{RDKit} with length 2048 and radius of 2. Fold 3.}
\label{fig:heatmap_coremd_train_vs_coremd_test_tani_3}
\end{figure}

\begin{figure}[ht]
    \centering
    \includegraphics[width=\linewidth]{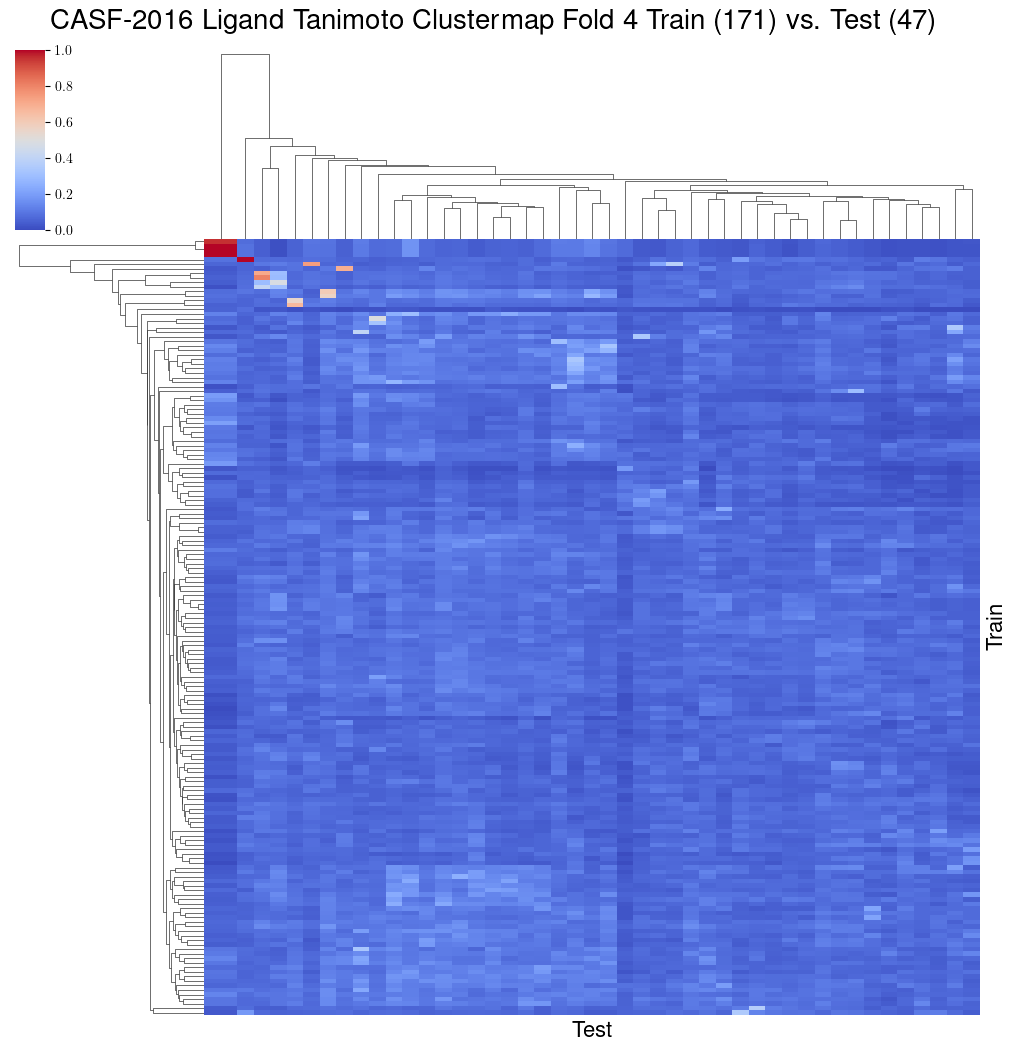}
    \caption{Clustered heatmap of the ligands from each PDB structure in the training set (rows) versus the ligands from the test set (columns). ECFP fingerprints are computed using RDKit~\cite{RDKit} with length 2048 and radius of 2. Fold 4.}
    \label{fig:heatmap_coremd_train_vs_coremd_test_tani_4}
\end{figure}

\begin{figure}[ht]
    \centering
    \includegraphics[width=\linewidth]{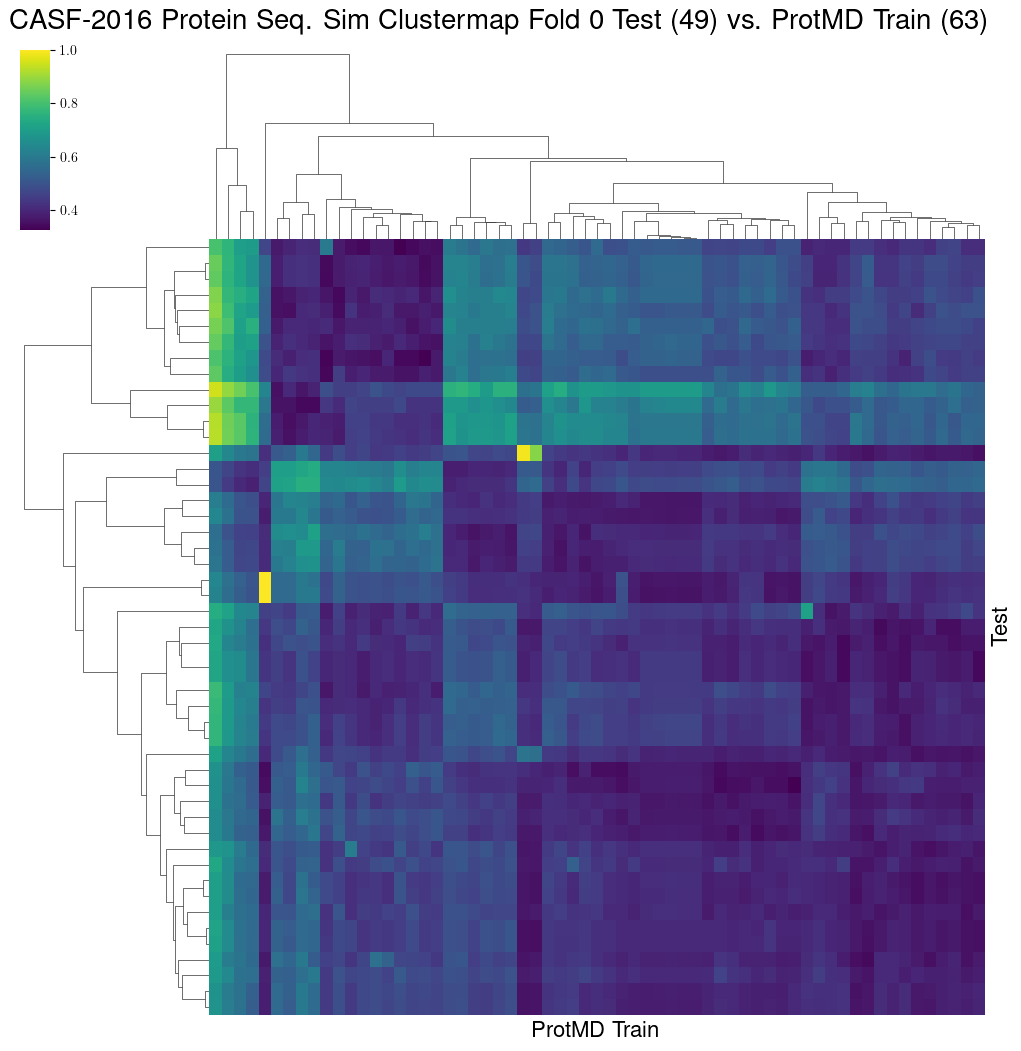}
    \caption{Clustered heatmap of the PDB structures contained with the ProtMD~\cite{Wu2022-bw} training set (columns) versus the PDB structures contained within each of the KFold split's test set (rows). Rows and columns are clustered according to protein sequence similarity. Fold 0.}
    \label{fig:heatmap_protmd_vs_coremd_seqsim_0}
\end{figure}

\begin{figure}[ht]
    \centering
    \includegraphics[width=\linewidth]{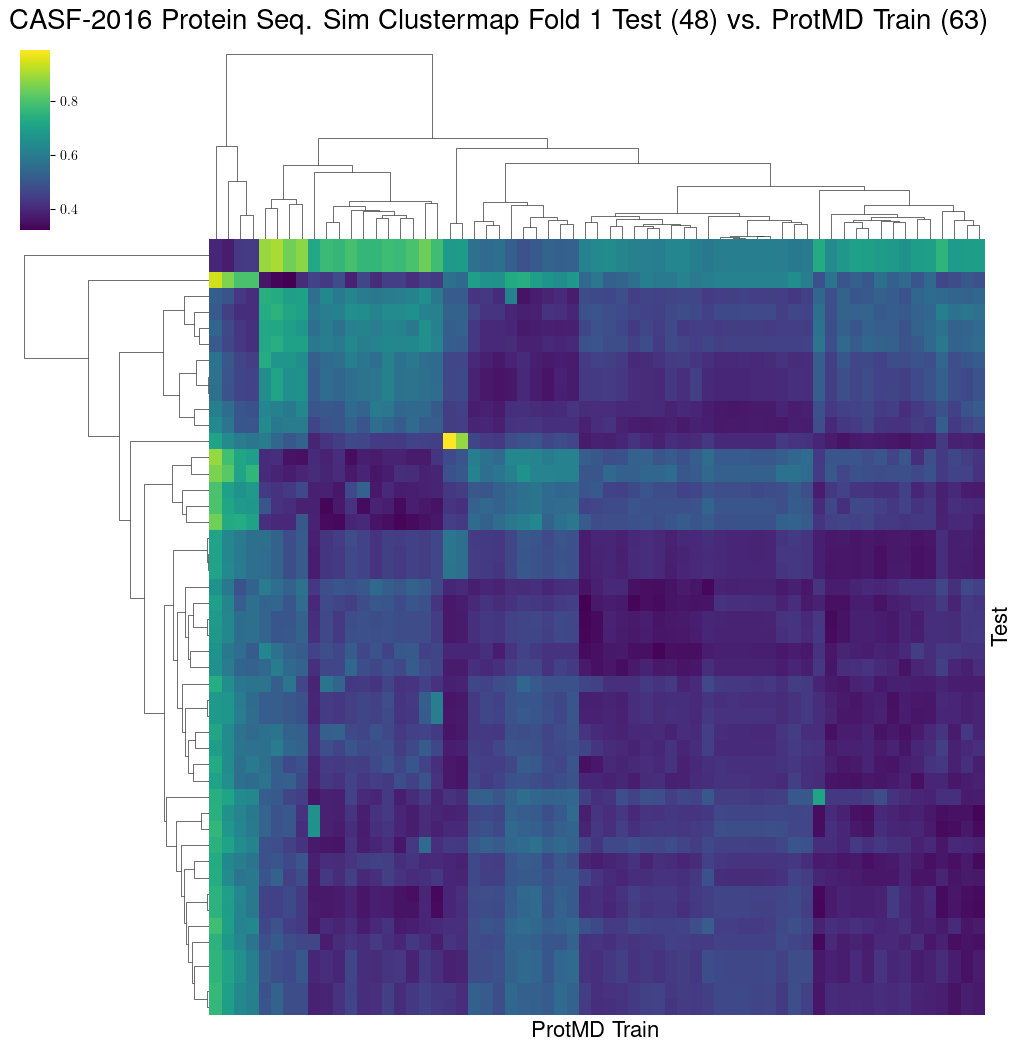}
    \caption{Clustered heatmap of the PDB structures contained with the ProtMD~\cite{Wu2022-bw} training set (columns) versus the PDB structures contained within each of the KFold split's test set (rows). Rows and columns are clustered according to protein sequence similarity. Fold 1.}
    \label{fig:heatmap_protmd_vs_coremd_seqsim_1}
\end{figure}

\begin{figure}[ht]
    \centering
    \includegraphics[width=\linewidth]{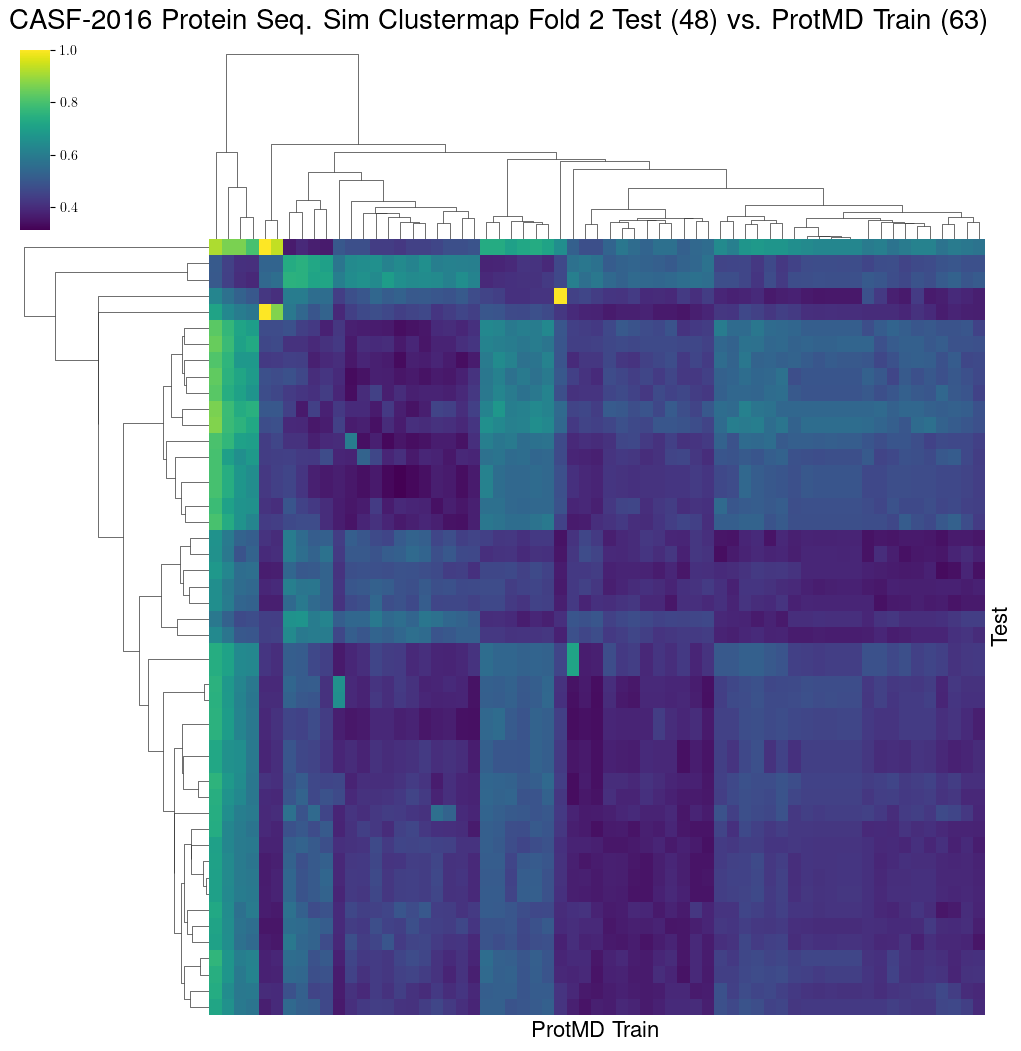}
    \caption{Clustered heatmap of the PDB structures contained with the ProtMD~\cite{Wu2022-bw} training set (columns) versus the PDB structures contained within each of the KFold split's test set (rows). Rows and columns are clustered according to protein sequence similarity. Fold 2.}
    \label{fig:heatmap_protmd_vs_coremd_seqsim_2}
\end{figure}

\begin{figure}[ht]
    \centering
    \includegraphics[width=\linewidth]{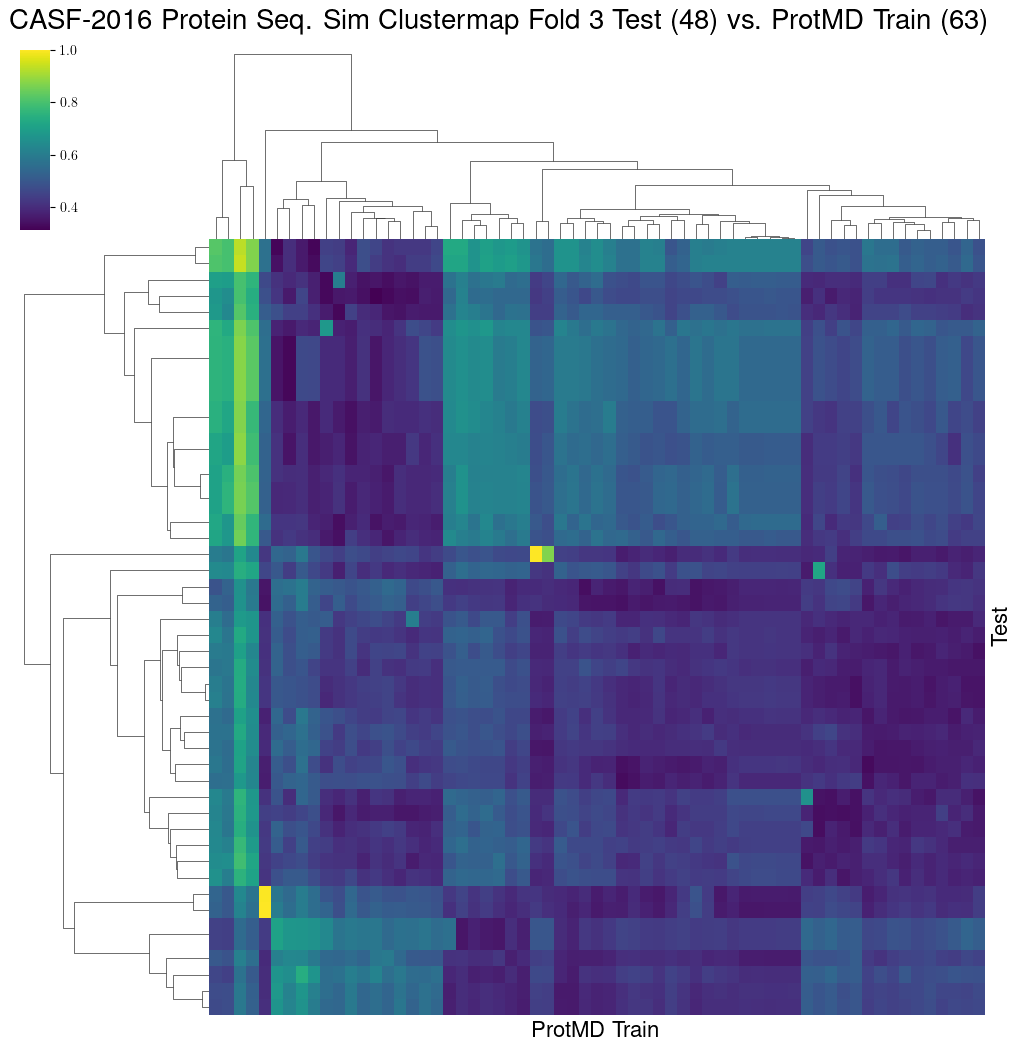}
    \caption{Clustered heatmap of the PDB structures contained with the ProtMD~\cite{Wu2022-bw} training set (columns) versus the PDB structures contained within each of the KFold split's test set (rows). Rows and columns are clustered according to protein sequence similarity. Fold 3.}
    \label{fig:heatmap_protmd_vs_coremd_seqsim_3}
\end{figure}

\begin{figure}[ht]
    \centering
    \includegraphics[width=\linewidth]{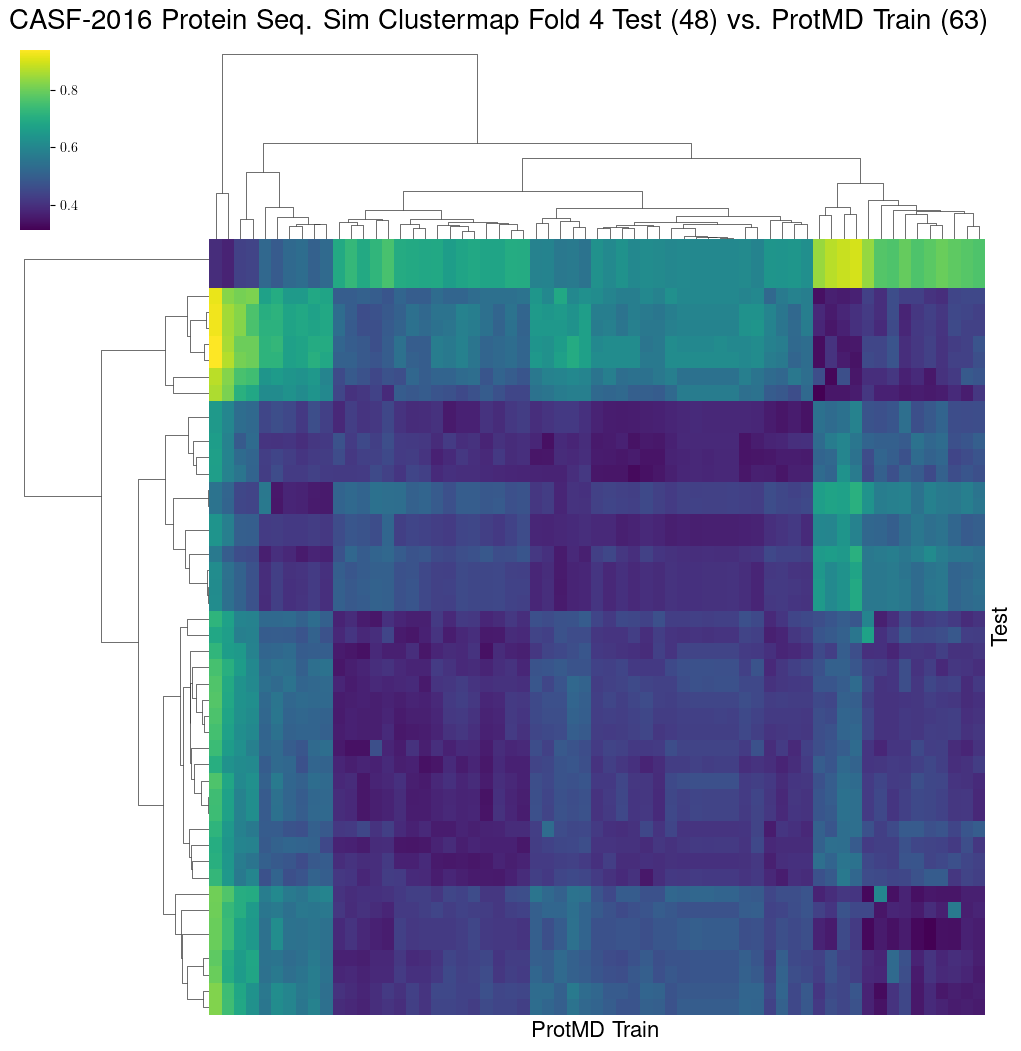}
    \caption{Clustered heatmap of the PDB structures contained with the ProtMD~\cite{Wu2022-bw} training set (columns) versus the PDB structures contained within each of the KFold split's test set (rows). Rows and columns are clustered according to protein sequence similarity. Fold 4.}
    \label{fig:heatmap_protmd_vs_coremd_seqsim_4}
\end{figure}

In Table \ref{tab:spgbsa_pred} we give the results for the SurGBSA model trained with each molecular encoder network. For fair comparison, we constrained each model to the same training PDB structure sets and use all available MD trajectories corresponding to the top 5 docking poses and the crystal structure. Our results show that the EGMN network, pretrained on molecular dynamics data~\cite{Wu2022-bw}, achieves the best overall performance, by a significant margin. Based on this observation, we extended the training of this model for an additional 500 epochs. Extended training significantly improves both the Pearson and Spearman correlations by 22.5\% and 24.5\% respectively. 
To account for possible information leakage of similar structures between each training and test set (5 in total), we computed the protein sequence identity between the PDB structures in each pair of splits (SI Figures~\ref{fig:heatmap_coremd_train_vs_coremd_test_seqsim_0}-\ref{fig:heatmap_coremd_train_vs_coremd_test_seqsim_4}.  While we observed a few instances of high similarity, the vast majority of training-testing pairs indicate lower sequence similarity. The high similarity in this context is also not accounting for the 3D conformations of the structure pairs. Additionally, we considered the Tanimoto similarity of the ligands for each PDB structure pair in the respective training/test splits (SI Figures~\ref{fig:heatmap_coremd_train_vs_coremd_test_tani_0}-\ref{fig:heatmap_coremd_train_vs_coremd_test_tani_4}). Our findings indicate persistent low similarity with only a small number of pairs with high similarity. ProtMD~\cite{Wu2022-bw} is a model trained to predict MD trajectory ordering and molecular movement between two time steps.  The model is evaluated as an additional pre-training step and we considered the possibility of data leakage from the ProtMD pretraining step (SI Figures~\ref{fig:heatmap_protmd_vs_coremd_seqsim_0}-\ref{fig:heatmap_protmd_vs_coremd_seqsim_4}). A low overall protein sequence similarity is observed between each of our test sets and the ProtMD pretraining set provided by the authors~\cite{Wu2022-bw}. This affirms the limited structure redundancy in the MD dataset, by leveraging the structure-similarity analysis used to produce the CASF set of PDBBind~\cite{Su2019-ni}, which is effective in representative sub-sampling of the available high-quality structures.

\begin{figure}[ht]
    \centering
   \includegraphics[width=\linewidth]{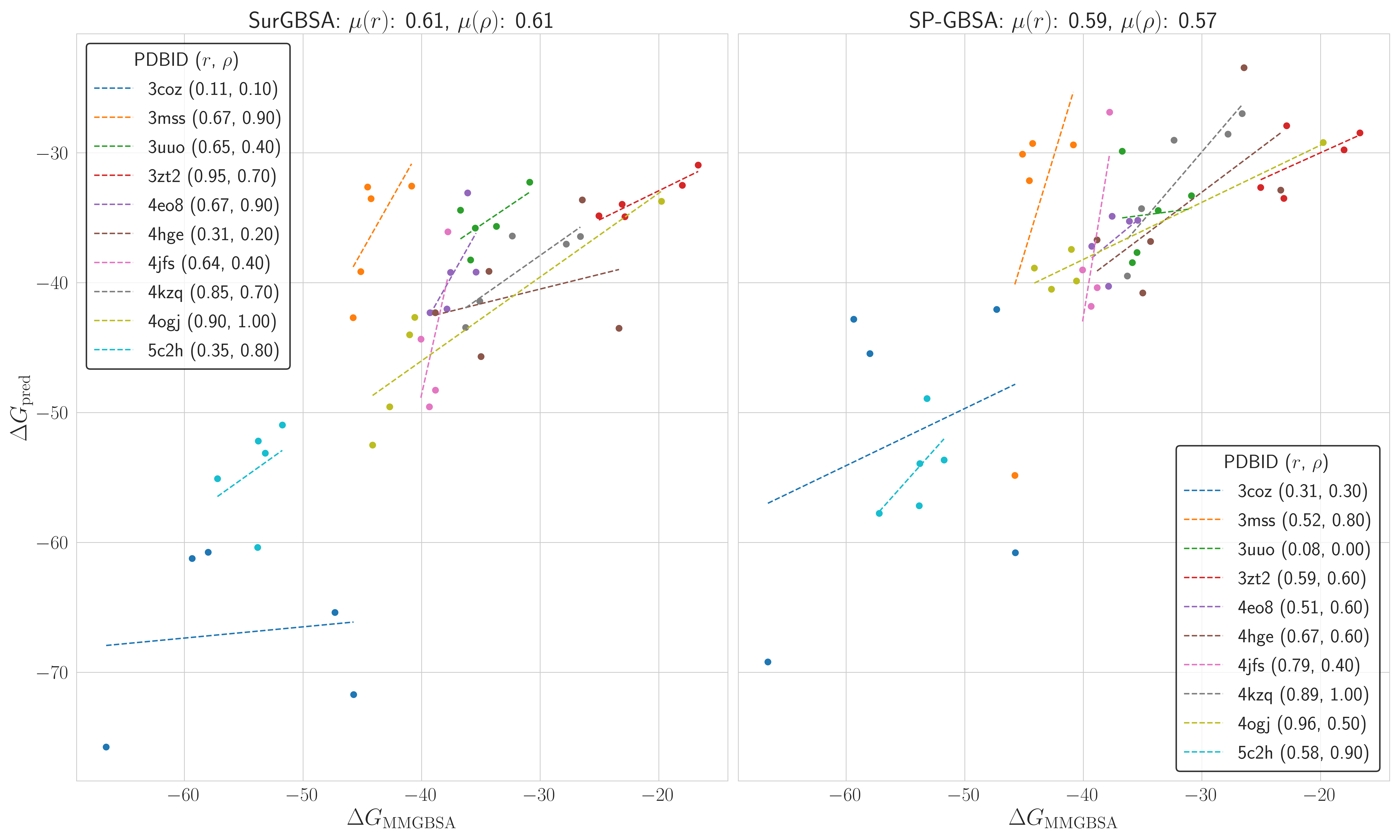}
    \caption{Scatterplots of the SurGBSA model and the single-point MMGBSA (SP-GBSA) calculation. The Pearson $r$ and Spearman $\rho$ are given for each approach using the top 5 docking poses of each PDB entry, averaged over all PDB entries. We randomly selected 2 PDB entries from each of the 5 folds, 10 in total. The difference in the metric values are fairly small, suggesting that SurGBSA is able to effectively replicate the predictions given by the single point MMGBSA calculation, at a fraction of the computational cost (Table \ref{tab:mmgbsa_cost}).}
    \label{fig:rescoring_scatter}
\end{figure}

In Figure~\ref{fig:rescoring_scatter}, we evaluated our best SurGBSA model and the single-point MMGBSA baseline (SP-GBSA) for accurate docking pose ranking using 10 randomly selected PDB structures. We use the ensemble-average MMGBSA as the ground truth for both methods. We note a high degree of agreement for each set of the trend lines between the SP-GBSA baseline and our SurGBSA model. In several cases, 3/10 for Pearson and 5/10 for Spearman, the ranking produced by SurGBSA improves upon the ranking according to SP-GBSA. Further, when considering the metrics averaged over the PDB structures, SurGBSA shows a slight improvement over the performance of the SP-GBSA baseline which was used as the training target. We believe that this may be attributed to the model effectively learning generalizable features across heterogeneous PDB structures and their conformational dynamics coupled with training on a large collection of MMGBSA energetics that may enable accurate interpolation for test structures supported by a robust representation.

\begin{figure}[ht]
        \centering
        \includegraphics[width=\linewidth]{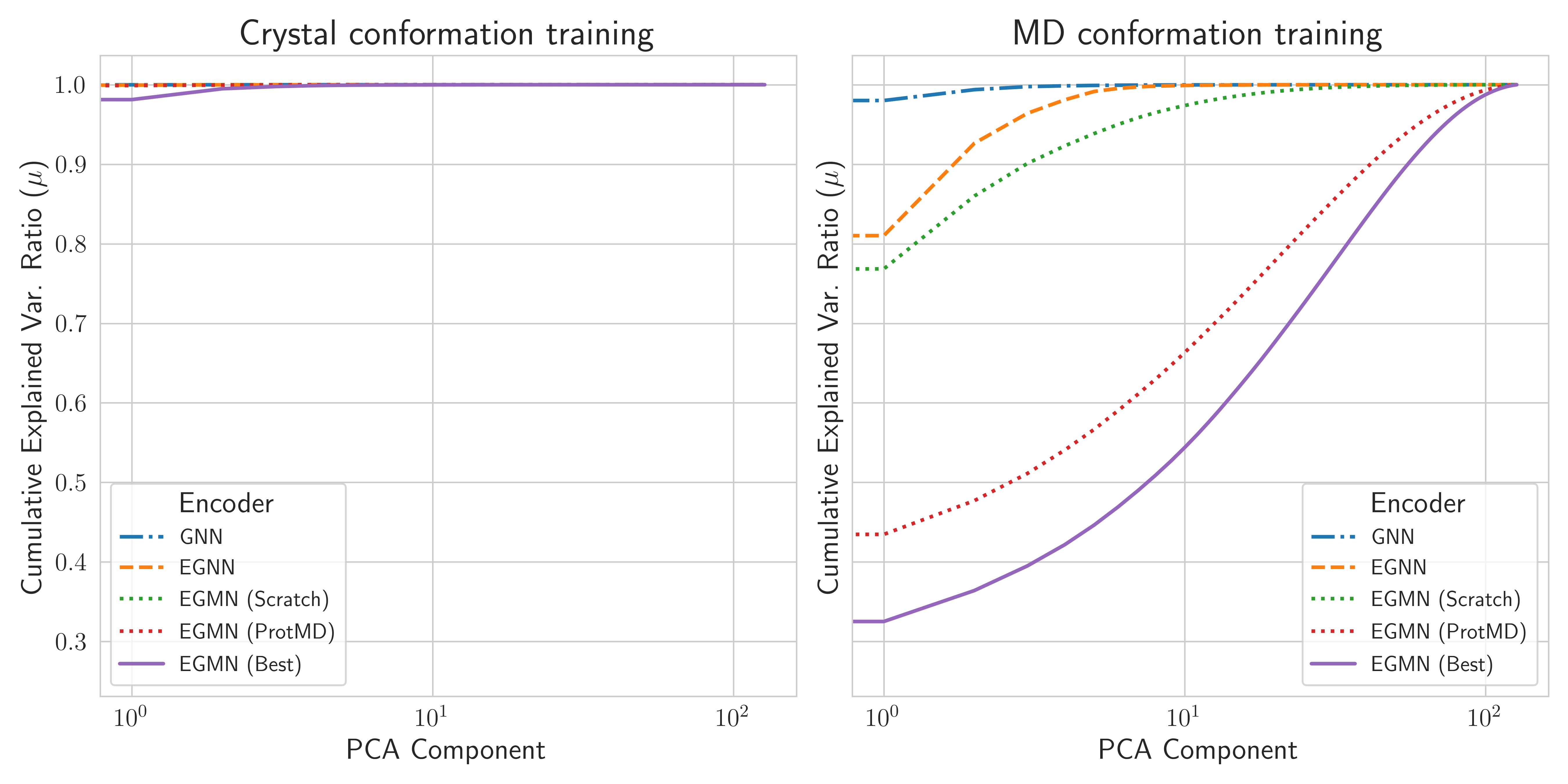}
        \caption{Cumulative explained variance ratios of the PCA components corresponding to the extracted features of each model considered in our work. The docking pose structures are used as input to each encoder considered in this work and the output latent vectors are used as the basis for solving the PCA components. We compute the PCA components for each seed in each fold independently and average the explained variance ratios, represented by each line. The results for the encoder networks trained only using the crystal structure conformation are shown on the left and the encoder networks trained using all MD sampled conformations using all available docking and crystal structure poses are shown on the right. The PCA components for the MD trajectories are illustrated in SI Figures \ref{fig:pca_pdbid_gnn_scratch_md}-\ref{fig:pca_y_pred_egmn_best_md}.}
        \label{fig:pca_scree_plot}
    \end{figure}

\begin{figure}[ht]
    \centering
    \includegraphics[width=\linewidth]{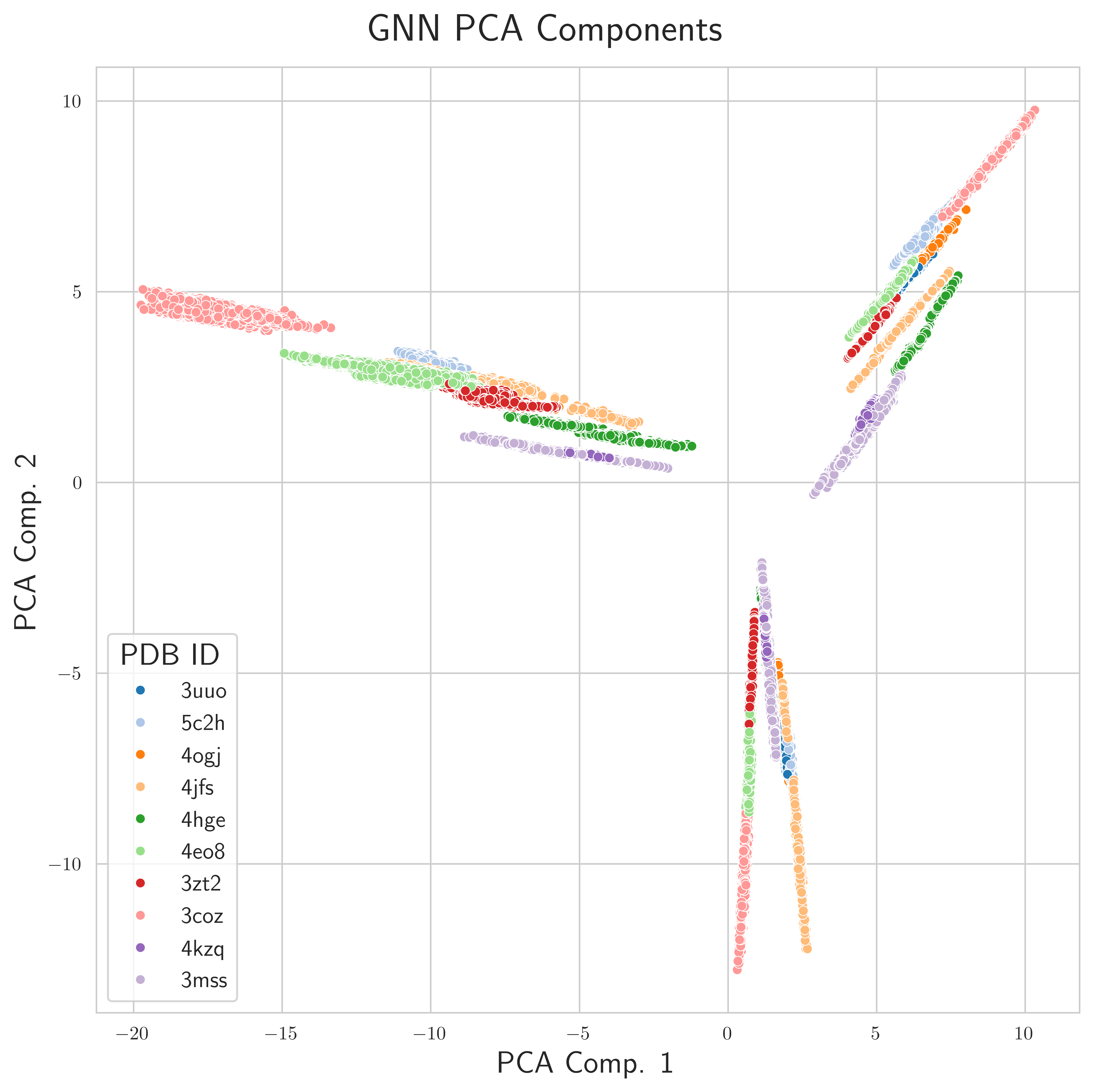}
    \caption{Scatter plot of the first two principal components calculated from the latent embeddings $h_G$ extracted from the GNN encoder, trained from scratch for a maximum of 100 epochs, for all test set MD frames corresponding to the top 5 docking poses. Points are colored according to the PDB ID of the input co-complex. Results are shown for 2 randomly selected PDB entries from each of the 5 test sets, 10 in total.}
    \label{fig:pca_pdbid_gnn_scratch_md}
\end{figure}

\begin{figure}[ht]
    \centering
    \includegraphics[width=\linewidth]{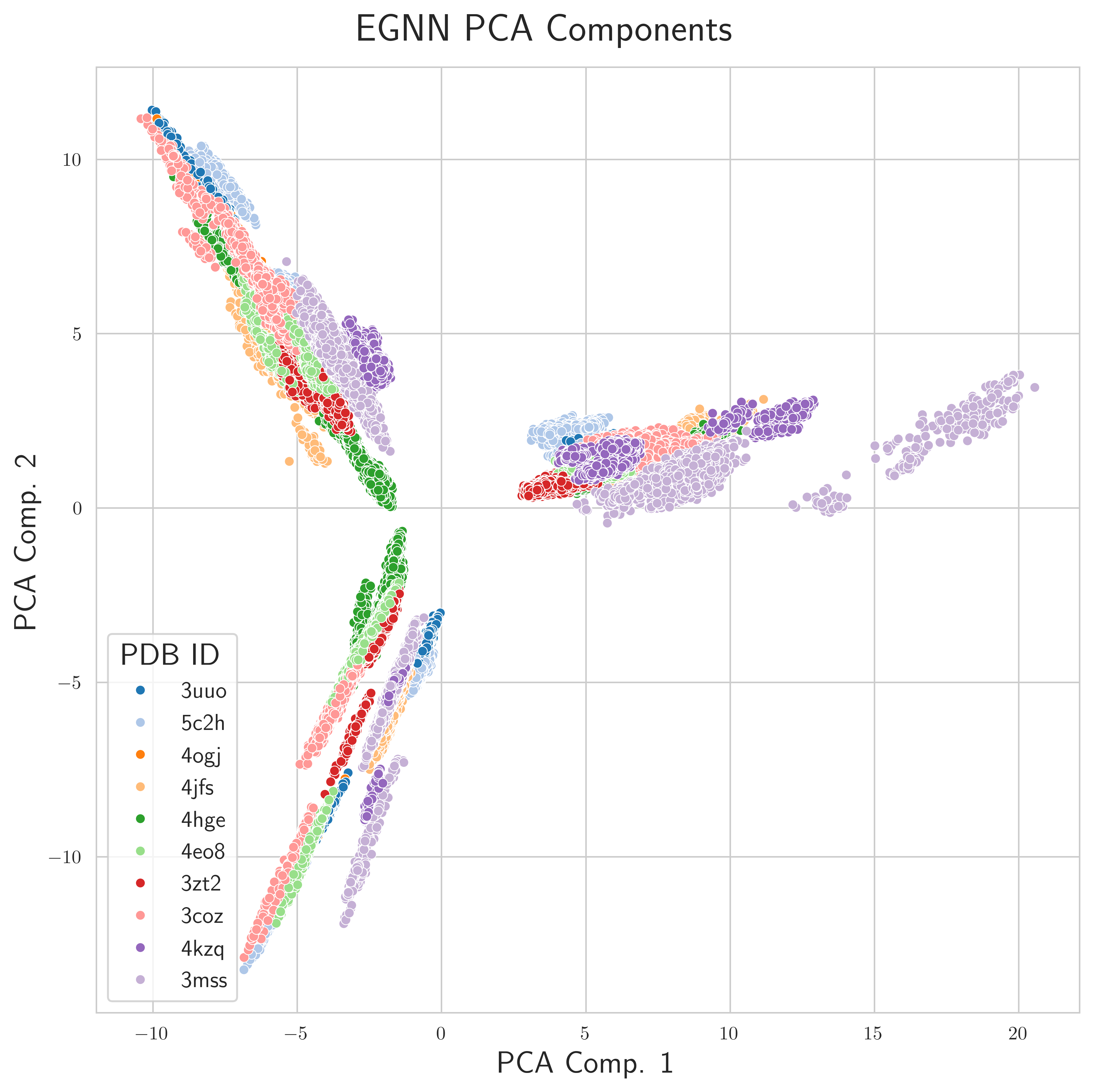}
    \caption{Scatter plot of the first two principal components calculated from the latent embeddings $h_G$ extracted from the EGNN encoder, trained from scratch for a maximum of 100 epochs, for all test set MD frames corresponding to the top 5 docking poses. Points are colored according to the PDB ID of the input co-complex. Results are shown for 2 randomly selected PDB entries from each of the 5 test sets, 10 in total.}
    \label{fig:pca_pdbid_egnn_scratch_md}
\end{figure}

\begin{figure}[ht]
    \centering
    \includegraphics[width=\linewidth]{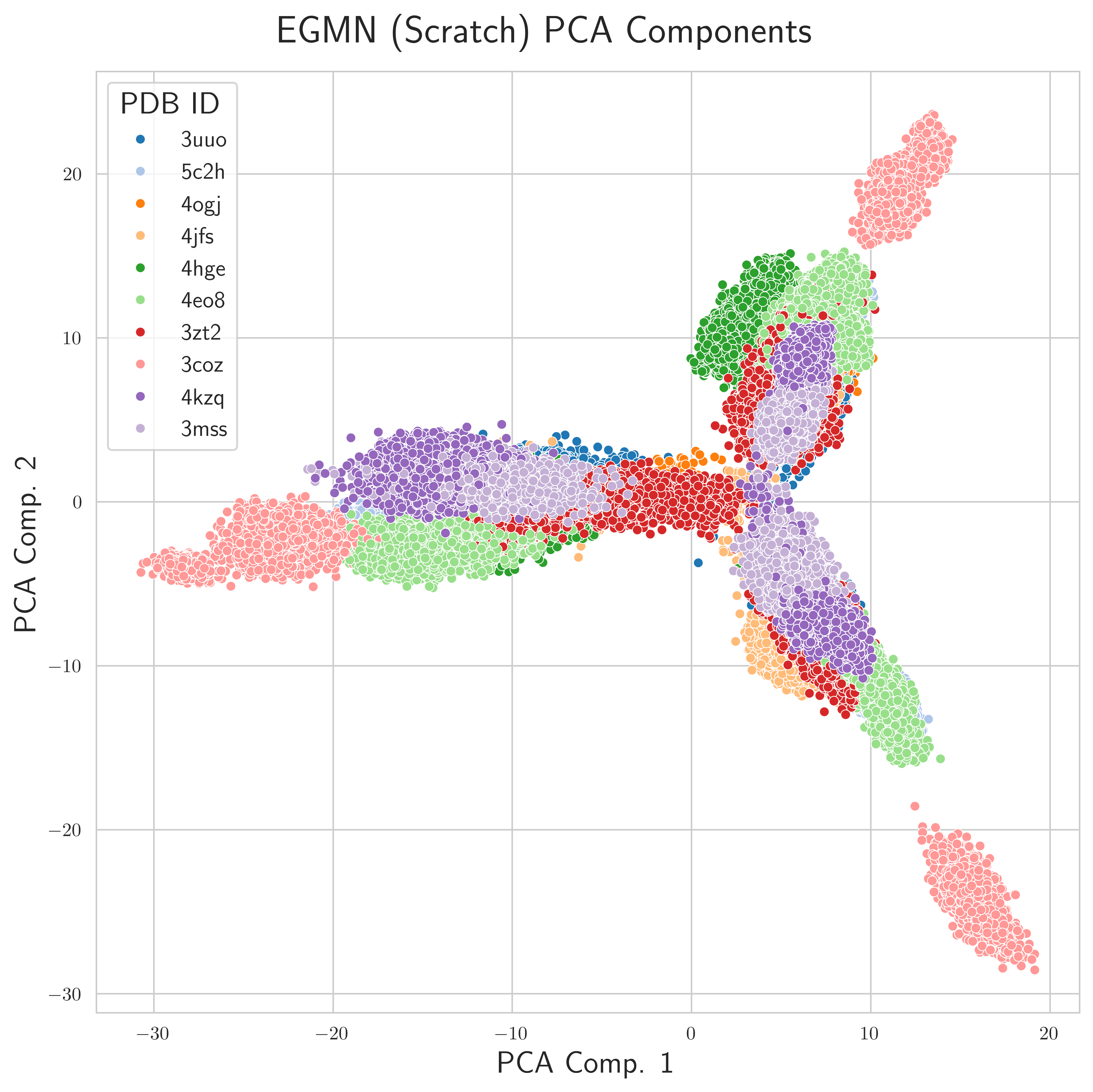}
    \caption{Scatter plot of the first two principal components calculated from the latent embeddings $h_G$ extracted from the EGMN encoder, trained from scratch for a maximum of 100 epochs, for all test set MD frames corresponding to the top 5 docking poses. Points are colored according to the PDB ID of the input co-complex. Results are shown for 2 randomly selected PDB entries from each of the 5 test sets, 10 in total.}
    \label{fig:pca_egmn_scratch_pdbid_md}
\end{figure}

\begin{figure}[ht]
    \centering
    \includegraphics[width=\linewidth]{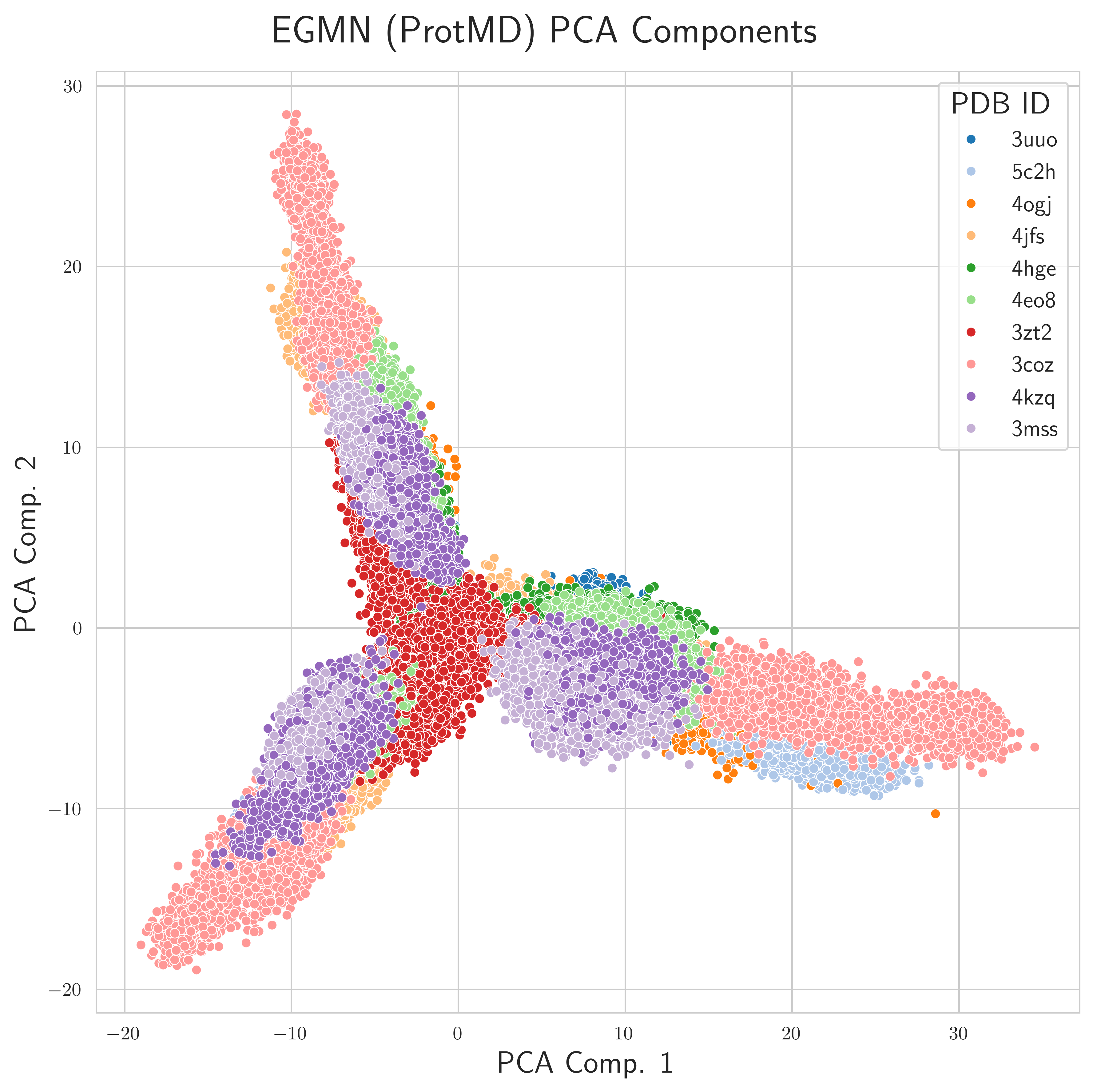}
    \caption{Scatter plot of the first two principal components calculated from the latent embeddings $h_G$ extracted from the pre-trained EGMN encoder, trained for a maximum of 100 epochs, for all test set MD frames corresponding to the top 5 docking poses. Points are colored according to the PDB ID of the input co-complex. Results are shown for 2 randomly selected PDB entries from each of the 5 test sets, 10 in total.}
    \label{fig:pca_egmn_protmd_pdbid_md}
\end{figure}

\begin{figure}[ht]
    \centering
    \includegraphics[width=\linewidth]{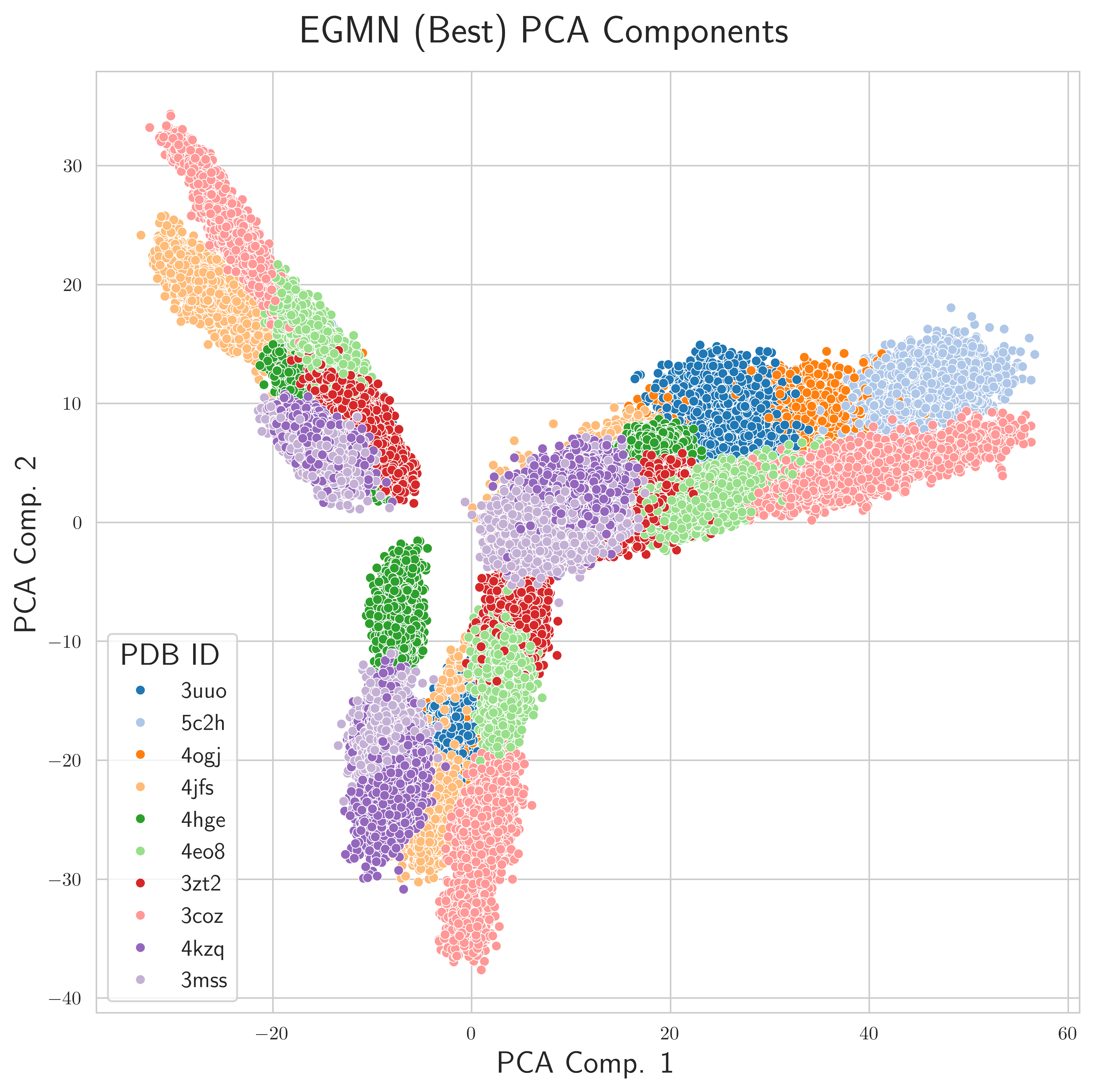}
    \caption{Scatter plot of the first two principal components calculated from the latent embeddings $h_G$ extracted from the best EGMN encoder, trained for a maximum of 600 epochs, for all test set MD frames corresponding to the top 5 docking poses. Points are colored according to the PDB ID of the input co-complex. Results are shown for 2 randomly selected PDB entries from each of the 5 test sets, 10 in total.}
    \label{fig:pca_egmn_best_pdbid_md}
\end{figure}

\begin{figure}[ht]
    \centering
    \includegraphics[width=\linewidth]{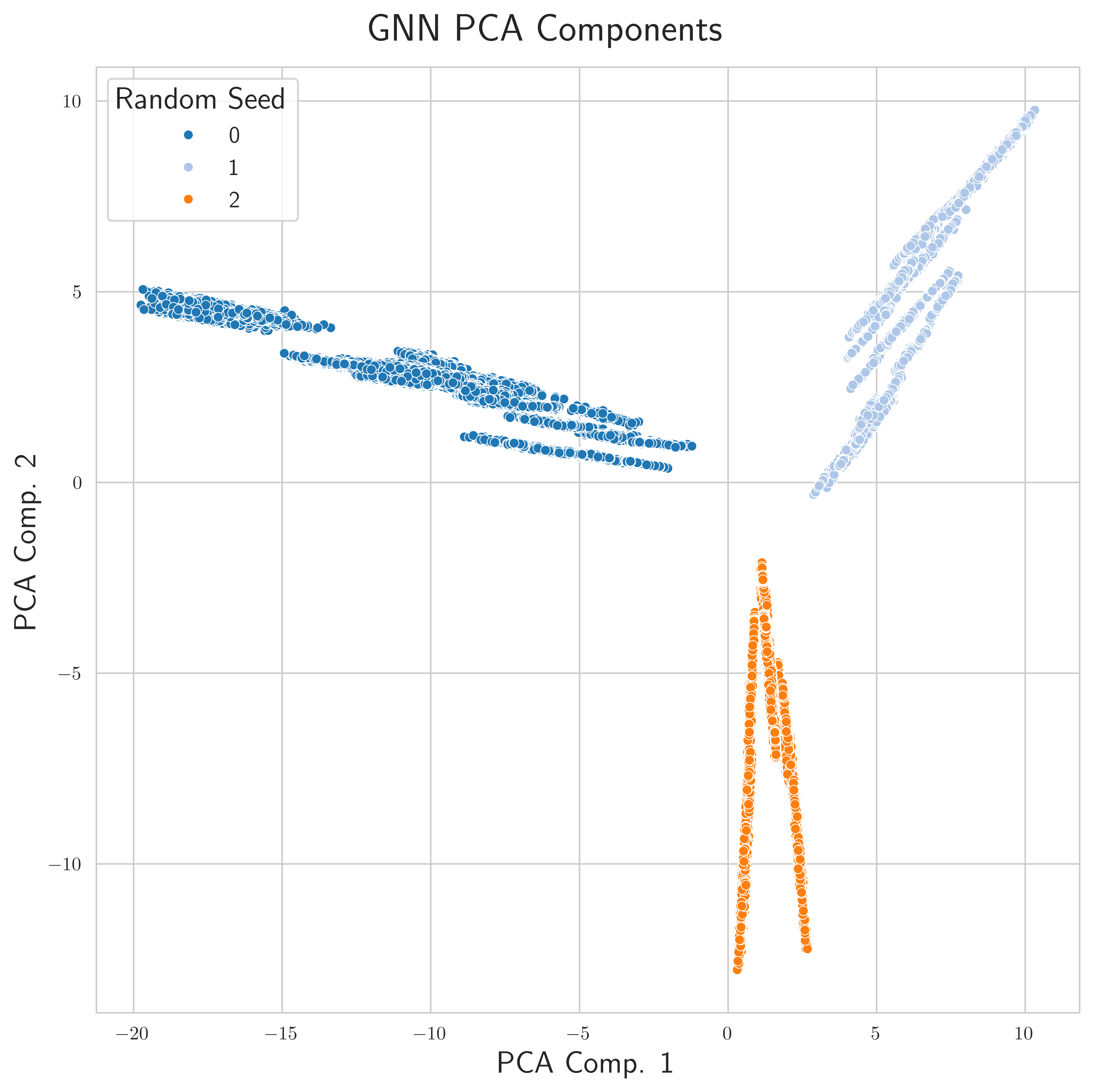}
    \caption{Scatter plot of the first two principal components calculated from the latent embeddings $h_G$ extracted from the GNN encoder, trained for a maximum of 100 epochs, for all test set MD frames corresponding to the top 5 docking poses. Points are colored according to the random seed used for training. Results are shown for 2 randomly selected PDB entries from each of the 5 test sets, 10 in total.}
    \label{fig:pca_gnn_scratch_seed_md}
\end{figure}

\begin{figure}[ht]
    \centering
    \includegraphics[width=\linewidth]{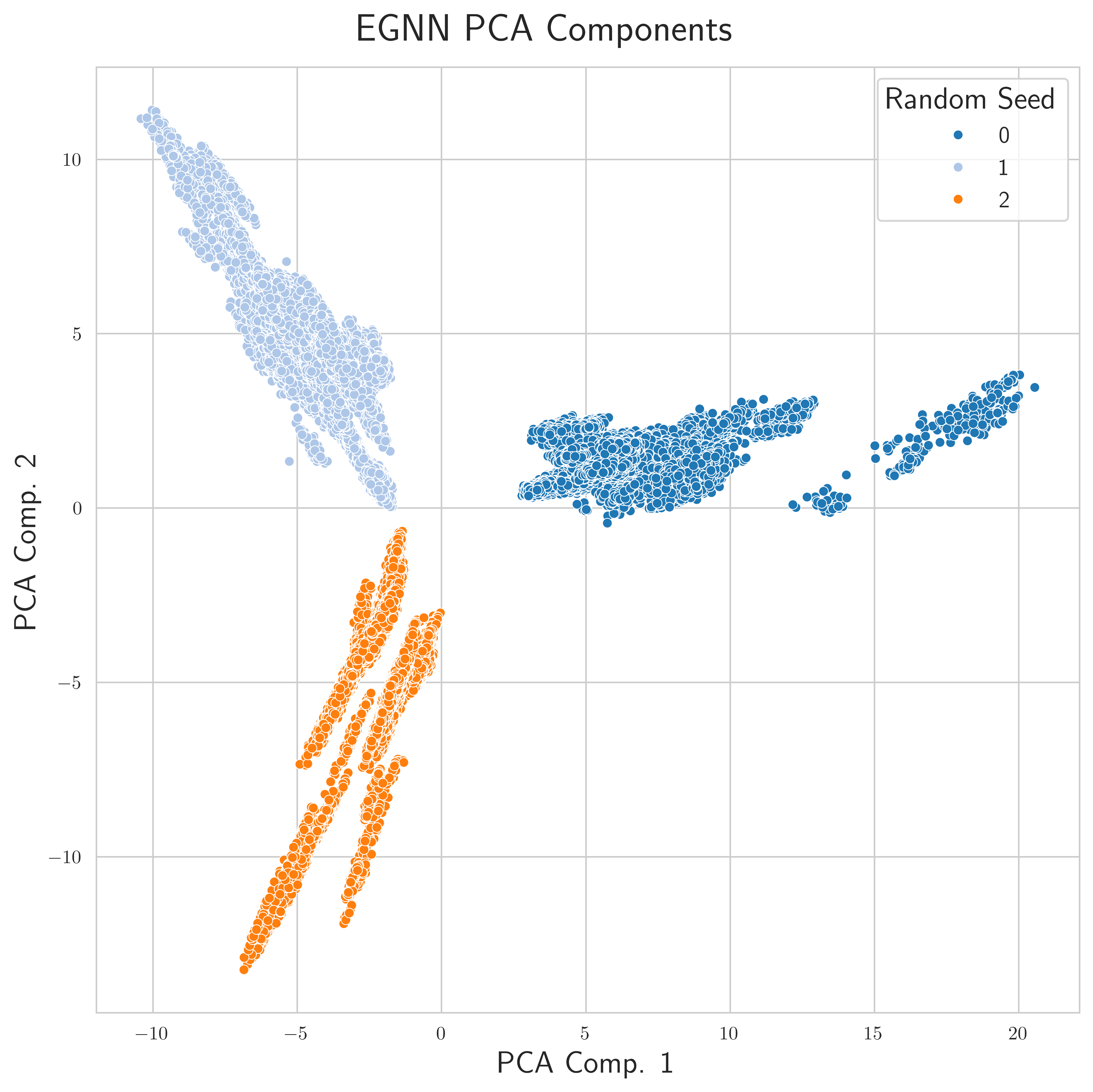}
    \caption{Scatter plot of the first two principal components calculated from the latent embeddings $h_G$ extracted from the EGNN encoder, trained from scratch for a maximum of 100 epochs, for all test set MD frames corresponding to the top 5 docking poses. Points are colored according to the random seed used for training. Results are shown for 2 randomly selected PDB entries from each of the 5 test sets, 10 in total.}
    \label{fig:pca_egnn_scratch_seed_md}
\end{figure}

\begin{figure}[ht]
    \centering
    \includegraphics[width=\linewidth]{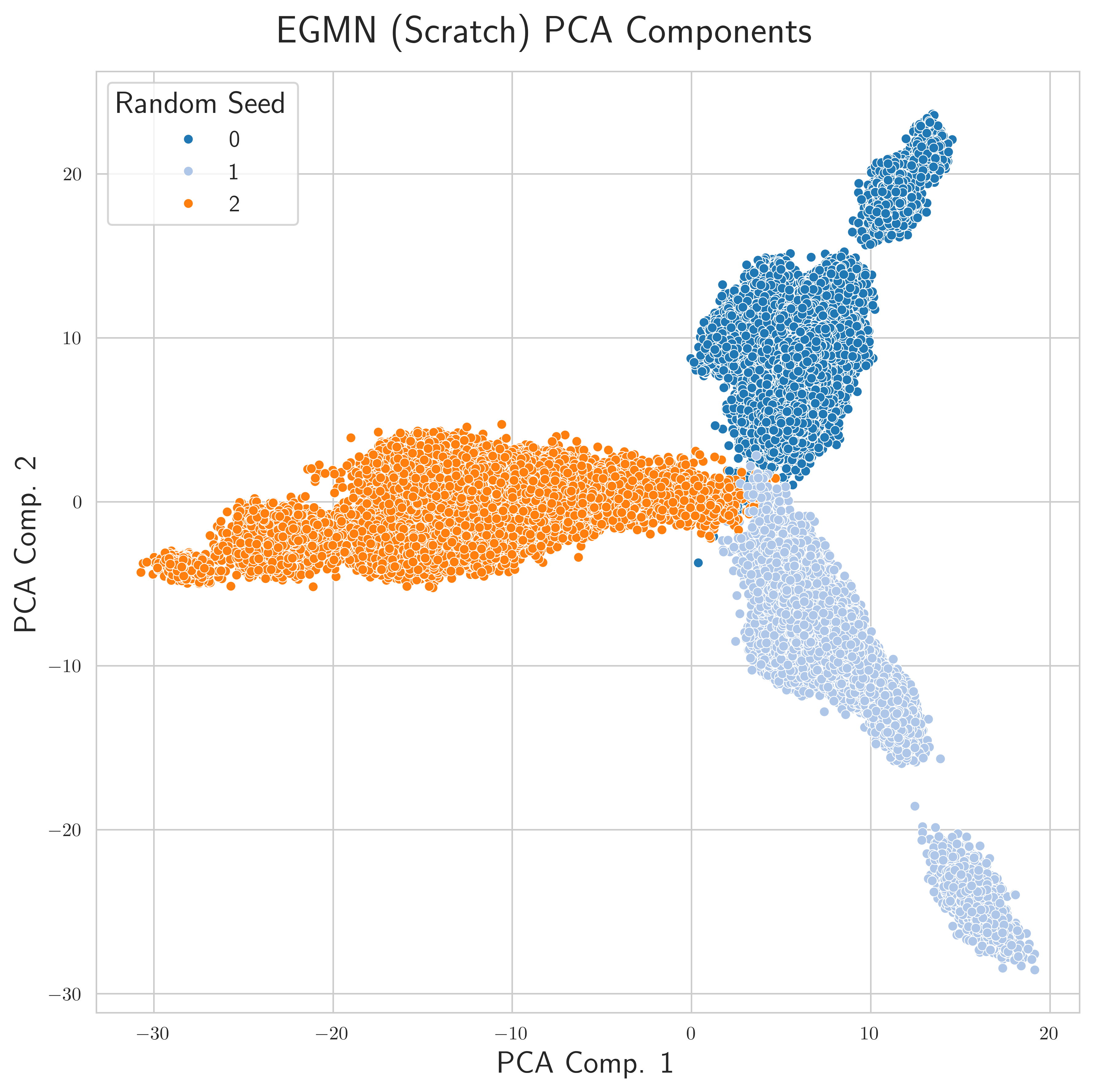}
    \caption{Scatter plot of the first two principal components calculated from the latent embeddings $h_G$ extracted from the EGMN encoder, trained from scratch for a maximum of 100 epochs, for all test set MD frames corresponding to the top 5 docking poses. Points are colored according to the random seed used for training. Results are shown for 2 randomly selected PDB entries from each of the 5 test sets, 10 in total.}
    \label{fig:pca_egmn_scratch_seed_md}
\end{figure}

\begin{figure}[ht]
    \centering
    \includegraphics[width=\linewidth]{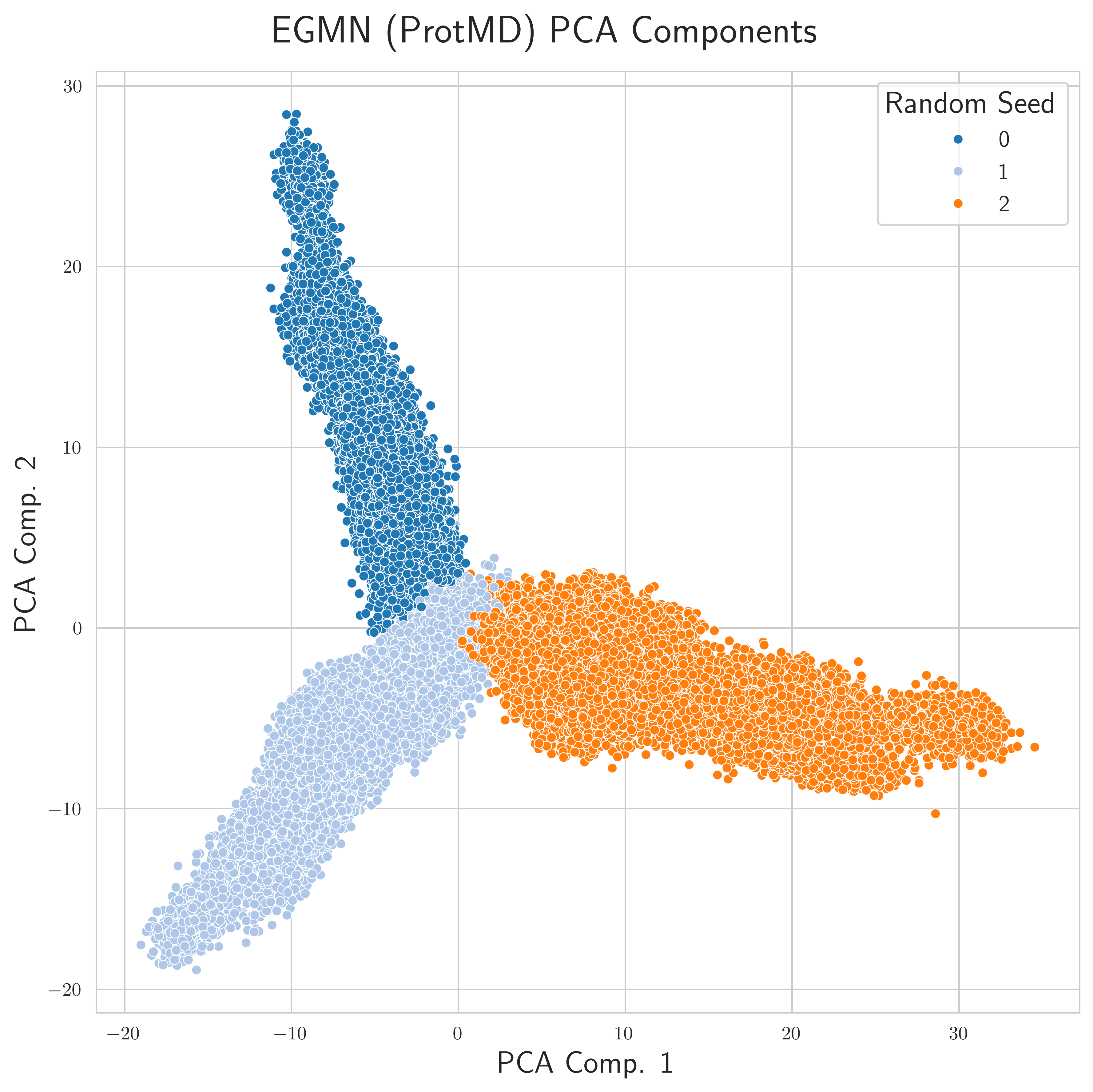}
    \caption{Scatter plot of the first two principal components calculated from the latent embeddings $h_G$ extracted from the pretrained EGMN encoder, trained for a maximum of 100 epochs, for all test set MD frames corresponding to the top 5 docking poses. Points are colored according to the random seed used for training. Results are shown for 2 randomly selected PDB entries from each of the 5 test sets, 10 in total.}
    \label{fig:pca_egmn_protmd_seed_md}
\end{figure}

\begin{figure}[ht]
    \centering
    \includegraphics[width=\linewidth]{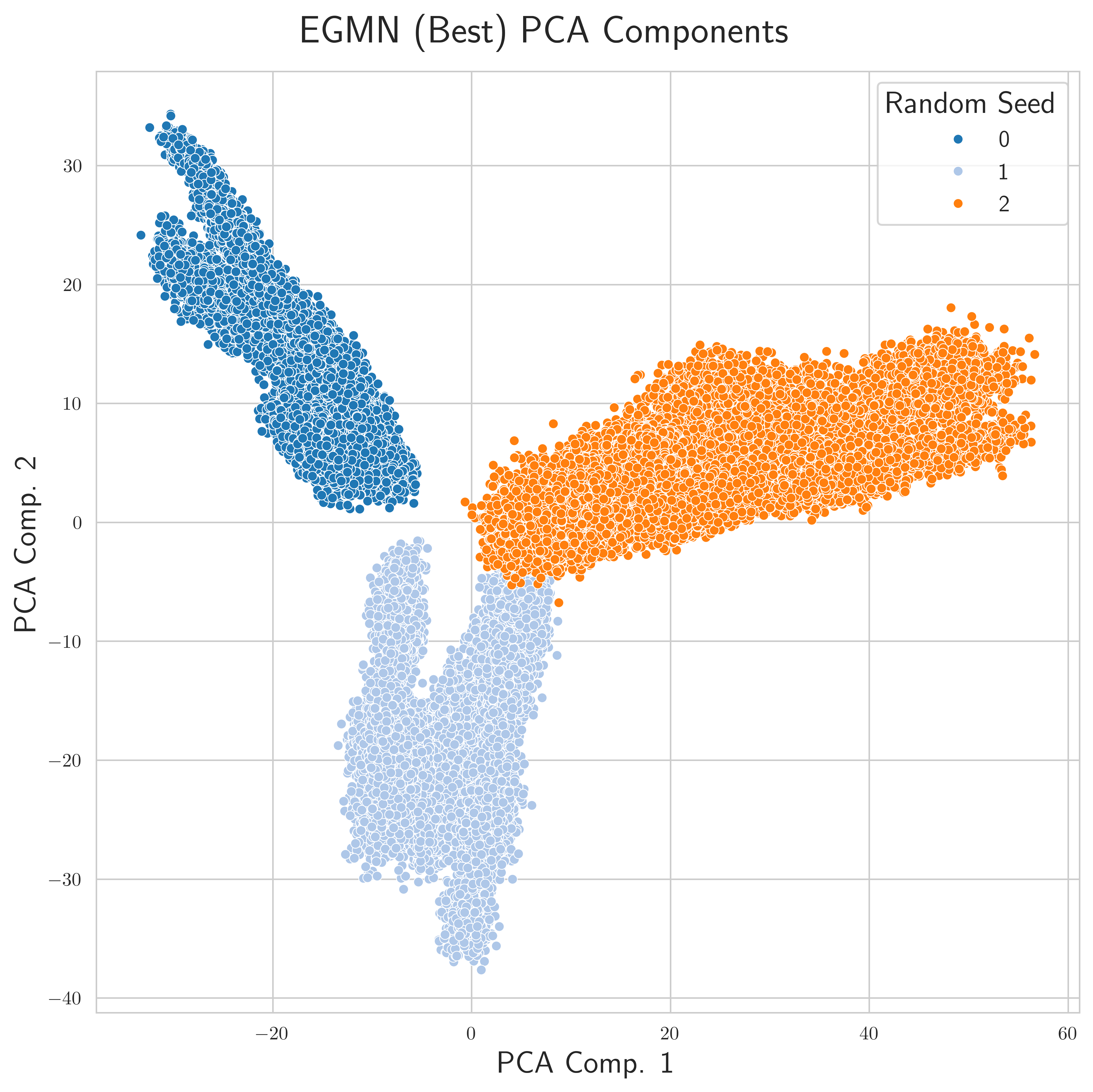}
    \caption{Scatter plot of the first two principal components calculated from the latent embeddings $h_G$ extracted from the best EGMN encoder, trained for a maximum of 600 epochs, for all test set MD frames corresponding to the top 5 docking poses. Points are colored according to the random seed used for training. Results are shown for 2 randomly selected PDB entries from each of the 5 test sets, 10 in total.}
    \label{fig:pca_egmn_best_seed_md}
\end{figure}

\begin{figure}[ht]
    \centering
    \includegraphics[width=\linewidth]{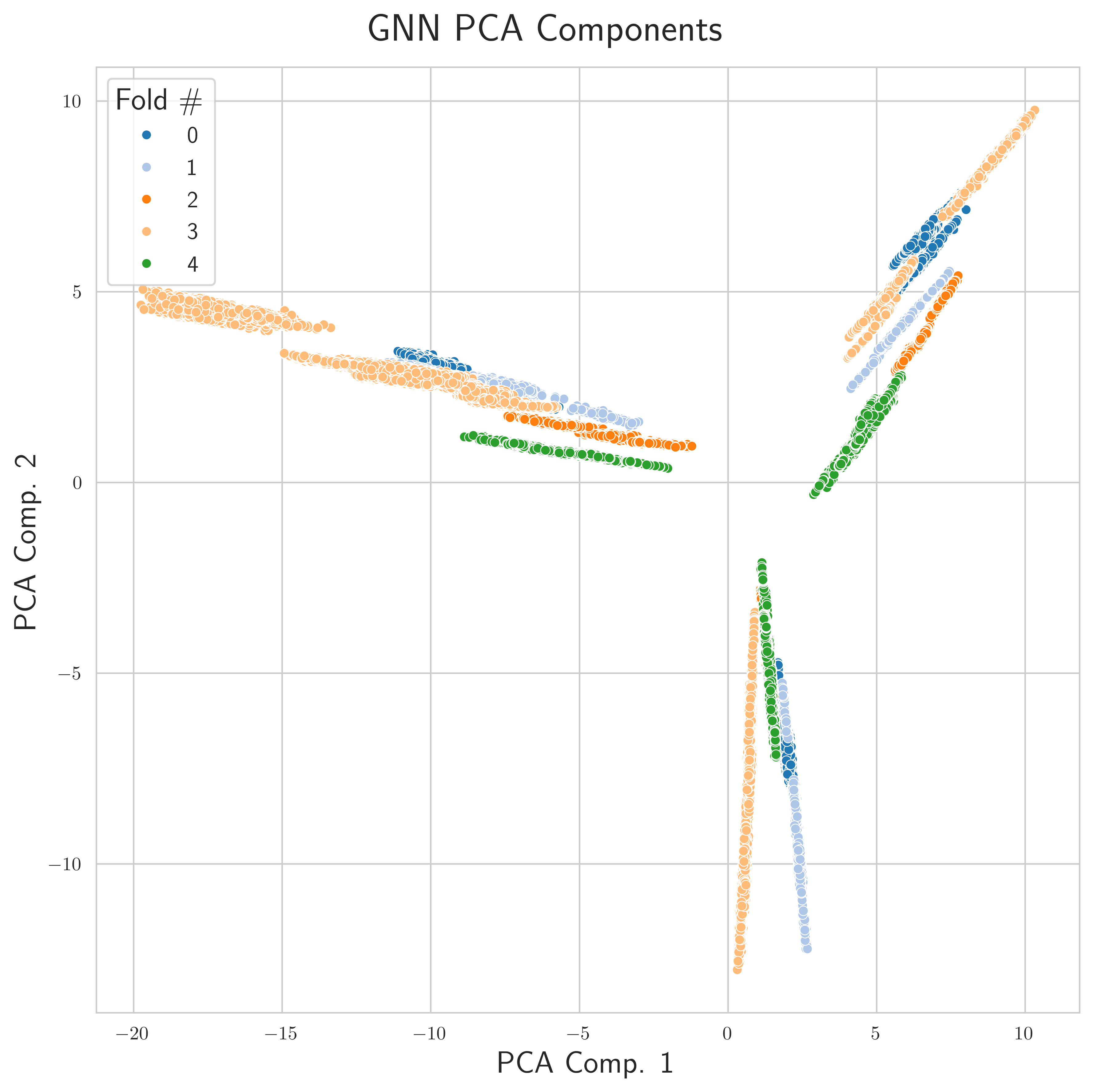}
    \caption{Scatter plot of the first two principal components calculated from the latent embeddings $h_G$ extracted from the GNN encoder, trained from scratch for a maximum of 100 epochs, for all test set MD frames corresponding to the top 5 docking poses. Points are colored according to the fold number of the respective test test of the input co-complex. Results are shown for 2 randomly selected PDB entries from each of the 5 test sets, 10 in total.}
    \label{fig:pca_gnn_scratch_fold_md}
\end{figure}

\begin{figure}[ht]
    \centering
    \includegraphics[width=\linewidth]{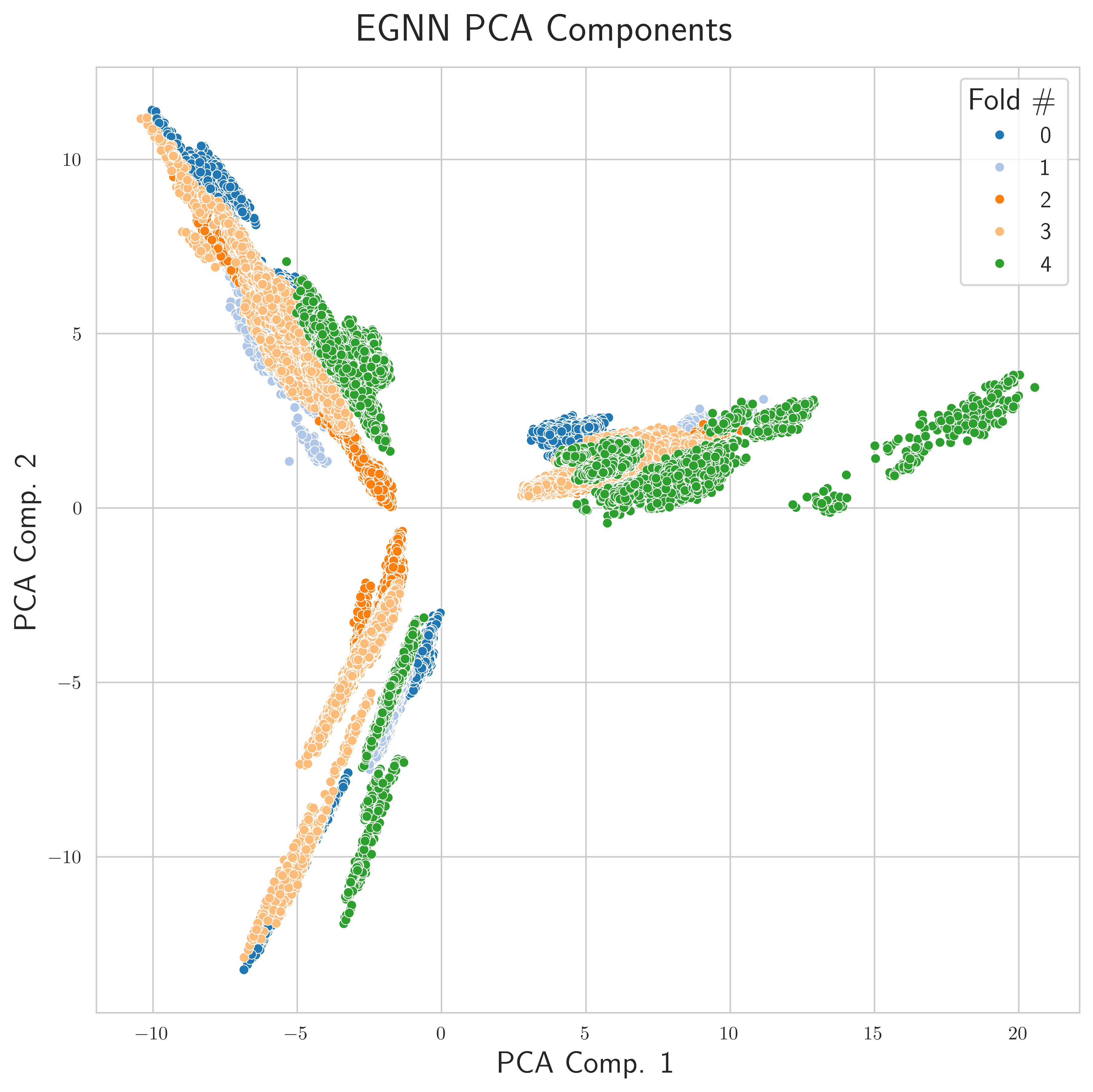}
    \caption{Scatter plot of the first two principal components calculated from the latent embeddings $h_G$ extracted from the EGNN encoder, trained from scratch for a maximum of 100 epochs, for all test set MD frames corresponding to the top 5 docking poses. Points are colored according to the fold number of the respective test test of the input co-complex. Results are shown for 2 randomly selected PDB entries from each of the 5 test sets, 10 in total.}
    \label{fig:pca_egnn_scratch_fold_md}
\end{figure}

\begin{figure}[ht]
    \centering
    \includegraphics[width=\linewidth]{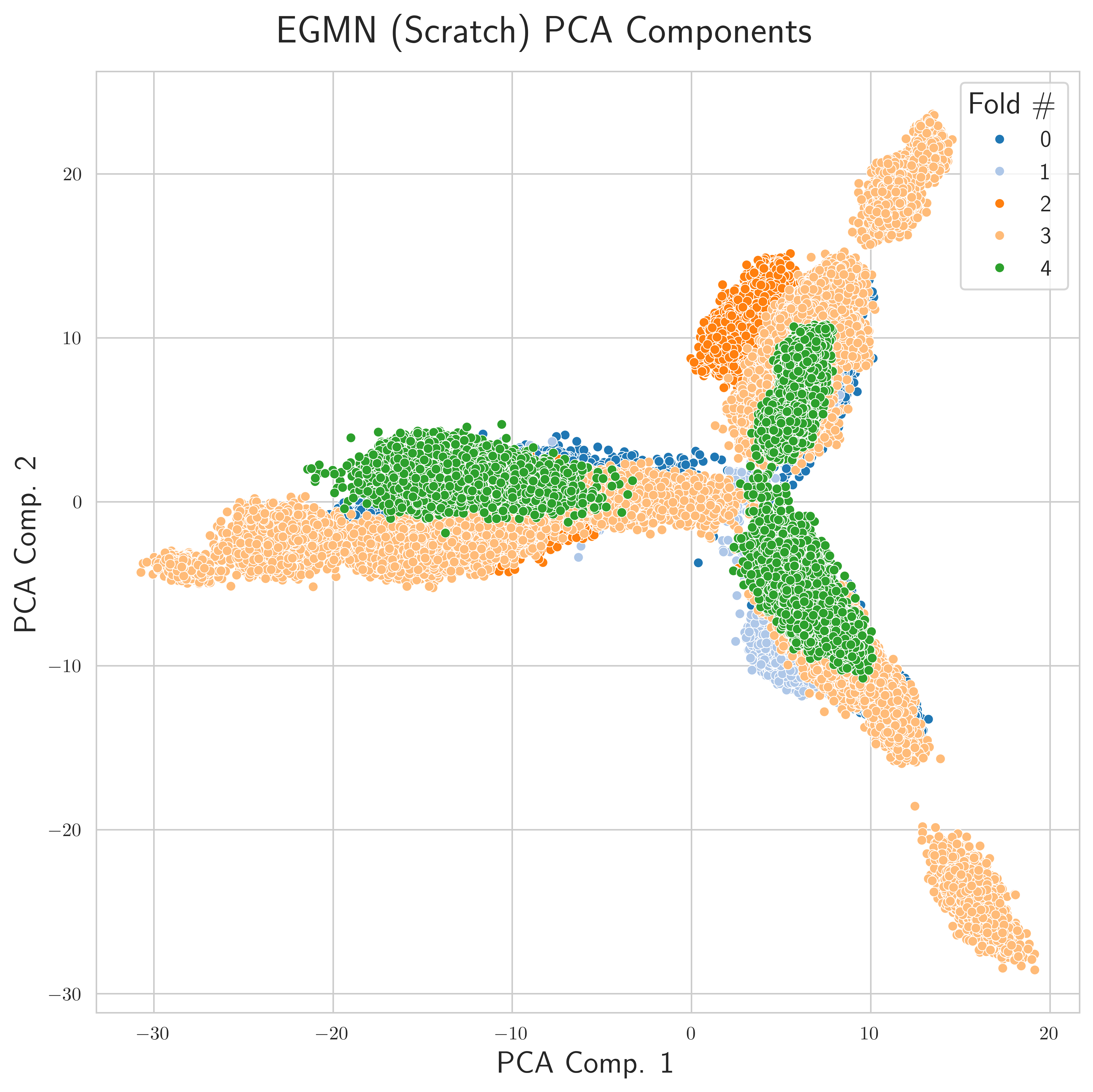}
    \caption{Scatter plot of the first two principal components calculated from the latent embeddings $h_G$ extracted from the EGMN encoder, trained from scratch for a maximum of 100 epochs, for all test set MD frames corresponding to the top 5 docking poses. Points are colored according to the fold number of the respective test test of the input co-complex. Results are shown for 2 randomly selected PDB entries from each of the 5 test sets, 10 in total.}
    \label{fig:pca_egmn_scratch_fold_md}
\end{figure}

\begin{figure}[ht]
    \centering
    \includegraphics[width=\linewidth]{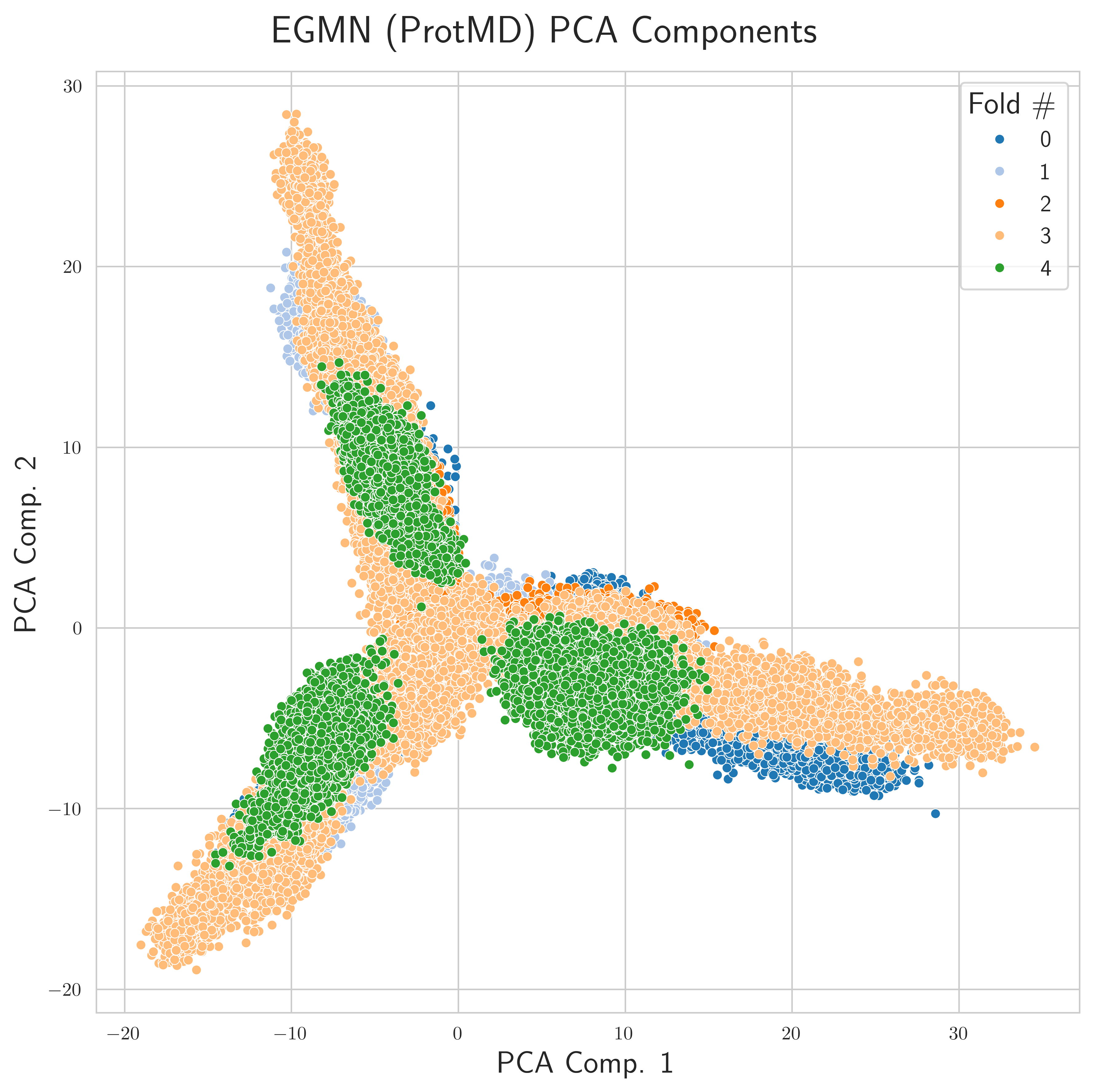}
    \caption{Scatter plot of the first two principal components calculated from the latent embeddings $h_G$ extracted from the pretrained EGMN encoder, trained for a maximum of 100 epochs, for all test set MD frames corresponding to the top 5 docking poses. Points are colored according to the fold number of the respective test test of the input co-complex. Results are shown for 2 randomly selected PDB entries from each of the 5 test sets, 10 in total.}
    \label{fig:pca_egmn_protmd_fold_md}
\end{figure}

\begin{figure}[ht]
    \centering
    \includegraphics[width=\linewidth]{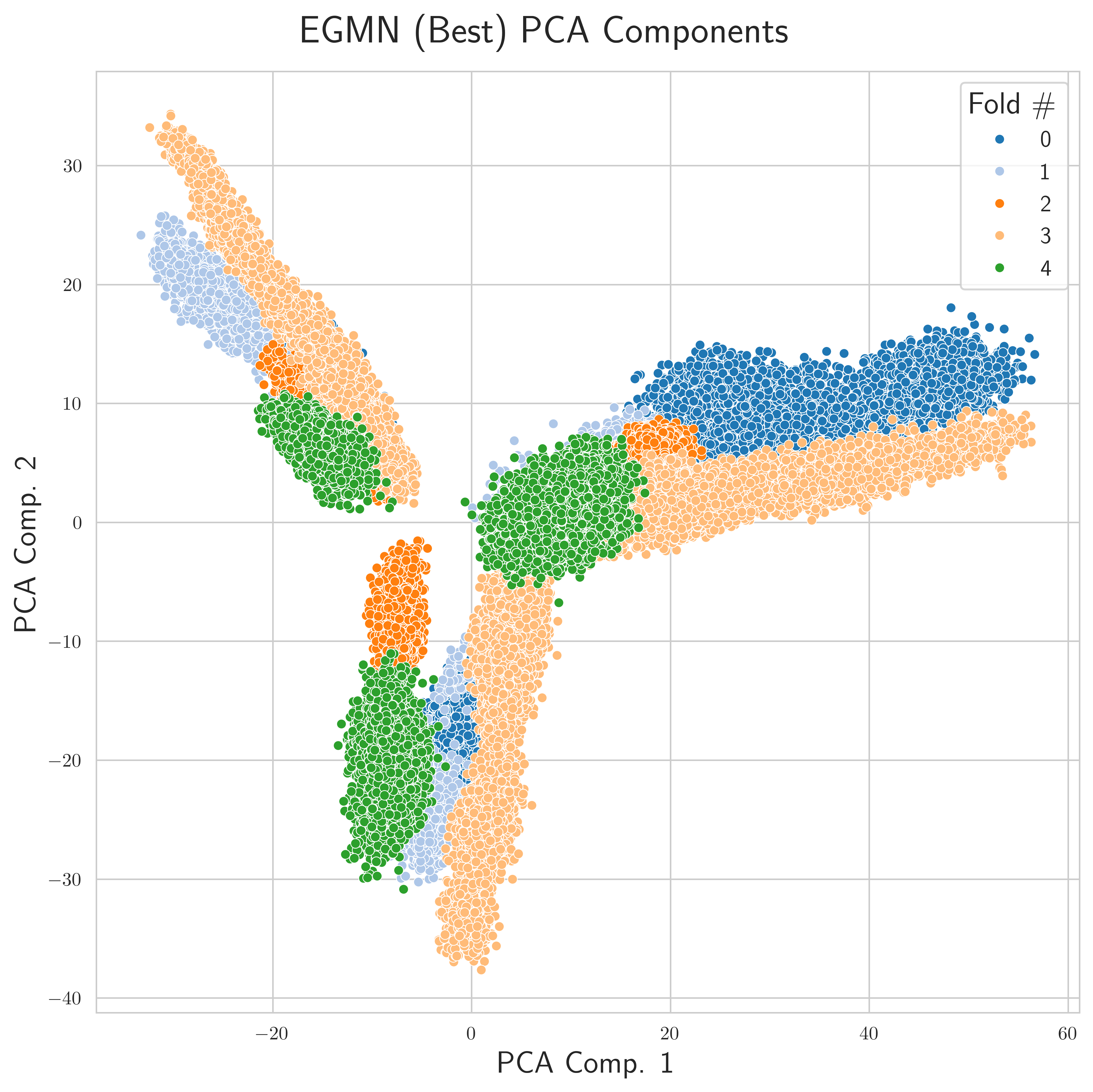}
    \caption{Scatter plot of the first two principal components calculated from the latent embeddings $h_G$ extracted from the best EGMN encoder, trained for a maximum of 600 epochs, for all test set MD frames corresponding to the top 5 docking poses. Points are colored according to the fold number of the respective test test of the input co-complex. Results are shown for 2 randomly selected PDB entries from each of the 5 test sets, 10 in total.}
    \label{fig:pca_egmn_best_fold_md}
\end{figure}

\begin{figure}[ht]
    \centering
    \includegraphics[width=\linewidth]{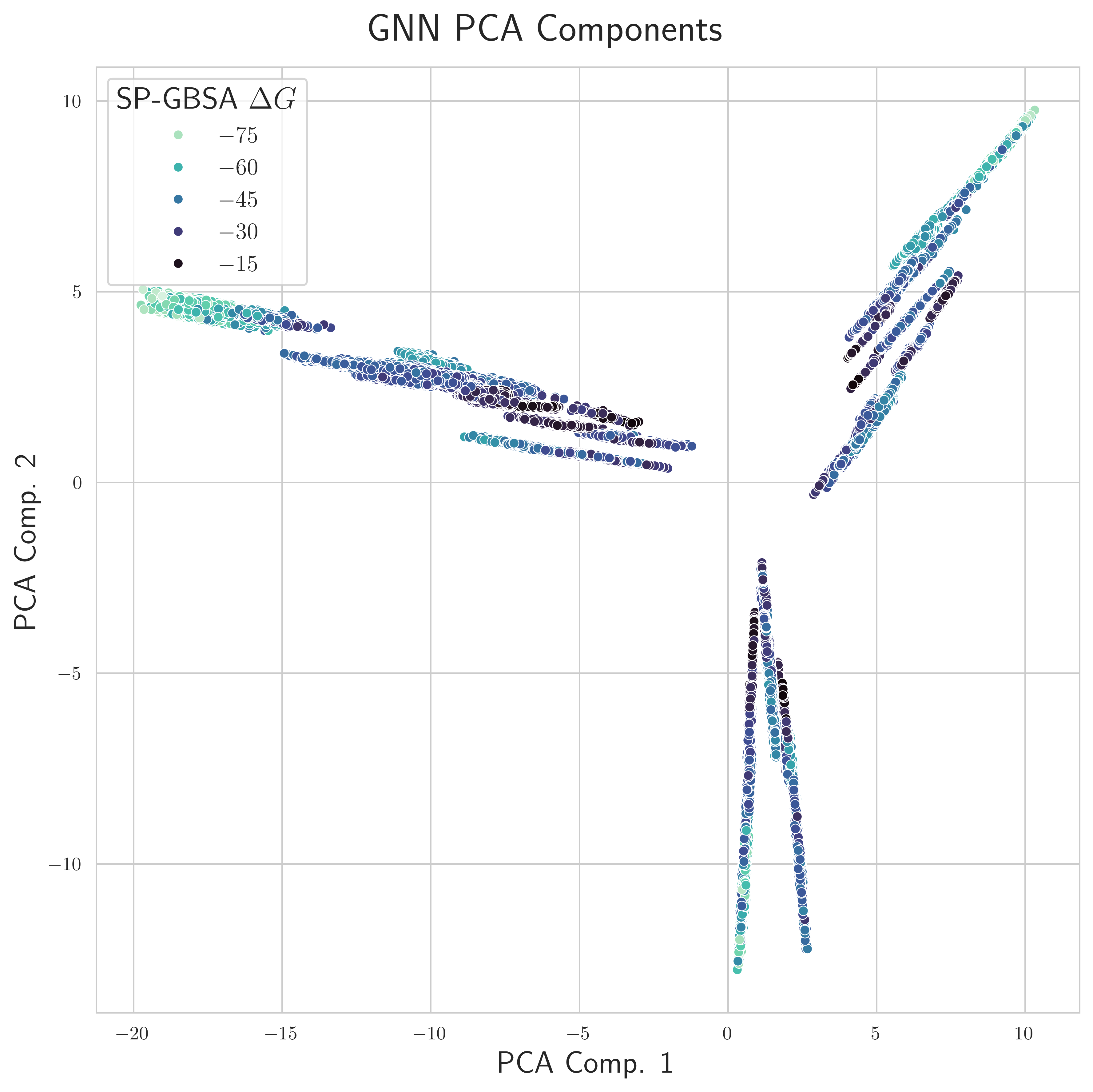}
    \caption{Scatter plot of the first two principal components calculated from the latent embeddings $h_G$ extracted from the GNN encoder, trained from scratch for a maximum of 100 epochs ,for all test set MD frames corresponding to the top 5 docking poses. Points are colored according to the single-point MMGBSA (SP-GBSA) scores of each frame. Results are shown for 2 randomly selected PDB entries from each of the 5 test sets, 10 in total.}
    \label{fig:pca_y_true_gnn_scratch_md}
\end{figure}

\begin{figure}[ht]
    \centering
    \includegraphics[width=\linewidth]{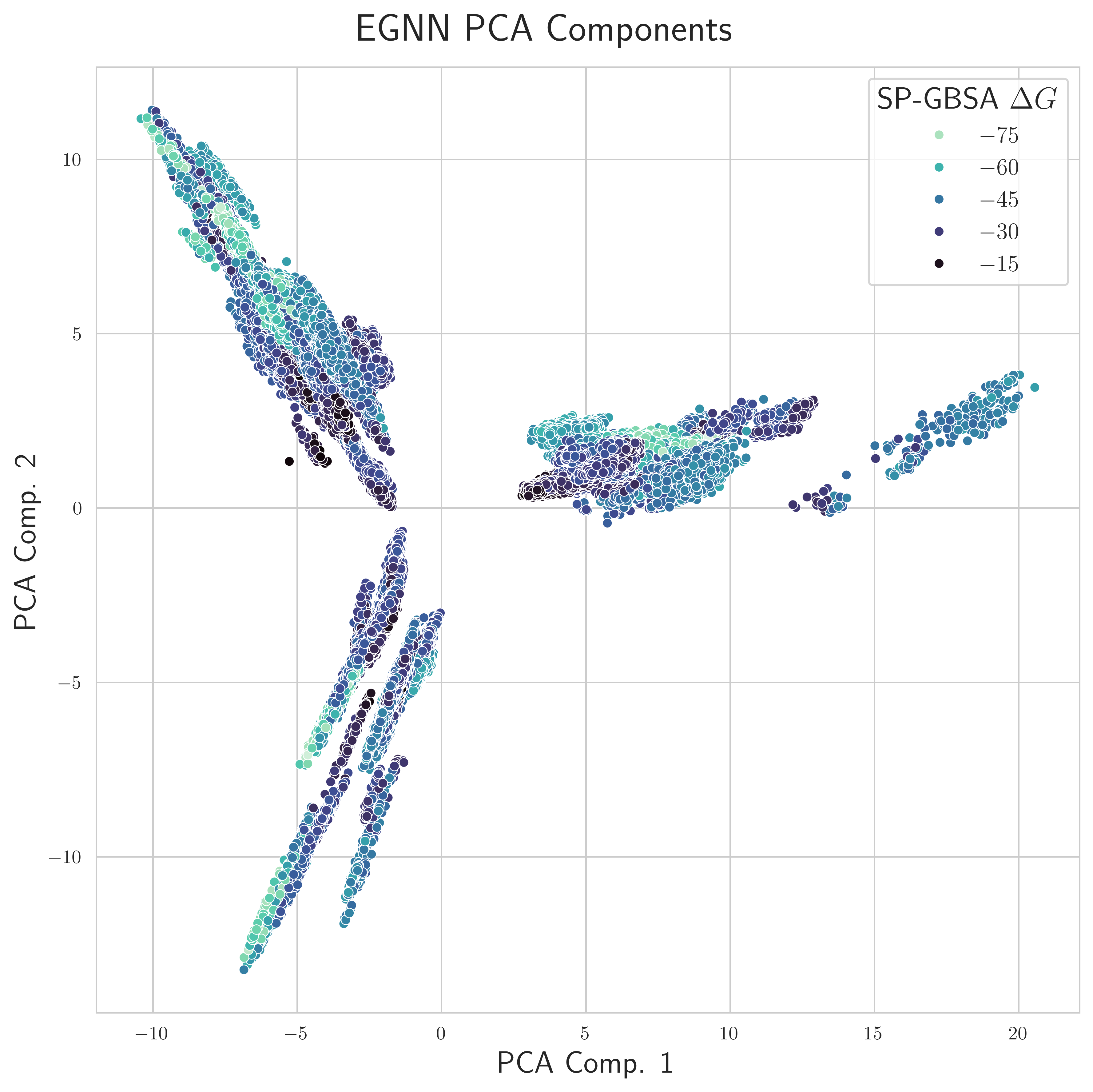}
    \caption{Scatter plot of the first two principal components calculated from the latent embeddings $h_G$ extracted from the EGNN encoder, trained from scratch for a maximum of 100 epochs, for all test set MD frames corresponding to the top 5 docking poses. Points are colored according to the single-point MMGBSA (SP-GBSA) scores of each frame. Results are shown for 2 randomly selected PDB entries from each of the 5 test sets, 10 in total.}
    \label{fig:pca_y_true_egnn_scratch_md}
\end{figure}

\begin{figure}[ht]
    \centering
    \includegraphics[width=\linewidth]{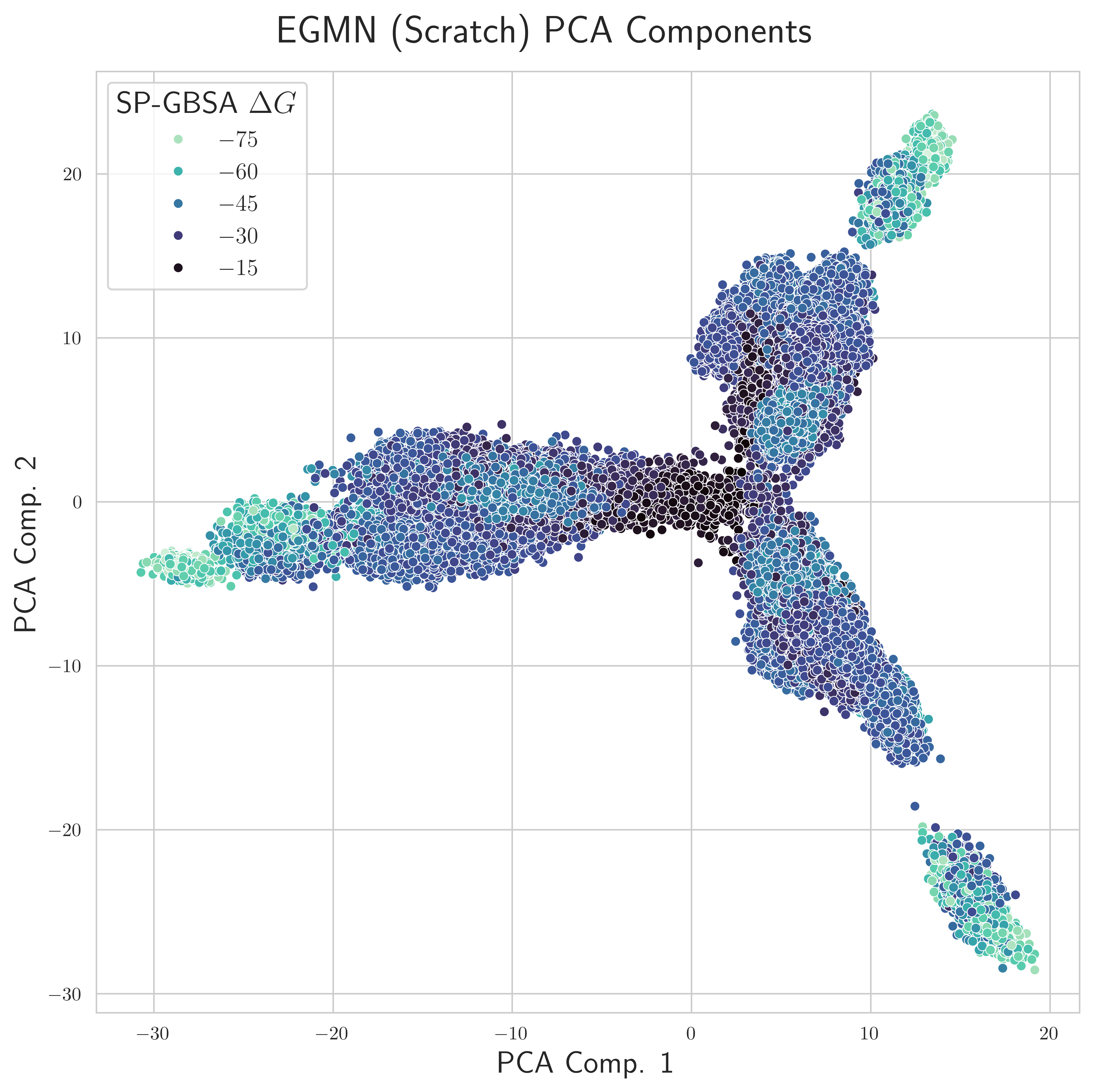}
    \caption{Scatter plot of the first two principal components calculated from the latent embeddings $h_G$ extracted from the EGMN encoder, trained from scratch for a maximum of 100 epochs, for all test set MD frames corresponding to the top 5 docking poses. Points are colored according to the single-point MMGBSA (SP-GBSA) scores of each frame. Results are shown for 2 randomly selected PDB entries from each of the 5 test sets, 10 in total.}
    \label{fig:pca_y_true_egmn_scratch_md}
\end{figure}

\begin{figure}[ht]
    \centering
    \includegraphics[width=\linewidth]{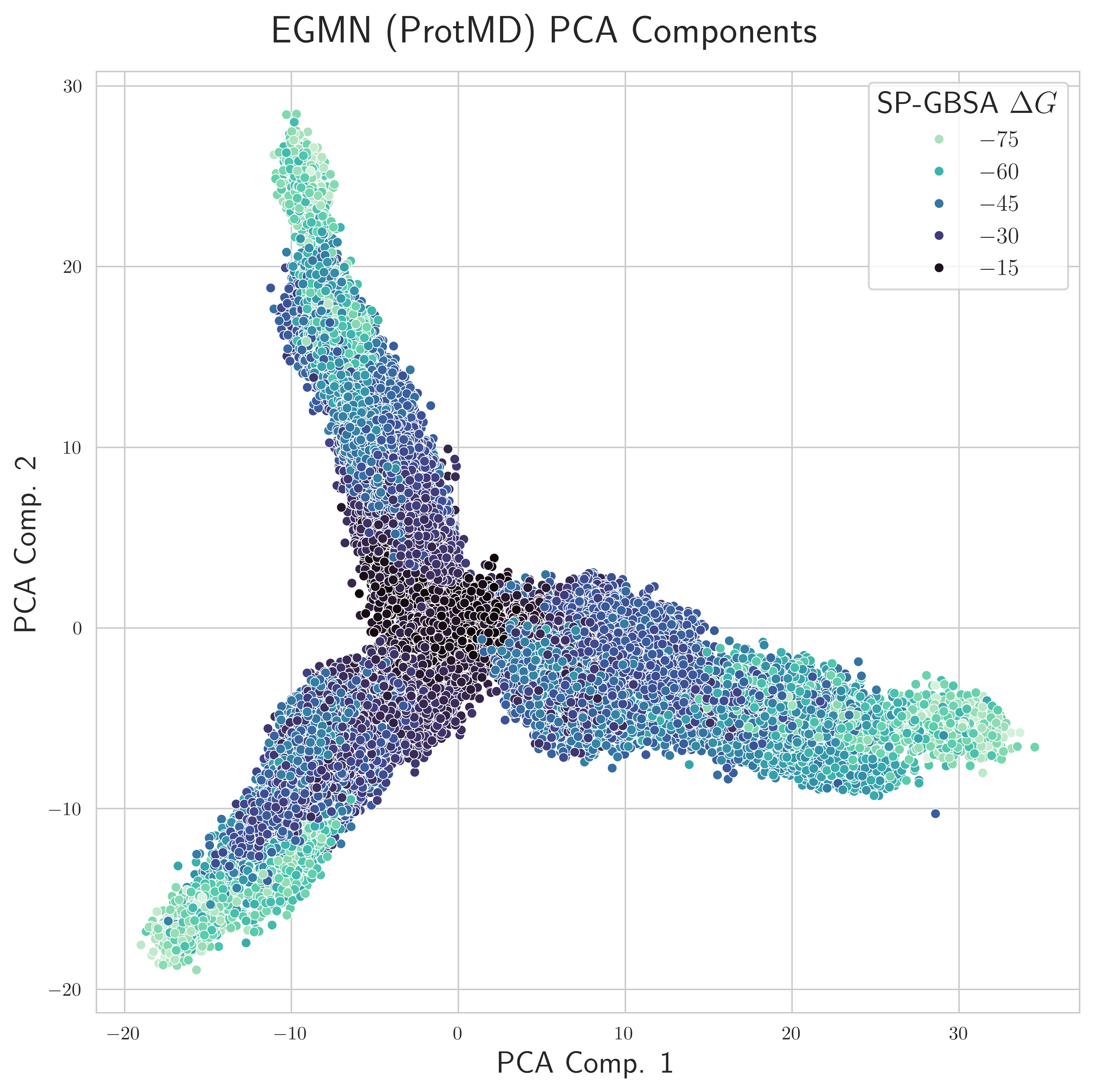}
    \caption{Scatter plot of the first two principal components calculated from the latent embeddings $h_G$ extracted from the pre-trained EGMN encoder, trained for a maximum of 100 epochs, for all test set MD frames corresponding to the top 5 docking poses. Points are colored according to the single-point MMGBSA (SP-GBSA) scores of each frame. Results are shown for 2 randomly selected PDB entries from each of the 5 test sets, 10 in total.}
    \label{fig:pca_y_true_egmn_protmd_md}
\end{figure}

\begin{figure}[ht]
    \centering
    \includegraphics[width=\linewidth]{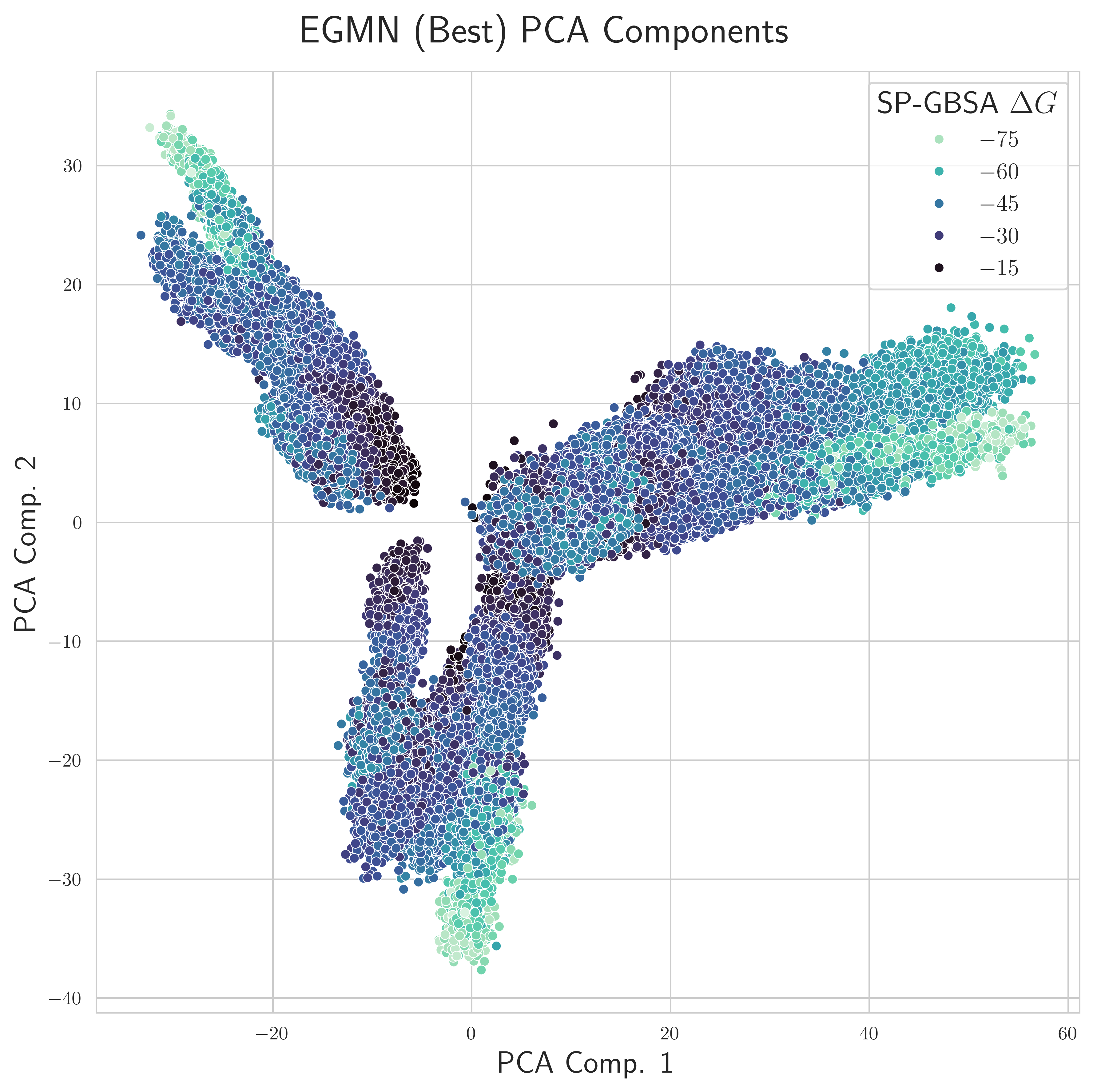}
    \caption{Scatter plot of the first two principal components calculated from the latent embeddings $h_G$ extracted from the best EGMN encoder, trained for a maximum of 600 epochs, for all test set MD frames corresponding to the top 5 docking poses. Points are colored according to the single-point MMGBSA (SP-GBSA) scores of each frame. Results are shown for 2 randomly selected PDB entries from each of the 5 test sets, 10 in total.}
    \label{fig:pca_y_true_egmn_best_md}
\end{figure}

\begin{figure}[ht]
    \centering
    \includegraphics[width=\linewidth]{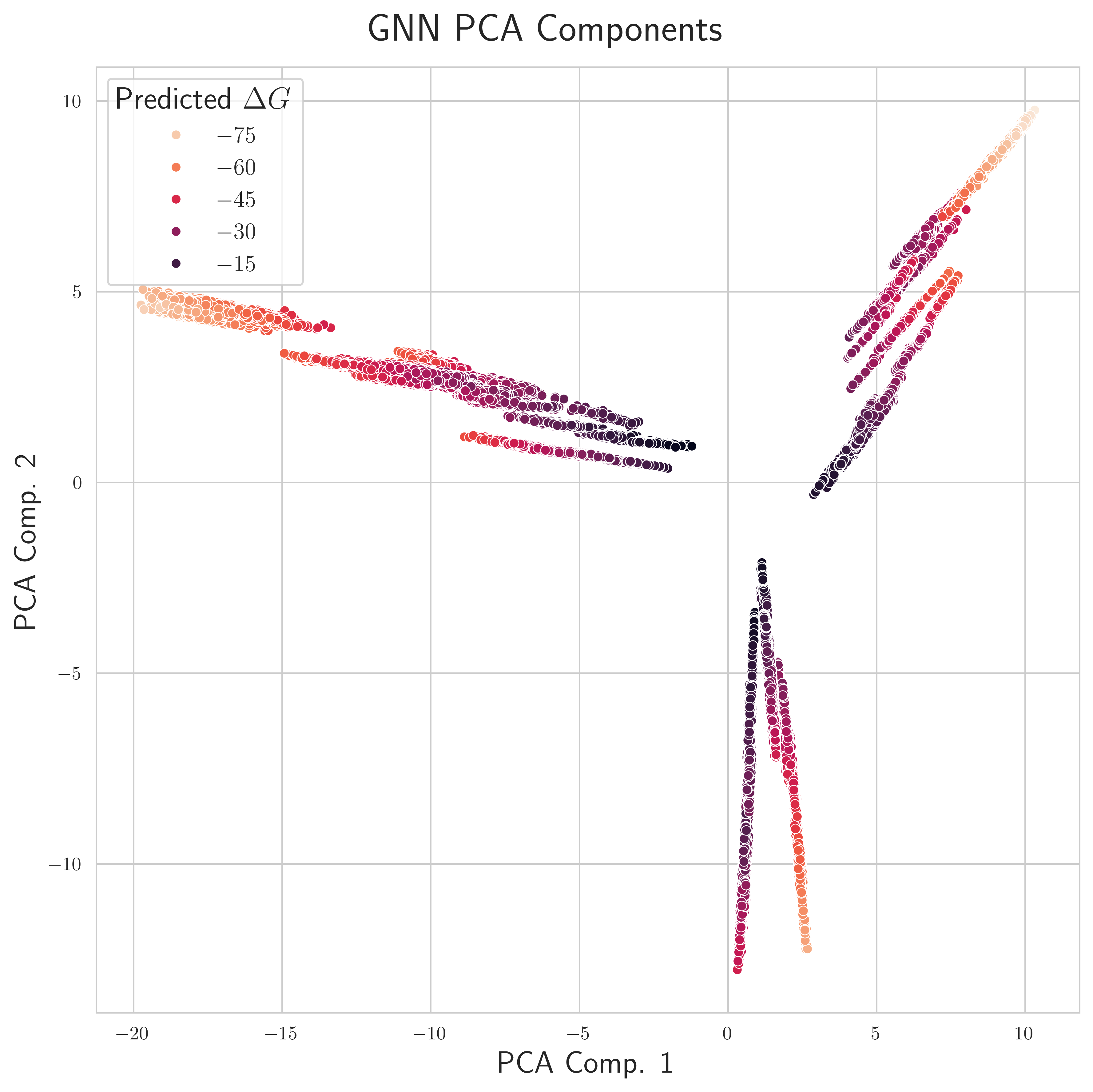}
    \caption{Scatter plot of the first two principal components calculated from the latent embeddings $h_G$ extracted from the GNN encoder, trained from scratch for a maximum of 100 epochs, for all test set MD frames corresponding to the top 5 docking poses. Points are colored according to the predicted SP-GBSA score of the input co-complex. Results are shown for 2 randomly selected PDB entries from each of the 5 test sets, 10 in total.}
    \label{fig:pca_y_pred_gnn_scratch_md}
\end{figure}

\begin{figure}[ht]
    \centering
    \includegraphics[width=\linewidth]{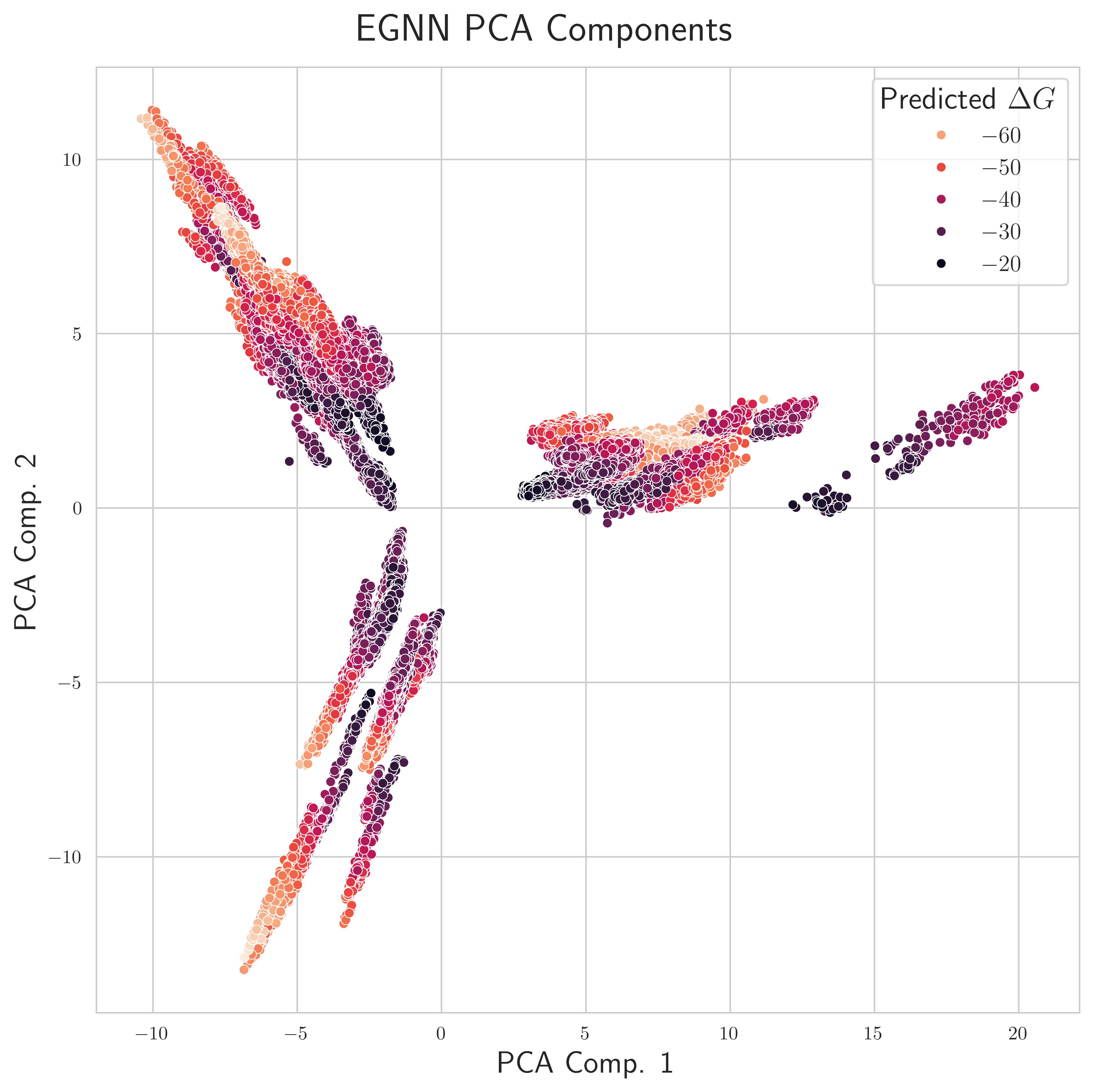}
    \caption{Scatter plot of the first two principal components calculated from the latent embeddings $h_G$ extracted from the EGNN encoder, trained from scratch for a maximum of 100 epochs, for all test set MD frames corresponding to the top 5 docking poses. Points are colored according to the predicted SP-GBSA score of the input co-complex. Results are shown for 2 randomly selected PDB entries from each of the 5 test sets, 10 in total.}
    \label{fig:pca_y_pred_egnn_scratch_md}
\end{figure}

\begin{figure}[ht]
    \centering
    \includegraphics[width=\linewidth]{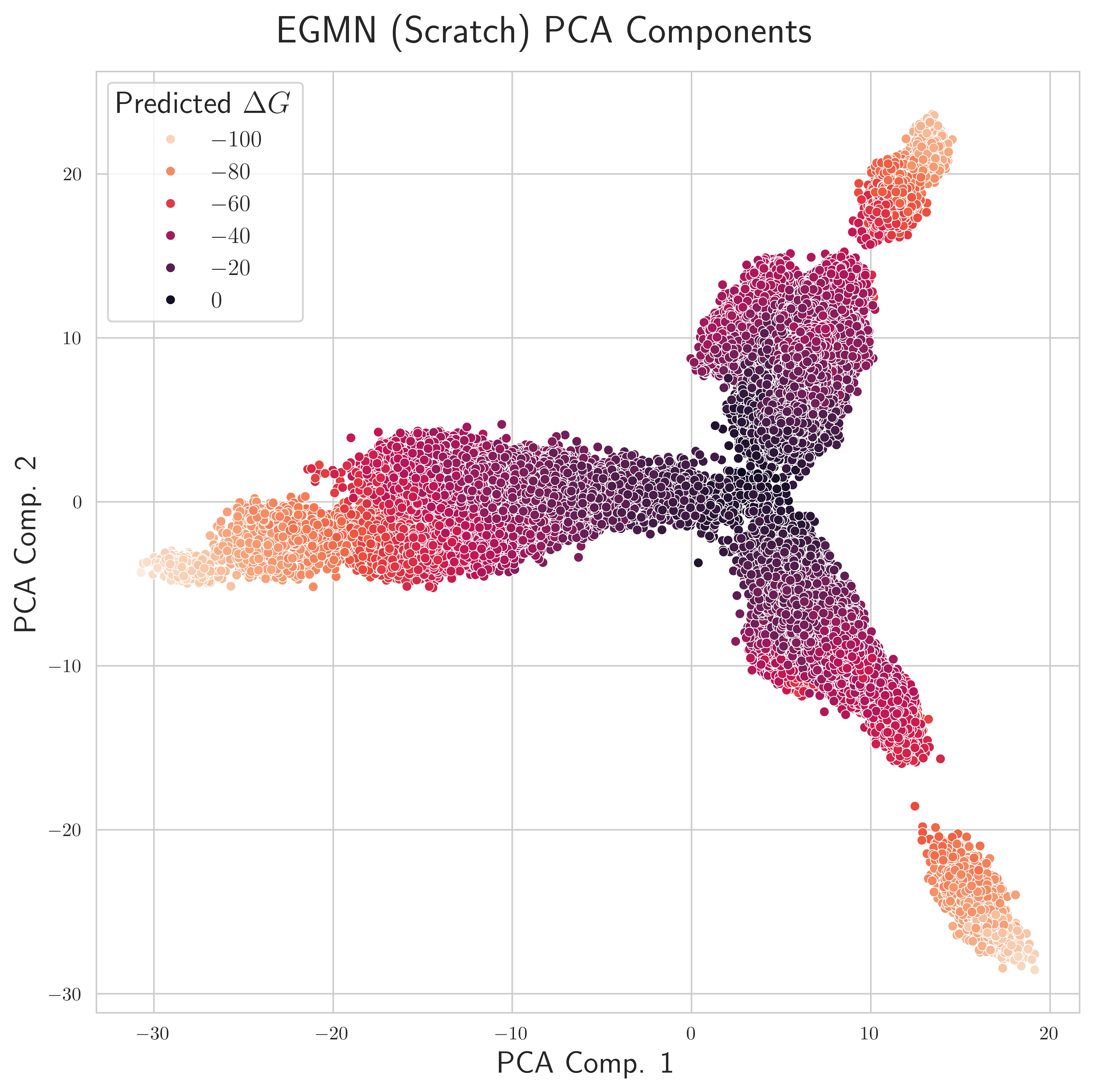}
    \caption{Scatter plot of the first two principal components calculated from the latent embeddings $h_G$ extracted from the EGMN encoder, trained from scratch for a maximum of 100 epochs, for all test set MD frames corresponding to the top 5 docking poses. Points are colored according to the predicted SP-GBSA score of the input co-complex. Results are shown for 2 randomly selected PDB entries from each of the 5 test sets, 10 in total.}
    \label{fig:pca_y_pred_egmn_scratch_md}
\end{figure}

\begin{figure}[ht]
    \centering
    \includegraphics[width=\linewidth]{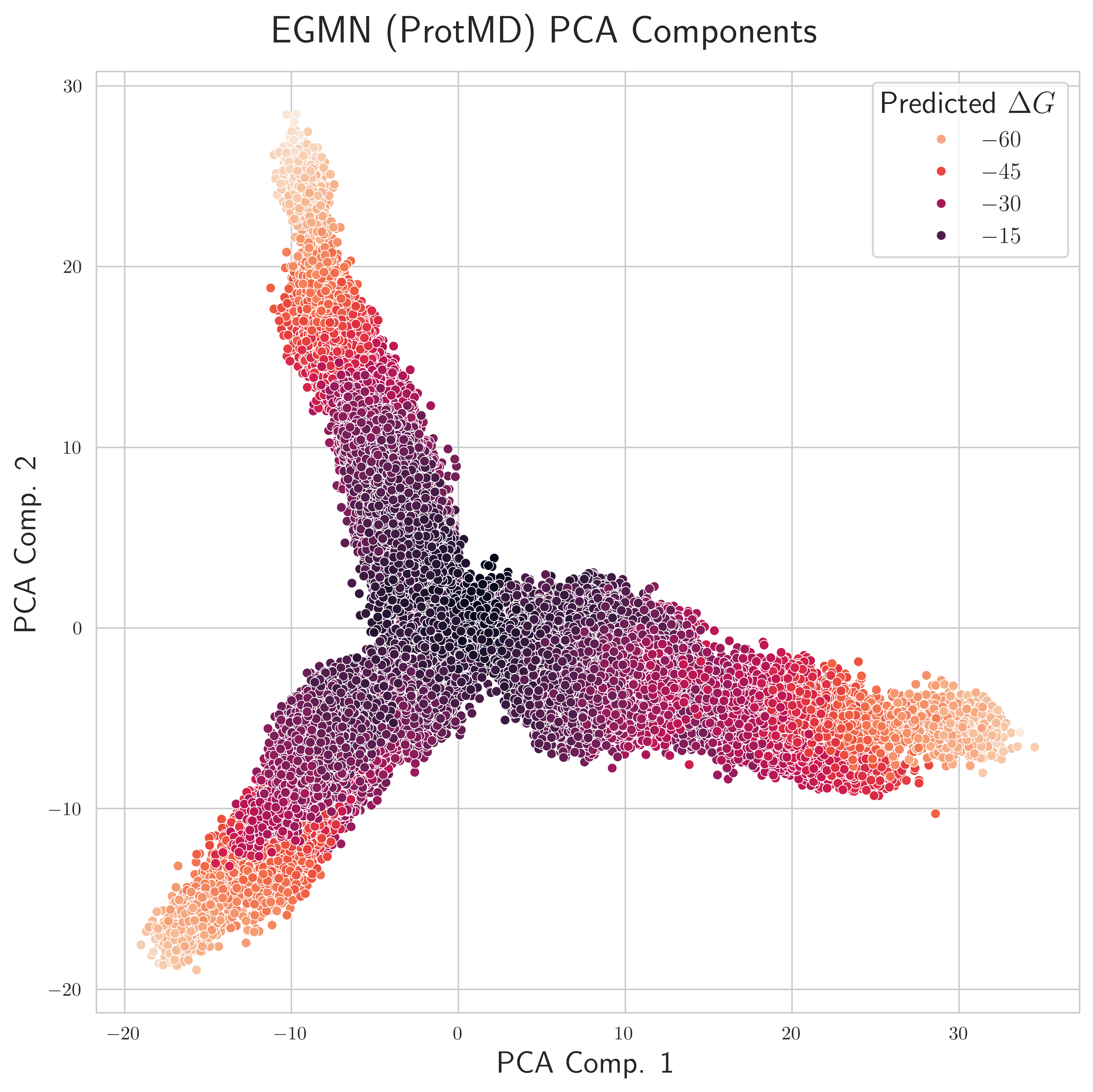}
    \caption{Scatter plot of the first two principal components calculated from the latent embeddings $h_G$ extracted from the pre-trained EGMN encoder, trained for a maximum of 100 epochs, for all test set MD frames corresponding to the top 5 docking poses. Points are colored according to the predicted SP-GBSA score of the input co-complex. Results are shown for 2 randomly selected PDB entries from each of the 5 test sets, 10 in total.}
    \label{fig:pca_y_pred_egmn_protmd_md}
\end{figure}

\begin{figure}[ht]
    \centering
    \includegraphics[width=\linewidth]{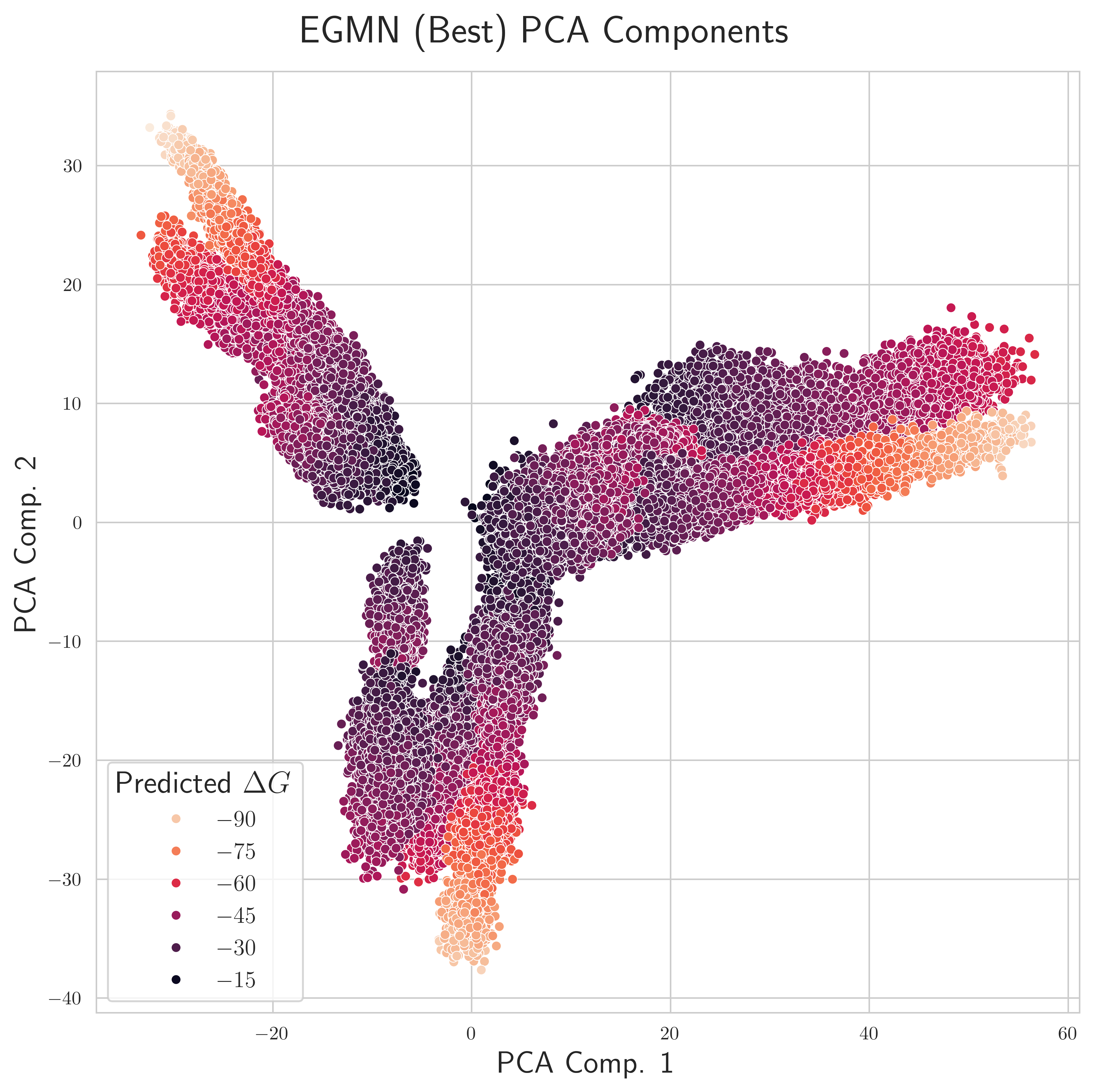}
    \caption{Scatter plot of the first two principal components calculated from the latent embeddings $h_G$ extracted from the best EGMN encoder, trained for a maximum of 600 epochs, for all test set MD frames corresponding to the top 5 docking poses. Points are colored according to the predicted SP-GBSA score of the input co-complex. Results are shown for 2 randomly selected PDB entries from each of the 5 test sets, 10 in total.}
    \label{fig:pca_y_pred_egmn_best_md}
\end{figure}

In order to probe the learned representations produced by each molecular graph encoder module, we used Principal Components Analysis (PCA). The intuition for using this approach is that we assume that the model has, over the course of the pre-training and MMGBSA training steps, performed much of the work in learning a non-linear projection of the molecular graphs to their graph embeddings in a lower dimensional space($d=128$). 
By applying linear dimensionality reduction to the learned embeddings, we can assess whether co-complexes naturally separate into interpretable groups, evidence of a robust representation. Figure~\ref{fig:pca_scree_plot} displays the cumulative distributions of the proportion of variance captured by each principal component.
We sought to understand, for each molecular graph encoder module, whether or not training using only crystal structures versus training using MD data (top 5 docking poses and crystal) has an impact on the distribution of explained variance. For example, a representation space for which each orthogonal PCA projection explains exactly the same variance would be an indication of a robust representation. For each test set, we chose the set of the static structures for the top 5 docking poses (i.e. frame 0) for each PDB structure, extracted their graph embeddings $\rmh^{\mathcal{G}}$, then performed PCA and collected the explained variance ratios. We averaged the ratios across all 15 of our independent trials for each model hyperparameter configuration (5 folds, 3 seeds). Each distribution sums approximately to 1 with infinitesimal error. We observe in the case of training using only crystal structures and the corresponding SP-GBSA energy, nearly all of the encoder modules we considered contain most of the explained variance in a single dimension, with the exception of our best EGMN encoder which is slightly below the other models for the first component. When considering the models trained on MD trajectories, we note a stark difference in the shapes of the curves that indicates that the explained variance is more uniformly spread throughout the dimensions, although the models we trained from scratch still exhibit a large proportion of the variance explained by the first few components. The order of the curves precisely follows the dimension of model complexity, ranging from the most simple GNN to the most complex physics-informed EGMN encoders. Our extended training of the pre-trained EGMN shows the best result, with the first component responsible for only ~30\% of the variance and with the remaining variance more uniformly spread relative to all other encoder modules. We visualized the PCA components for each encoder module in SI Figures~\ref{fig:pca_pdbid_gnn_scratch_md}-\ref{fig:pca_y_pred_egmn_best_md}.

\begin{figure}[ht]
    \centering
    \includegraphics[width=\linewidth]{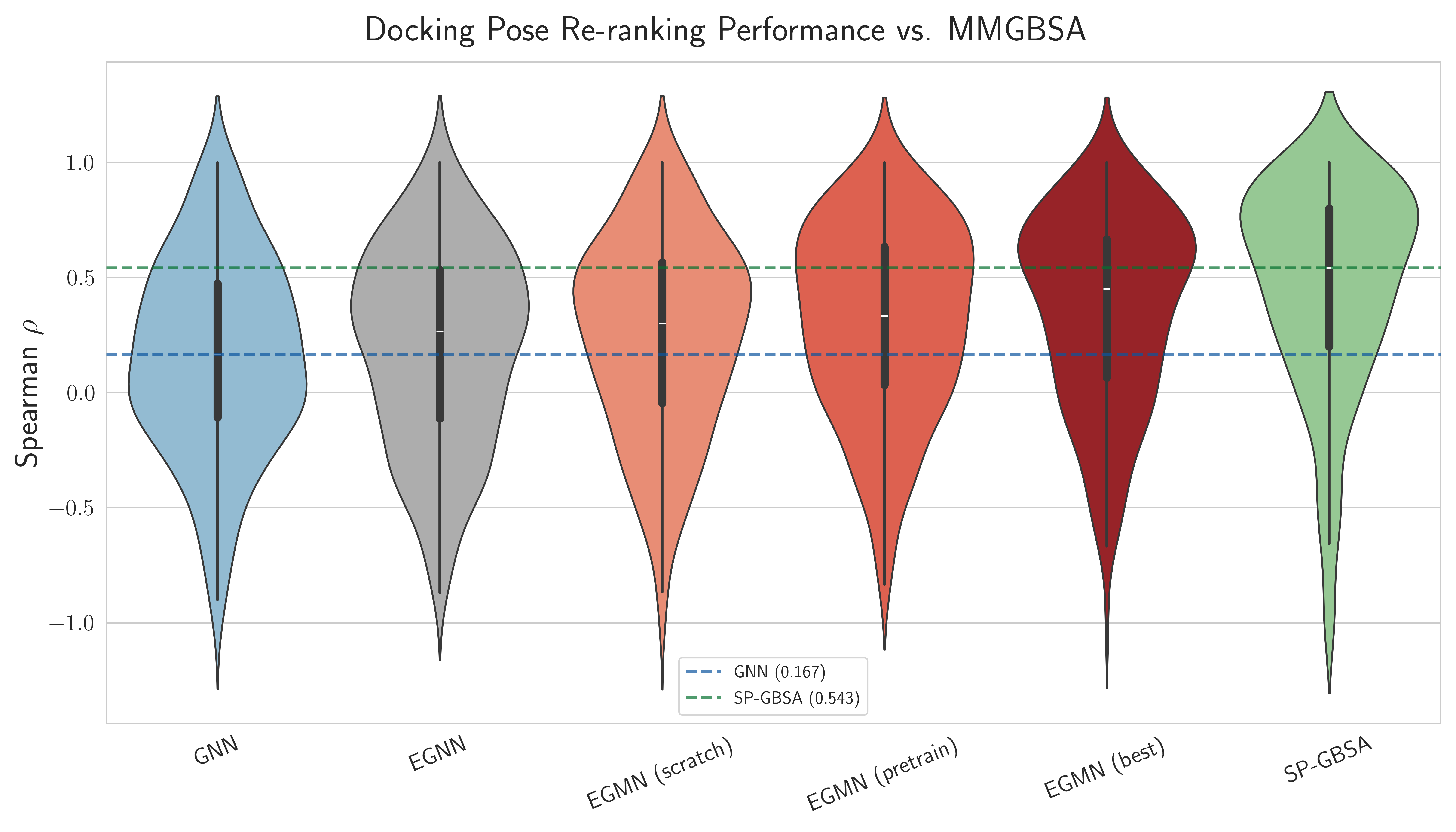}
    \caption{Distributions of the Spearman $\rho$ ranking correlations using the top 5 docking poses for each PDB structure with the MMGBSA score using the MD simulations for each pose. Each metric is computed using the 5 docking poses for each PDB structure separately. Distributions are given across all PDB entries from each of the test sets from the 5-fold splits. Dark dashed lines represent the median $\rho$ for the GNN (blue) and SP-GBSA (Green) baseline models.}
    \label{fig:spgbsa_vs_surgbsa_dist}
\end{figure}

\begin{table}[ht]
    \centering
    \begin{tabular}{c|c|c|c}
        \textbf{Model} & \textbf{Spearman} $\rho$ & $\Delta\%_{\text{SP-GBSA}}$ & $\Delta\%_{\text{GNN}}$\\ \hline
        
        GNN & .17 & -68.5 & 1.0\\
        EGNN & .27 & -50.0 & 58.8 \\
        EGMN (scratch) & .30 & -44.4 & 76.5\\
        EGMN (pretrain) & .33 & -38.9 & 94.1 \\
        EGMN (best) & .45 & \textbf{-16.7} & \textbf{164.7} \\ \hline
        SP-GBSA & .54 & 1.0 & 217.6\\

    \end{tabular}
    \caption{Median re-scoring performance on the test using the top 5 docking poses and their corresponding MMGBSA scores as ground truth. $\Delta\%_{\text{SP-GBSA}}$ and $\Delta\%_{\text{GNN}}$ represent the percentage change in the median docking pose re-ranking performance versus SP-GBSA and the GNN as the respective baseline models. Our model using best EGMN encoder experiences the lowest ranking performance degradation of -16.7\% versus the SP-GBSA model, an improvement of 36.4\% compared to the next best model. Considering the GNN as the baseline, our model using the best EGMN encoder demonstrates an improvement of 164.7\% in docking pose re-ranking performance. Results for the best encoder network used in SurGBSA model development is shown in bold.}
    \label{tab:median_ranking_docking}
\end{table}

\begin{table}[ht]
    \centering
    \begin{tabular}{c|c}
        \textbf{Model} & \textbf{Correct Top Pose \%} \\ \hline
        
        GNN & 30.0 \\
        EGNN & 33.1 \\
        EGMN (scratch) & 38.0\\
        EGMN (pretrain) & 40.1 \\
        EGMN (best) & \textbf{44.7} \\ \hline
        SP-GBSA & 45.1\\

    \end{tabular}
    \caption{Comparison of each model for correctly identifying the top MMGBSA ranked pose for each PDB structure. We show additionally the SP-GBSA model performance as the baseline approach for efficient pose re-ranking. We observed a trend showing increasing model complexity consistently improved ranking performance. Our best model based on the EGMN encoder achieves the best overall ranking performance for top pose identification among the neural network approaches while comparing favorably versus the SP-GBSA baseline, where the difference performance is only $-0.4\%$, the difference of only a single ranking.}
    \label{tab:correct_pose}
\end{table}

In Figure~\ref{fig:spgbsa_vs_surgbsa_dist} and Table~\ref{tab:median_ranking_docking} we detail the comparison in docking pose re-ranking with respect to the MMGBSA score based on all docking pose MD snapshots versus SurGBSA, using the various molecular encoder networks considered in this work, and the SP-GBSA baseline. Although our results show that SP-GBSA achieves the highest mean Spearman $\rho$ taken over all PDB structures in each test set, the difference in this ranking performance versus our best SurGBSA model is $-16.7\%$. Compared to using our baseline GNN encoder module, the improvement by our best SurGBSA model is significant at $164.7\%$.
Lastly, in Table \ref{tab:correct_pose} we give the results for correct pose identification. For each PDB structure, we presented each model with the five corresponding docking poses we generated. The poses are then ranked according to both the model prediction and to their ensemble-averaged MMGBSA energies. If the model top ranking agrees with the MMGBSA ranking, then we count this as a success. We again note a trend in terms of model complexity with our best overall method using the physics-informed EGMN extended training run. Compared to the baseline scoring method SP-GBSA, the difference is only a single test sample.

\subsection{Efficiency of SurGBSA}\label{subsec:surgbsa_eff}

\begin{table}[ht]
    \centering
    \begin{tabular}{c|c|c|c}
         \textbf{Method} &
         \textbf{Host} & \textbf{Avg. Scoring (s)} & \textbf{Speedup}\\ \hline

         SP-GBSA~\cite{Miller2012-tx}  & CPU (Dane) & 8.378 & 1.0\\
        SurGBSA & AMD MI300A (Tuolumne) & 0.002 & 4,189.0\\
        SurGBSA & Nvidia H100 (Matrix) & 0.0003 & 27,926.7 \\

    \end{tabular}
        \caption{Average scoring latency normalized per docking pose. Single-point MMGBSA is denoted as SP-GBSA. To collect the SP-GBSA latency, we extract the timing information from the MMPBSA.py script provided in AmberTools~\cite{Miller2012-tx}. MMGBSA represents the timing information collected from the MMPBSA.py.MPI script. For SP-GBSA, we measured the timing over all poses in our dataset and report the average. For MMGBSA, we used the full trajectories and all available CPU cores on Dane (112), and we report the average over all poses in our dataset. SurGBSA latency measurements measure the overhead of the forward pass through the model to compute a prediction, normalized to give the unit cost per conformation. SurGBSA scoring on the H100 is approximately 6.7$\times$ faster than the MI300A. When comparing to SP-GBSA, SurGBSA using the H100 GPU achieves a speedup of approximately 27,927$\times$.}

    \label{tab:mmgbsa_cost}
\end{table}

The MMGBSA score itself is relatively expensive compared to docking scoring functions as it requires MD simulation to sample conformations to improve its accuracy~\cite{Wang2019-ja}. In Section~\ref{subsec:surgbsa_result} we discussed the ability to accurately recapitulate the MMGBSA score through a physics-informed surrogate model. Next, the surrogate model's computational efficiency is compared to the MMGBSA score. In Table \ref{tab:mmgbsa_cost} we provide performance analysis collected on HPC clusters (Dane, Matrix, and Tuolumne). We used the MMPBSA.py~\cite{Miller2012-tx} implementation included in the Amber 22 MD simulation toolkit.
We only considered the Generalized Born (GB) solvation model as it is more efficient and serves as a more competitive baseline while scaling reasonably to the full dataset we consider in this work. 
As the MMGBSA calculation is parallelized over the frame dimension, SP-GBSA can only take advantage of a single core of the CPU per frame. Our results show that for the latency required to score a single docking pose versus the most competitive baseline, SP-GBSA, our SurGBSA model can achieve a speed-up improvement of approximately 27,927$\times$. 
Given the improved efficiency in conjunction with docking pose ranking performance as compared to the SP-GBSA and MMGBSA baselines, we believe that SurGBSA is capable of serving as a scalable method for docking pose re-scoring that can be used to supplement more expensive methods, including MMGBSA based on MD, by providing an efficient and comparatively accurate predictor. SurGBSA does not require MD for inference; thus, it can be applied to significantly larger datasets of docking poses by several orders of magnitude, improving the size of the tractable search space. This scalability allows SurGBSA to augment the more expensive MD-based scoring methods as a pre-processing step for ranking 3D structures of protein-ligand co-complexes.

\section{Conclusion}\label{sec:conclusion}


SurGBSA is a new surrogate model for physics-based scoring of binding affinity of protein-ligand co-complex structures. Extending previous physics-informed pre-training, the new model learns an accurate and efficient mapping between molecular structure phase space and the corresponding binding free energy surface using the MMGBSA score. We validate our approach using a rigorous train/test split of representative protein-ligand structures and cross-validation with multiple random seeds for each evaluated model architecture. Results demonstrate that the SurGBSA model, which combines ProtMD~\cite{Wu2022-bw} physics-informed pre-training with learning a mapping between molecular conformation and binding free energy produces a more robust representation compared to methods not trained on molecular dynamics trajectories. The SurGBSA model differs from the baseline physics-based single point (SP) GBSA calculation by a single ranking (i.e. $-0.4\%$), when evaluated on a challenging pose ranking problem, with approximately a 27,927$\times$ speedup. 

Future work will investigate improving the pre-training stage using recently published datasets such as ATLAS~\cite{Vander-Meersche2024-vs} which contains 1390 standardized MD simulations for a variety of PDB structures. Improving the inductive biases of the EGMN architecture to more efficiently model the molecular structure dynamics could improve scaling to larger structures. Adding new pre-training tasks and more training examples should continue to improve SurGBSA model accuracy in downstream applications. Physics-informed pretraining can contribute to improving the collection of predictive protein-ligand interaction models.

\section{Methods}\label{sec:methods}

In this section we first describe the MD dataset used as the basis for the SurGBSA model development. We then introduce background on the MMGBSA method and describe the tools used to compute the MMGBSA data used in this work. We then provide descriptions of each graph encoder network architecture and the MLP architecture used for learning the MMGBSA score based on the graph encoder output embeddings. Finally, we describe the model hyperparameter optimization and training.

\subsection{Dataset}\label{subsec_method_dataset}

We use the publicly available collection of protein-ligand co-complexes available in the PDBBind 2019. In order to collect a representative subsampling of the full PDBBind, which contains over 18k unique co-complex structures, we use the PDBBind Core set which is a subset of 285 structures that span the structural space present in the refined set. We successfully computed 5 docking poses for 244 structures out of the full 285, resulting in 6 structures for each of the 244 co-complexes. Thus we have 1464 total ``poses'' including the crystal structure pose. We used Amber to collect 10 nanosecond atomistic molecular dynamics simulations for each of the 1464 poses with snapshots collected every 10 picoseconds, resulting in 1000 frames of simulation per each of the 1464 poses. For each of the snapshots, we collected an MMGBSA score to estimate the binding free energy $\Delta G$. Thus, in total, our full MD dataset thus contains 1,464,000 frames of simulation with corresponding MMGBSA scores for each frame. 

To prepare the dataset for learning, we filter the simulation frames based upon the pocket-ligand co-complex structure to the starting frame, excluding all frames beyond \SI{4}{\angstrom} to account for instabilities in the simulations while retaining meaningful fluctuations and conformational changes. We also filter out frames with MMGBSA scores greater than or equal to -10 and with MMGBSA scores less than -100. We then scaled the MMGBSA scores by dividing by a constant factor of -10, as we found that it leads to better numerical stability during training. We only keep the atoms within a radius of 6 angstroms from the ligand and use a maximum of 600 atoms while we keep the full ligand in each frame. We used the MDAnalysis~\cite{Michaud-Agrawal2011-bp, Gowers2016-nc} python library to process the MD simulations. We aligned each trajectory with the respective crystal structure simulation and collected the RMSD for each frame with respect to the crystal pose. We filter out all water molecules, metals, and ions. To measure generalization performance, we considered a K-fold split of the core set complexes with $k=5$, the default value of the scikit-learn utility, resulting in approximately 80\% of the data used for training and 20\% for testing. In total, after filtering based on the MMGBSA score and pocket-ligand co-complex starting point RMSD, we retain a total of 1,070,272 frames of simulation out of the 1.4 million ($76\%$) collected from the MD simulations. We encode the atom types and positions as input for the models.

\subsection{MD Simulations}
We followed a standardized protocol for all MD simulations in this work, using the Amber22~\cite{amber2022} MD simulation software package. For each PDB structure, we used the conveyorLC~\cite{Zhang2022-jx} parallel docking program to generate ligand poses, selecting the top 5 poses for each PDB co-complex according to the Vina scoring function. Then for each pose, we used Sander to perform energy minimization for 1000 steps. The minimized structures were then heated to 300K under the NVT ensemble using a restraint on the protein backbone $\text{C-}\alpha$ atoms of 1.0 kcal/mol for 200ps. The systems were then equilibrated to reach a pressure of 1 atm over 4.8 ns, using a weaker restraint on the $\text{C-}\alpha$ atoms of 0.25 kcal/mol. Finally, we performed unrestrained production MD under the NPT ensemble for 10ns, with snapshots collected every 10 ps.

\subsection{MMGBSA}\label{subsec:mmgbsa}
The MMGBSA method is an end-point free energy method that is more computationally efficient compared to Alchemical free energy (AFE) methods while being more accurate than docking scoring functions~\cite{Wang2019-ja}. Due to these properties, MMGBSA has found application for evaluating docking poses and predicting binding affinities~\cite{Wang2019-ja}. 
For our baselines, we use the MMPBSA.py script provided in the AmberTools22 software package~\cite{amber2022}. We calculated single-point MMGBSA scores for each simulation frame in our dataset. We formed two distinct MMGBSA baselines; (1) SP-GBSA: uses the docked or crystal pose and the corresponding single-point MMGBSA score, and (2) MMGBSA which is based on the average of the single-point MMGBSA scores for each docking pose or crystal structure MD trajectory. SP-GBSA serves to function as a realistic efficient approximation to conserve computational resources when selecting ligands and their poses for further MD simulation. MMGBSA serves to represent the most accurate version of the re-scoring function based on all available conformational dynamics. For the MMGBSA calculations, the Generalized Born implicit solvent model 5 was performed in pure water with salt contribution of 0 (igb=5, saltcon=0.0).

\subsection{Graph Neural Networks}
\label{subsec:background_gnn}
Graph Neural Networks~\cite{Gilmer2017-ul,Bruna2013-kf,Kipf2016-sp} operate on undirected structured graph data and are a natural choice for molecular representation learning~\cite{Gilmer2017-ul, Duvenaud2015-kl, Wang2022-hg}. We based our GNN on the implementation provided by \citet{Satorras2021-tw}.
Given a graph $\mathcal{G} = (\mathcal{V}, \mathcal{E})$ and node embeddings $\rmh^l =\{\rmh_0^l, \dots, \rmh_{M-1}^l\}$ where $\rmh^l_i \in \mathbb{R}^n$, the Graph Convolution Layer (GCL) computes a permutation equivariant transformation on the node embeddings $\rmh^{l+1}=\text{GCL}[\rmh^{l}, \mathcal{E}]$. The GCL is defined by the following equations:
\begin{align}
\label{eq:gnn_edge_message}
\rmm_{ij} &= \phi_e(\rmh_i^l, \rmh_j^l, a_{ij}) \\
\label{eq:gnn_message_pool}
\rmm_i &= \sum_{j \in \mathcal{N}(i)} \rmm_{i j} \\
\label{eq:gnn_node_message}
    \rmh_i^{l+1} &= \phi_h(\rmh_i^l, \rmm_i)
\end{align}

where $\mathcal{N}(i)$ denotes the neighbors of node $v_i$, $\phi_e$ and $\phi_h$ are learned differentiable functions. The GNN is non-equivariant~\cite{Satorras2021-tw}, which means that the GNN is sample inefficient for processing 3D molecular structure data as it will fail to generalize over rotations and translations of the molecules~\cite{Satorras2021-tw}.

\subsection{E(n) Equivariant Graph Neural Networks}\label{subsec:background_egnn}

The Equivariant Graph Convolutional Layer (EGCL)~\cite{Satorras2021-tw} improves the GCL by introducing $\text{E}(n)$ equivariant operations, which improves sample efficiency when processing different conformations of the same molecule(s) as compared to the GNN. We based our EGCL on the implementation provided by \citet{Satorras2021-tw}. Given the set of node embeddings $\rmh^l$, edge information $\mathcal{E}=(e_{ij})$, coordinate embeddings $\rmx^l =\{\rmx_0^l, \dots, \rmx_{M-1}^l\}$ where $\rmx_i^{l} \in \mathbb{R}^3$, the EGCL layer computes a transformation on the node embeddings and coordinates $\rmh_i^{l+1}, \rmx_i^{l+1} = \text{EGCL}[\rmh_i^l, \rmx_i^l, \mathcal{E}]$. The EGCL is defined by the following equations:
\begin{align}\label{eq:method_edge}
\rmm_{ij} &=\phi_{e}\left(\rmh_{i}^{l}, \rmh_{j}^{l},\left\|\rmx_{i}^{l}-\rmx_{j}^{l}\right\|^{2}, a_{i j}\right) \\ \label{eq:method_coords}
\rmx_{i}^{l+1} &=\rmx_{i}^{l}+ C\sum_{j \neq i}\left(\rmx_{i}^{l}-\rmx_{j}^{l}\right) \phi_{x}\left(\rmm_{ij}\right) \\
 \label{eq:method_agg}
\rmm_{i} &=\sum_{j \in \mathcal{N}(i)} \rmm_{ij} \\ 
\label{eq:method_node}
\rmh_{i}^{l+1} &=\phi_{h}\left(\rmh_{i}^l, \rmm_{i}\right)
\end{align}

where $C$ is a scaling constant, $\phi_x$ is a learned differentiable function, and $\left\|\rmx_{i}^{l}-\rmx_{j}^{l}\right\|^{2}$ is the relative squared distance between $v_i$ and $v_j$. From the output of the EGCL layer, the node embeddings $\rmh^{l+1}$ are $\text{E}(n)$ invariant and the node coordinates $\rmx^{l+1}$ are $\text{E}(n)$ equivariant to $\rmx^l$. The EGNN improves sample efficiency in contrast to the GNN in that the equivariant properties allow for the model to recognize different conformations of the same molecule, thus improving generalization under rotation and translations.

\subsection{Equivariant Graph Matching Networks}\label{subsec:egmn}

The Equivariant
Graph Matching Network (EGMN) architecture implemented by \citet{Wu2022-bw} is closely related to the Independent E(3)-Equivariant Graph Matching Network (IEGMNs)~\cite{Ganea2021-tj}. \citet{Wu2022-bw} treat the time dimension in MD simulations by considering each time-dependent conformation as distinct molecular graphs.
Specifically, the EGMN assumes distinct graphs $\mathcal{G}=(\mathcal{G}_1$,  $\mathcal{G}_2)$ and additional cross-graph edges $\Ecal_{12}$. The EGMN takes as input the set of node embeddings and coordinates at time $t$ and layer $l$ then computes the transformation $\rmh_1^{(t),l+1}$, $\rmx_1^{(t),l+1}, \rmh_2^{(t), l+1}, \rmx_2^{(t), l+1} = \text{EGMN}(\rmh_1^{(t), l}, \rmx_1^{(t), l}, \rmh_2^{(t), l}, \rmx_2^{(t), l})$ defined by the following equations:

\begin{align}
    \mb_{ij} &= \phi_e(\hb_i^{(t), l},\hb_j^{(t), l},x_{ij}^{(t)}), ~\forall e_{ij}\in\Ecal_{1}^{(t)}\cup\Ecal_{2}^{(t)} \\
\boldsymbol{\mu}_{ij} &= a_{ij}\hb_j^{(t), l} \cdot \phi_d(x_{ij}^{(t), l}), ~\forall e_{ij} \in \Ecal_{12}^{(t)}\\
    \mb_i &= \sum_{j \in \mathcal{N}(i)}\mb_{ij}
    \\
\boldsymbol{\mu}_i &= \sum_{j\in\mathcal{N}(i)}\boldsymbol{\mu}_{ij}, ~\forall e_{ij}\in \Ecal_{12}^{(t)}
    \\
    \xb_{i}^{(t), l+1} &= x_{i}^{(t), l} + \sum_{j \in \Ncal(i)}\big(x_{i}^{(t), l} - x_{j}^{(t), l}\big) \phi_{x}(i,j)\\
     \hb_i^{(t),l+1} &= \phi_{h}(\hb_{i}^{(t), l}, \mb_i, \boldsymbol{\mu}_{i})
\end{align}

where $x_{ij}^{(t)} = \exp(-\norm{\xb_i^{(t), l}-\xb_j^{(t), l}}^2/ \sigma))$ corresponds to L2 normalized euclidean distances between nodes $i$ and $j$ in a given graph $\mathcal{G}$. $\phi_h$, $\phi_e$, $\phi_d$, and $\phi_d$ are learnable parameterized functions. Specifically, $\phi_x = \phi_m(m_{ij})$ if $e_{ij} \in \Ecal_{1}^{(t)} \cup \Ecal_{2}^{(t)}$ otherwise $\phi_x = \phi_{\mu} (\mu_{ij})$ when $e_{ij} \in \Ecal_{12}^{(t)}$, where $\phi_m$ and $\phi_\mu$ are learnable parameterized functions specific to inter-graph or intra-graph edge types.
The attention coefficient $a_{ij}$ between the nodes $v_i$ and $v_j$ is calculated as:
\vspace{-0.1cm}
\begin{align}
    a_{ij} &= \frac{\text{exp}(\langle \psi^q(\hb_i^{(t), l}),\psi^k(\hb_j^{(t), l})\rangle)}{\sum_{j'}\text{exp}(\langle \psi^q( \hb_i^{(t), l}), \psi^k(\hb_{j'}^{(t), l})\rangle)},
\end{align}

where $\psi^q$ and $\psi^k$ are trainable neural networks.

\subsection{Supervised models}\label{subsec:supervised_models}

To learn the MMGBSA score, we first aggregated the learned embeddings extracted from each graph encoder network:

\begin{align}
    \rmh^{\mathcal{L}}, \rmx^{\mathcal{L}} = \text{GraphEncoder}[\rmh^{0}, \rmx^{0}, \Ecal; ~\mathcal{L}]
\end{align}

where $L$ is the number of graph convolution layers and $\rmh^0$, $\rmx^0$ are the initial node embeddings and coordinates, respectively. In the case of the GNN encoder, we treat $\rmx^L=\rmx^{0}$ as it does not output a coordinate transformation. We discard $\rmx^{L}$ and keep only the node embeddings $\rmh^{L}$ for the rest of the computation. To compute the embedding of the graph $\rmh^{G}$, we use average pooling over the atomic dimension:

\begin{align}\label{eq:avg_pool}
    \rmh^{G} = \frac{1}{\vert \mathcal{V} \vert} \sum_{i \in \mathcal{V}} \rmh^{\mathcal{L}}_i
\end{align}

where $\mathcal{V} = \{\mathcal{V}_1 \cup \mathcal{V}_2\}$.
Then, we use an MLP network to map $\rmh^{G}$ to the scalar output:

\begin{align}
    \hat{y} = W_{1}(\sigma(W_{0}\rmh^{G} + b_0)) + b_1
\end{align}

where $W_0$ and $W_1$ are weight matrices, $b_0$ and $b_1$ are bias vectors, and $\sigma$ is the ReLU non-linearity.

\subsection{Model training and optimization}\label{subsec:model_train_optim}

We considered a grid search over various settings of the hyperparameters. We considered batch sizes of 64, 128, and 256. We considered learning rates of $1e^{-3}$ and $1e^{-4}$. We train all models using the AdamW optimizer with no weight decay in order to constrain the hyperparameter optimization and focus our study on the learning dynamics induced by the choice of learning rate and batch size instead. We considered 5 distinct random splits of the CASF-2016 complexes, which represent a representative sub-sampling of the PDBBind 2019 refined set structures. We treat the graphs from different frames independently, i.e. as separate graphs and train all layers of the constituent molecular graph encoder and MMGBSA prediction head using the MSE loss. We trained a model, independently, for each unique configuration of hyperparameters on each of the 5 folds. Further, we replicated each experiment using 3 random seeds, thus for each configuration of hyperparameters we have a total of 15 independent training runs. In total, we conducted 5400 training runs for our study using the PyTorch~\cite{Paszke2019-us} Distributed Data-Parallel (DDP) backend (v. 2.5.1), AMD ROCm 6.2, on the Tuloumne HPC cluster equipped with the AMD MI300A Accelerated Processing Unit (APU)~\cite{amd_mi300a}. In contrast to the GPU accelerator where the CPU and GPU access disjoint memory spaces, the APU combines on-chip memory resources that are shared across both the CPU and GPU, improving memory bandwidth between the processing units~\cite{amd_mi300a}. For training runs using only static information (i.e. crystal structure and docking poses) we used 1 node with 4 total APUs per job. For training runs using the MD conformations, we used 4 nodes per job with a total of 16 APUs per job. We trained all models for a maximum of 100 epochs, regardless of the dataset. Using the best model configuration after 100 epochs, the ProtMD pre-trained EGMN encoder network, we extended training to 600 epochs and report this combination as our best SurGBSA model. Training our best model for 600 epochs took approximately 68.5 APU hours per device ($\sim3$ days). We report the results using the models with the hyperparameters that achieve the best average spearman correlation, averaged by PDB structure, on the test set.

\bibliography{bibliography}


\begin{thebibliography}{50}
\ifx \bisbn   \undefined \def \bisbn  #1{ISBN #1}\fi
\ifx \binits  \undefined \def \binits#1{#1}\fi
\ifx \bauthor  \undefined \def \bauthor#1{#1}\fi
\ifx \batitle  \undefined \def \batitle#1{#1}\fi
\ifx \bjtitle  \undefined \def \bjtitle#1{#1}\fi
\ifx \bvolume  \undefined \def \bvolume#1{\textbf{#1}}\fi
\ifx \byear  \undefined \def \byear#1{#1}\fi
\ifx \bissue  \undefined \def \bissue#1{#1}\fi
\ifx \bfpage  \undefined \def \bfpage#1{#1}\fi
\ifx \blpage  \undefined \def \blpage #1{#1}\fi
\ifx \burl  \undefined \def \burl#1{\textsf{#1}}\fi
\ifx \doiurl  \undefined \def \doiurl#1{\url{https://doi.org/#1}}\fi
\ifx \betal  \undefined \def \betal{\textit{et al.}}\fi
\ifx \binstitute  \undefined \def \binstitute#1{#1}\fi
\ifx \binstitutionaled  \undefined \def \binstitutionaled#1{#1}\fi
\ifx \bctitle  \undefined \def \bctitle#1{#1}\fi
\ifx \beditor  \undefined \def \beditor#1{#1}\fi
\ifx \bpublisher  \undefined \def \bpublisher#1{#1}\fi
\ifx \bbtitle  \undefined \def \bbtitle#1{#1}\fi
\ifx \bedition  \undefined \def \bedition#1{#1}\fi
\ifx \bseriesno  \undefined \def \bseriesno#1{#1}\fi
\ifx \blocation  \undefined \def \blocation#1{#1}\fi
\ifx \bsertitle  \undefined \def \bsertitle#1{#1}\fi
\ifx \bsnm \undefined \def \bsnm#1{#1}\fi
\ifx \bsuffix \undefined \def \bsuffix#1{#1}\fi
\ifx \bparticle \undefined \def \bparticle#1{#1}\fi
\ifx \barticle \undefined \def \barticle#1{#1}\fi
\bibcommenthead
\ifx \bconfdate \undefined \def \bconfdate #1{#1}\fi
\ifx \botherref \undefined \def \botherref #1{#1}\fi
\ifx \url \undefined \def \url#1{\textsf{#1}}\fi
\ifx \bchapter \undefined \def \bchapter#1{#1}\fi
\ifx \bbook \undefined \def \bbook#1{#1}\fi
\ifx \bcomment \undefined \def \bcomment#1{#1}\fi
\ifx \oauthor \undefined \def \oauthor#1{#1}\fi
\ifx \citeauthoryear \undefined \def \citeauthoryear#1{#1}\fi
\ifx \endbibitem  \undefined \def \endbibitem {}\fi
\ifx \bconflocation  \undefined \def \bconflocation#1{#1}\fi
\ifx \arxivurl  \undefined \def \arxivurl#1{\textsf{#1}}\fi
\csname PreBibitemsHook\endcsname

\bibitem[\protect\citeauthoryear{Jumper et~al.}{2021}]{Jumper2021-xh}
\begin{barticle}
\bauthor{\bsnm{Jumper}, \binits{J.}},
\bauthor{\bsnm{Evans}, \binits{R.}},
\bauthor{\bsnm{Pritzel}, \binits{A.}},
\bauthor{\bsnm{Green}, \binits{T.}},
\bauthor{\bsnm{Figurnov}, \binits{M.}},
\bauthor{\bsnm{Ronneberger}, \binits{O.}},
\bauthor{\bsnm{Tunyasuvunakool}, \binits{K.}},
\bauthor{\bsnm{Bates}, \binits{R.}},
\bauthor{\bsnm{Žídek}, \binits{A.}},
\bauthor{\bsnm{Potapenko}, \binits{A.}},
\bauthor{\bsnm{Bridgland}, \binits{A.}},
\bauthor{\bsnm{Meyer}, \binits{C.}},
\bauthor{\bsnm{Kohl}, \binits{S.A.A.}},
\bauthor{\bsnm{Ballard}, \binits{A.J.}},
\bauthor{\bsnm{Cowie}, \binits{A.}},
\bauthor{\bsnm{Romera-Paredes}, \binits{B.}},
\bauthor{\bsnm{Nikolov}, \binits{S.}},
\bauthor{\bsnm{Jain}, \binits{R.}},
\bauthor{\bsnm{Adler}, \binits{J.}},
\bauthor{\bsnm{Back}, \binits{T.}},
\bauthor{\bsnm{Petersen}, \binits{S.}},
\bauthor{\bsnm{Reiman}, \binits{D.}},
\bauthor{\bsnm{Clancy}, \binits{E.}},
\bauthor{\bsnm{Zielinski}, \binits{M.}},
\bauthor{\bsnm{Steinegger}, \binits{M.}},
\bauthor{\bsnm{Pacholska}, \binits{M.}},
\bauthor{\bsnm{Berghammer}, \binits{T.}},
\bauthor{\bsnm{Bodenstein}, \binits{S.}},
\bauthor{\bsnm{Silver}, \binits{D.}},
\bauthor{\bsnm{Vinyals}, \binits{O.}},
\bauthor{\bsnm{Senior}, \binits{A.W.}},
\bauthor{\bsnm{Kavukcuoglu}, \binits{K.}},
\bauthor{\bsnm{Kohli}, \binits{P.}},
\bauthor{\bsnm{Hassabis}, \binits{D.}}:
\batitle{Highly accurate protein structure prediction with {AlphaFold}}.
\bjtitle{Nature}
\bvolume{596}(\bissue{7873}),
\bfpage{583}--\blpage{589}
(\byear{2021})
\end{barticle}
\endbibitem

\bibitem[\protect\citeauthoryear{Baek et~al.}{2021}]{Baek2021-wq}
\begin{barticle}
\bauthor{\bsnm{Baek}, \binits{M.}},
\bauthor{\bsnm{DiMaio}, \binits{F.}},
\bauthor{\bsnm{Anishchenko}, \binits{I.}},
\bauthor{\bsnm{Dauparas}, \binits{J.}},
\bauthor{\bsnm{Ovchinnikov}, \binits{S.}},
\bauthor{\bsnm{Lee}, \binits{G.R.}},
\bauthor{\bsnm{Wang}, \binits{J.}},
\bauthor{\bsnm{Cong}, \binits{Q.}},
\bauthor{\bsnm{Kinch}, \binits{L.N.}},
\bauthor{\bsnm{Schaeffer}, \binits{R.D.}},
\bauthor{\bsnm{Millán}, \binits{C.}},
\bauthor{\bsnm{Park}, \binits{H.}},
\bauthor{\bsnm{Adams}, \binits{C.}},
\bauthor{\bsnm{Glassman}, \binits{C.R.}},
\bauthor{\bsnm{DeGiovanni}, \binits{A.}},
\bauthor{\bsnm{Pereira}, \binits{J.H.}},
\bauthor{\bsnm{Rodrigues}, \binits{A.V.}},
\bauthor{\bsnm{Dijk}, \binits{A.A.}},
\bauthor{\bsnm{Ebrecht}, \binits{A.C.}},
\bauthor{\bsnm{Opperman}, \binits{D.J.}},
\bauthor{\bsnm{Sagmeister}, \binits{T.}},
\bauthor{\bsnm{Buhlheller}, \binits{C.}},
\bauthor{\bsnm{Pavkov-Keller}, \binits{T.}},
\bauthor{\bsnm{Rathinaswamy}, \binits{M.K.}},
\bauthor{\bsnm{Dalwadi}, \binits{U.}},
\bauthor{\bsnm{Yip}, \binits{C.K.}},
\bauthor{\bsnm{Burke}, \binits{J.E.}},
\bauthor{\bsnm{Garcia}, \binits{K.C.}},
\bauthor{\bsnm{Grishin}, \binits{N.V.}},
\bauthor{\bsnm{Adams}, \binits{P.D.}},
\bauthor{\bsnm{Read}, \binits{R.J.}},
\bauthor{\bsnm{Baker}, \binits{D.}}:
\batitle{Accurate prediction of protein structures and interactions using a
  three-track neural network}.
\bjtitle{Science}
\bvolume{373}(\bissue{6557}),
\bfpage{871}--\blpage{876}
(\byear{2021})
\end{barticle}
\endbibitem

\bibitem[\protect\citeauthoryear{Abramson et~al.}{2024}]{Abramson2024-xu}
\begin{barticle}
\bauthor{\bsnm{Abramson}, \binits{J.}},
\bauthor{\bsnm{Adler}, \binits{J.}},
\bauthor{\bsnm{Dunger}, \binits{J.}},
\bauthor{\bsnm{Evans}, \binits{R.}},
\bauthor{\bsnm{Green}, \binits{T.}},
\bauthor{\bsnm{Pritzel}, \binits{A.}},
\bauthor{\bsnm{Ronneberger}, \binits{O.}},
\bauthor{\bsnm{Willmore}, \binits{L.}},
\bauthor{\bsnm{Ballard}, \binits{A.J.}},
\bauthor{\bsnm{Bambrick}, \binits{J.}},
\bauthor{\bsnm{Bodenstein}, \binits{S.W.}},
\bauthor{\bsnm{Evans}, \binits{D.A.}},
\bauthor{\bsnm{Hung}, \binits{C.-C.}},
\bauthor{\bsnm{O'Neill}, \binits{M.}},
\bauthor{\bsnm{Reiman}, \binits{D.}},
\bauthor{\bsnm{Tunyasuvunakool}, \binits{K.}},
\bauthor{\bsnm{Wu}, \binits{Z.}},
\bauthor{\bsnm{Žemgulytė}, \binits{A.}},
\bauthor{\bsnm{Arvaniti}, \binits{E.}},
\bauthor{\bsnm{Beattie}, \binits{C.}},
\bauthor{\bsnm{Bertolli}, \binits{O.}},
\bauthor{\bsnm{Bridgland}, \binits{A.}},
\bauthor{\bsnm{Cherepanov}, \binits{A.}},
\bauthor{\bsnm{Congreve}, \binits{M.}},
\bauthor{\bsnm{Cowen-Rivers}, \binits{A.I.}},
\bauthor{\bsnm{Cowie}, \binits{A.}},
\bauthor{\bsnm{Figurnov}, \binits{M.}},
\bauthor{\bsnm{Fuchs}, \binits{F.B.}},
\bauthor{\bsnm{Gladman}, \binits{H.}},
\bauthor{\bsnm{Jain}, \binits{R.}},
\bauthor{\bsnm{Khan}, \binits{Y.A.}},
\bauthor{\bsnm{Low}, \binits{C.M.R.}},
\bauthor{\bsnm{Perlin}, \binits{K.}},
\bauthor{\bsnm{Potapenko}, \binits{A.}},
\bauthor{\bsnm{Savy}, \binits{P.}},
\bauthor{\bsnm{Singh}, \binits{S.}},
\bauthor{\bsnm{Stecula}, \binits{A.}},
\bauthor{\bsnm{Thillaisundaram}, \binits{A.}},
\bauthor{\bsnm{Tong}, \binits{C.}},
\bauthor{\bsnm{Yakneen}, \binits{S.}},
\bauthor{\bsnm{Zhong}, \binits{E.D.}},
\bauthor{\bsnm{Zielinski}, \binits{M.}},
\bauthor{\bsnm{Žídek}, \binits{A.}},
\bauthor{\bsnm{Bapst}, \binits{V.}},
\bauthor{\bsnm{Kohli}, \binits{P.}},
\bauthor{\bsnm{Jaderberg}, \binits{M.}},
\bauthor{\bsnm{Hassabis}, \binits{D.}},
\bauthor{\bsnm{Jumper}, \binits{J.M.}}:
\batitle{Accurate structure prediction of biomolecular interactions with
  {AlphaFold} 3}.
\bjtitle{Nature}
\bvolume{630}(\bissue{8016}),
\bfpage{493}--\blpage{500}
(\byear{2024})
\end{barticle}
\endbibitem

\bibitem[\protect\citeauthoryear{Xu et~al.}{2025}]{Xu2025-ih}
\begin{barticle}
\bauthor{\bsnm{Xu}, \binits{Z.}},
\bauthor{\bsnm{Ren}, \binits{F.}},
\bauthor{\bsnm{Wang}, \binits{P.}},
\bauthor{\bsnm{Cao}, \binits{J.}},
\bauthor{\bsnm{Tan}, \binits{C.}},
\bauthor{\bsnm{Ma}, \binits{D.}},
\bauthor{\bsnm{Zhao}, \binits{L.}},
\bauthor{\bsnm{Dai}, \binits{J.}},
\bauthor{\bsnm{Ding}, \binits{Y.}},
\bauthor{\bsnm{Fang}, \binits{H.}},
\bauthor{\bsnm{Li}, \binits{H.}},
\bauthor{\bsnm{Liu}, \binits{H.}},
\bauthor{\bsnm{Luo}, \binits{F.}},
\bauthor{\bsnm{Meng}, \binits{Y.}},
\bauthor{\bsnm{Pan}, \binits{P.}},
\bauthor{\bsnm{Xiang}, \binits{P.}},
\bauthor{\bsnm{Xiao}, \binits{Z.}},
\bauthor{\bsnm{Rao}, \binits{S.}},
\bauthor{\bsnm{Satler}, \binits{C.}},
\bauthor{\bsnm{Liu}, \binits{S.}},
\bauthor{\bsnm{Lv}, \binits{Y.}},
\bauthor{\bsnm{Zhao}, \binits{H.}},
\bauthor{\bsnm{Chen}, \binits{S.}},
\bauthor{\bsnm{Cui}, \binits{H.}},
\bauthor{\bsnm{Korzinkin}, \binits{M.}},
\bauthor{\bsnm{Gennert}, \binits{D.}},
\bauthor{\bsnm{Zhavoronkov}, \binits{A.}}:
\batitle{A generative {AI}-discovered {TNIK} inhibitor for idiopathic pulmonary
  fibrosis: a randomized phase {2a} trial}.
\bjtitle{Nat. Med.}
\bvolume{31}(\bissue{8}),
\bfpage{2602}--\blpage{2610}
(\byear{2025})
\end{barticle}
\endbibitem

\bibitem[\protect\citeauthoryear{Zhang et~al.}{2025}]{Zhang2025-ma}
\begin{barticle}
\bauthor{\bsnm{Zhang}, \binits{K.}},
\bauthor{\bsnm{Yang}, \binits{X.}},
\bauthor{\bsnm{Wang}, \binits{Y.}},
\bauthor{\bsnm{Yu}, \binits{Y.}},
\bauthor{\bsnm{Huang}, \binits{N.}},
\bauthor{\bsnm{Li}, \binits{G.}},
\bauthor{\bsnm{Li}, \binits{X.}},
\bauthor{\bsnm{Wu}, \binits{J.C.}},
\bauthor{\bsnm{Yang}, \binits{S.}}:
\batitle{Artificial intelligence in drug development}.
\bjtitle{Nat. Med.}
\bvolume{31}(\bissue{1}),
\bfpage{45}--\blpage{59}
(\byear{2025})
\end{barticle}
\endbibitem

\bibitem[\protect\citeauthoryear{Wohlwend et~al.}{2024}]{Wohlwend2024-ag}
\begin{botherref}
\oauthor{\bsnm{Wohlwend}, \binits{J.}},
\oauthor{\bsnm{Corso}, \binits{G.}},
\oauthor{\bsnm{Passaro}, \binits{S.}},
\oauthor{\bsnm{Reveiz}, \binits{M.}},
\oauthor{\bsnm{Leidal}, \binits{K.}},
\oauthor{\bsnm{Swiderski}, \binits{W.}},
\oauthor{\bsnm{Portnoi}, \binits{T.}},
\oauthor{\bsnm{Chinn}, \binits{I.}},
\oauthor{\bsnm{Silterra}, \binits{J.}},
\oauthor{\bsnm{Jaakkola}, \binits{T.}},
\oauthor{\bsnm{Barzilay}, \binits{R.}}:
Boltz-1: Democratizing biomolecular interaction modeling.
bioRxiv
(2024)
{\href{https://arxiv.org/abs/2024.11.19.624167}{{bioRxiv:2024.11.19.624167}}}
\end{botherref}
\endbibitem

\bibitem[\protect\citeauthoryear{Passaro et~al.}{2025}]{Passaro2025-ci}
\begin{botherref}
\oauthor{\bsnm{Passaro}, \binits{S.}},
\oauthor{\bsnm{Corso}, \binits{G.}},
\oauthor{\bsnm{Wohlwend}, \binits{J.}},
\oauthor{\bsnm{Reveiz}, \binits{M.}},
\oauthor{\bsnm{Thaler}, \binits{S.}},
\oauthor{\bsnm{Somnath}, \binits{V.R.}},
\oauthor{\bsnm{Getz}, \binits{N.}},
\oauthor{\bsnm{Portnoi}, \binits{T.}},
\oauthor{\bsnm{Roy}, \binits{J.}},
\oauthor{\bsnm{Stark}, \binits{H.}},
\oauthor{\bsnm{Kwabi-Addo}, \binits{D.}},
\oauthor{\bsnm{Beaini}, \binits{D.}},
\oauthor{\bsnm{Jaakkola}, \binits{T.}},
\oauthor{\bsnm{Barzilay}, \binits{R.}}:
Boltz-2: Towards accurate and efficient binding affinity prediction.
bioRxiv
(2025)
{\href{https://arxiv.org/abs/2025.06.14.659707}{{bioRxiv:2025.06.14.659707}}}
\end{botherref}
\endbibitem

\bibitem[\protect\citeauthoryear{Corso et~al.}{2022}]{Corso2022-hd}
\begin{botherref}
\oauthor{\bsnm{Corso}, \binits{G.}},
\oauthor{\bsnm{Stärk}, \binits{H.}},
\oauthor{\bsnm{Jing}, \binits{B.}},
\oauthor{\bsnm{Barzilay}, \binits{R.}},
\oauthor{\bsnm{Jaakkola}, \binits{T.}}:
{DiffDock}: Diffusion steps, twists, and turns for molecular docking.
arXiv [q-bio.BM]
(2022)
{[q-bio.BM]}
\end{botherref}
\endbibitem

\bibitem[\protect\citeauthoryear{Lu et~al.}{2024}]{Lu2024-xh}
\begin{barticle}
\bauthor{\bsnm{Lu}, \binits{W.}},
\bauthor{\bsnm{Zhang}, \binits{J.}},
\bauthor{\bsnm{Huang}, \binits{W.}},
\bauthor{\bsnm{Zhang}, \binits{Z.}},
\bauthor{\bsnm{Jia}, \binits{X.}},
\bauthor{\bsnm{Wang}, \binits{Z.}},
\bauthor{\bsnm{Shi}, \binits{L.}},
\bauthor{\bsnm{Li}, \binits{C.}},
\bauthor{\bsnm{Wolynes}, \binits{P.G.}},
\bauthor{\bsnm{Zheng}, \binits{S.}}:
\batitle{{DynamicBind}: predicting ligand-specific protein-ligand complex
  structure with a deep equivariant generative model}.
\bjtitle{Nat. Commun.}
\bvolume{15}(\bissue{1}),
\bfpage{1071}
(\byear{2024})
\end{barticle}
\endbibitem

\bibitem[\protect\citeauthoryear{Stepniewska-Dziubinska
  et~al.}{2018}]{Stepniewska-Dziubinska2018-fo}
\begin{barticle}
\bauthor{\bsnm{Stepniewska-Dziubinska}, \binits{M.M.}},
\bauthor{\bsnm{Zielenkiewicz}, \binits{P.}},
\bauthor{\bsnm{Siedlecki}, \binits{P.}}:
\batitle{Development and evaluation of a deep learning model for protein-ligand
  binding affinity prediction}.
\bjtitle{Bioinformatics}
\bvolume{34}(\bissue{21}),
\bfpage{3666}--\blpage{3674}
(\byear{2018})
\end{barticle}
\endbibitem

\bibitem[\protect\citeauthoryear{Jiménez et~al.}{2018}]{Jimenez2018-ei}
\begin{barticle}
\bauthor{\bsnm{Jiménez}, \binits{J.}},
\bauthor{\bsnm{Škalič}, \binits{M.}},
\bauthor{\bsnm{Martínez-Rosell}, \binits{G.}},
\bauthor{\bsnm{De~Fabritiis}, \binits{G.}}:
\batitle{{KDEEP}: {Protein–Ligand} absolute binding affinity prediction via
  {3D}-convolutional neural networks}.
\bjtitle{J. Chem. Inf. Model.}
\bvolume{58}(\bissue{2}),
\bfpage{287}--\blpage{296}
(\byear{2018})
\end{barticle}
\endbibitem

\bibitem[\protect\citeauthoryear{Jones et~al.}{2021}]{Jones2021-ar}
\begin{barticle}
\bauthor{\bsnm{Jones}, \binits{D.}},
\bauthor{\bsnm{Kim}, \binits{H.}},
\bauthor{\bsnm{Zhang}, \binits{X.}},
\bauthor{\bsnm{Zemla}, \binits{A.}},
\bauthor{\bsnm{Stevenson}, \binits{G.}},
\bauthor{\bsnm{Bennett}, \binits{W.F.D.}},
\bauthor{\bsnm{Kirshner}, \binits{D.}},
\bauthor{\bsnm{Wong}, \binits{S.E.}},
\bauthor{\bsnm{Lightstone}, \binits{F.C.}},
\bauthor{\bsnm{Allen}, \binits{J.E.}}:
\batitle{Improved protein-ligand binding affinity prediction with
  structure-based deep fusion inference}.
\bjtitle{J. Chem. Inf. Model.}
\bvolume{61}(\bissue{4}),
\bfpage{1583}--\blpage{1592}
(\byear{2021})
\end{barticle}
\endbibitem

\bibitem[\protect\citeauthoryear{Volkov et~al.}{2022}]{Volkov2022-vc}
\begin{botherref}
\oauthor{\bsnm{Volkov}, \binits{M.}},
\oauthor{\bsnm{Turk}, \binits{J.-A.}},
\oauthor{\bsnm{Drizard}, \binits{N.}},
\oauthor{\bsnm{Martin}, \binits{N.}},
\oauthor{\bsnm{Hoffmann}, \binits{B.}},
\oauthor{\bsnm{Gaston-Mathé}, \binits{Y.}},
\oauthor{\bsnm{Rognan}, \binits{D.}}:
On the frustration to predict binding affinities from {Protein–Ligand}
  structures with deep neural networks.
J. Med. Chem.
(2022)
\end{botherref}
\endbibitem

\bibitem[\protect\citeauthoryear{Wang}{2024}]{Wang2024-lx}
\begin{barticle}
\bauthor{\bsnm{Wang}, \binits{H.}}:
\batitle{Prediction of protein-ligand binding affinity via deep learning
  models}.
\bjtitle{Brief. Bioinform.}
\bvolume{25}(\bissue{2}),
\bfpage{081}
(\byear{2024})
\end{barticle}
\endbibitem

\bibitem[\protect\citeauthoryear{Fuchs et~al.}{2020}]{Fuchs2020-cx}
\begin{botherref}
\oauthor{\bsnm{Fuchs}, \binits{F.B.}},
\oauthor{\bsnm{Worrall}, \binits{D.E.}},
\oauthor{\bsnm{Fischer}, \binits{V.}},
\oauthor{\bsnm{Welling}, \binits{M.}}:
{SE}(3)-transformers: {3D} roto-translation equivariant attention networks.
arXiv [cs.LG]
(2020)
{[cs.LG]}
\end{botherref}
\endbibitem

\bibitem[\protect\citeauthoryear{Satorras et~al.}{2021}]{Satorras2021-tw}
\begin{botherref}
\oauthor{\bsnm{Satorras}, \binits{V.G.}},
\oauthor{\bsnm{Hoogeboom}, \binits{E.}},
\oauthor{\bsnm{Welling}, \binits{M.}}:
{E}(n) equivariant graph neural networks.
arXiv [cs.LG]
(2021)
{[cs.LG]}
\end{botherref}
\endbibitem

\bibitem[\protect\citeauthoryear{Bommasani et~al.}{2021}]{Bommasani2021-uo}
\begin{botherref}
\oauthor{\bsnm{Bommasani}, \binits{R.}},
\oauthor{\bsnm{Hudson}, \binits{D.A.}},
\oauthor{\bsnm{Adeli}, \binits{E.}},
\oauthor{\bsnm{Altman}, \binits{R.}},
\oauthor{\bsnm{Arora}, \binits{S.}},
\oauthor{\bsnm{Arx}, \binits{S.}},
\oauthor{\bsnm{Bernstein}, \binits{M.S.}},
\oauthor{\bsnm{Bohg}, \binits{J.}},
\oauthor{\bsnm{Bosselut}, \binits{A.}},
\oauthor{\bsnm{Brunskill}, \binits{E.}},
\oauthor{\bsnm{Brynjolfsson}, \binits{E.}},
\oauthor{\bsnm{Buch}, \binits{S.}},
\oauthor{\bsnm{Card}, \binits{D.}},
\oauthor{\bsnm{Castellon}, \binits{R.}},
\oauthor{\bsnm{Chatterji}, \binits{N.}},
\oauthor{\bsnm{Chen}, \binits{A.}},
\oauthor{\bsnm{Creel}, \binits{K.}},
\oauthor{\bsnm{Davis}, \binits{J.Q.}},
\oauthor{\bsnm{Demszky}, \binits{D.}},
\oauthor{\bsnm{Donahue}, \binits{C.}},
\oauthor{\bsnm{Doumbouya}, \binits{M.}},
\oauthor{\bsnm{Durmus}, \binits{E.}},
\oauthor{\bsnm{Ermon}, \binits{S.}},
\oauthor{\bsnm{Etchemendy}, \binits{J.}},
\oauthor{\bsnm{Ethayarajh}, \binits{K.}},
\oauthor{\bsnm{Fei-Fei}, \binits{L.}},
\oauthor{\bsnm{Finn}, \binits{C.}},
\oauthor{\bsnm{Gale}, \binits{T.}},
\oauthor{\bsnm{Gillespie}, \binits{L.}},
\oauthor{\bsnm{Goel}, \binits{K.}},
\oauthor{\bsnm{Goodman}, \binits{N.}},
\oauthor{\bsnm{Grossman}, \binits{S.}},
\oauthor{\bsnm{Guha}, \binits{N.}},
\oauthor{\bsnm{Hashimoto}, \binits{T.}},
\oauthor{\bsnm{Henderson}, \binits{P.}},
\oauthor{\bsnm{Hewitt}, \binits{J.}},
\oauthor{\bsnm{Ho}, \binits{D.E.}},
\oauthor{\bsnm{Hong}, \binits{J.}},
\oauthor{\bsnm{Hsu}, \binits{K.}},
\oauthor{\bsnm{Huang}, \binits{J.}},
\oauthor{\bsnm{Icard}, \binits{T.}},
\oauthor{\bsnm{Jain}, \binits{S.}},
\oauthor{\bsnm{Jurafsky}, \binits{D.}},
\oauthor{\bsnm{Kalluri}, \binits{P.}},
\oauthor{\bsnm{Karamcheti}, \binits{S.}},
\oauthor{\bsnm{Keeling}, \binits{G.}},
\oauthor{\bsnm{Khani}, \binits{F.}},
\oauthor{\bsnm{Khattab}, \binits{O.}},
\oauthor{\bsnm{Koh}, \binits{P.W.}},
\oauthor{\bsnm{Krass}, \binits{M.}},
\oauthor{\bsnm{Krishna}, \binits{R.}},
\oauthor{\bsnm{Kuditipudi}, \binits{R.}},
\oauthor{\bsnm{Kumar}, \binits{A.}},
\oauthor{\bsnm{Ladhak}, \binits{F.}},
\oauthor{\bsnm{Lee}, \binits{M.}},
\oauthor{\bsnm{Lee}, \binits{T.}},
\oauthor{\bsnm{Leskovec}, \binits{J.}},
\oauthor{\bsnm{Levent}, \binits{I.}},
\oauthor{\bsnm{Li}, \binits{X.L.}},
\oauthor{\bsnm{Li}, \binits{X.}},
\oauthor{\bsnm{Ma}, \binits{T.}},
\oauthor{\bsnm{Malik}, \binits{A.}},
\oauthor{\bsnm{Manning}, \binits{C.D.}},
\oauthor{\bsnm{Mirchandani}, \binits{S.}},
\oauthor{\bsnm{Mitchell}, \binits{E.}},
\oauthor{\bsnm{Munyikwa}, \binits{Z.}},
\oauthor{\bsnm{Nair}, \binits{S.}},
\oauthor{\bsnm{Narayan}, \binits{A.}},
\oauthor{\bsnm{Narayanan}, \binits{D.}},
\oauthor{\bsnm{Newman}, \binits{B.}},
\oauthor{\bsnm{Nie}, \binits{A.}},
\oauthor{\bsnm{Niebles}, \binits{J.C.}},
\oauthor{\bsnm{Nilforoshan}, \binits{H.}},
\oauthor{\bsnm{Nyarko}, \binits{J.}},
\oauthor{\bsnm{Ogut}, \binits{G.}},
\oauthor{\bsnm{Orr}, \binits{L.}},
\oauthor{\bsnm{Papadimitriou}, \binits{I.}},
\oauthor{\bsnm{Park}, \binits{J.S.}},
\oauthor{\bsnm{Piech}, \binits{C.}},
\oauthor{\bsnm{Portelance}, \binits{E.}},
\oauthor{\bsnm{Potts}, \binits{C.}},
\oauthor{\bsnm{Raghunathan}, \binits{A.}},
\oauthor{\bsnm{Reich}, \binits{R.}},
\oauthor{\bsnm{Ren}, \binits{H.}},
\oauthor{\bsnm{Rong}, \binits{F.}},
\oauthor{\bsnm{Roohani}, \binits{Y.}},
\oauthor{\bsnm{Ruiz}, \binits{C.}},
\oauthor{\bsnm{Ryan}, \binits{J.}},
\oauthor{\bsnm{Ré}, \binits{C.}},
\oauthor{\bsnm{Sadigh}, \binits{D.}},
\oauthor{\bsnm{Sagawa}, \binits{S.}},
\oauthor{\bsnm{Santhanam}, \binits{K.}},
\oauthor{\bsnm{Shih}, \binits{A.}},
\oauthor{\bsnm{Srinivasan}, \binits{K.}},
\oauthor{\bsnm{Tamkin}, \binits{A.}},
\oauthor{\bsnm{Taori}, \binits{R.}},
\oauthor{\bsnm{Thomas}, \binits{A.W.}},
\oauthor{\bsnm{Tramèr}, \binits{F.}},
\oauthor{\bsnm{Wang}, \binits{R.E.}},
\oauthor{\bsnm{Wang}, \binits{W.}},
\oauthor{\bsnm{Wu}, \binits{B.}},
\oauthor{\bsnm{Wu}, \binits{J.}},
\oauthor{\bsnm{Wu}, \binits{Y.}},
\oauthor{\bsnm{Xie}, \binits{S.M.}},
\oauthor{\bsnm{Yasunaga}, \binits{M.}},
\oauthor{\bsnm{You}, \binits{J.}},
\oauthor{\bsnm{Zaharia}, \binits{M.}},
\oauthor{\bsnm{Zhang}, \binits{M.}},
\oauthor{\bsnm{Zhang}, \binits{T.}},
\oauthor{\bsnm{Zhang}, \binits{X.}},
\oauthor{\bsnm{Zhang}, \binits{Y.}},
\oauthor{\bsnm{Zheng}, \binits{L.}},
\oauthor{\bsnm{Zhou}, \binits{K.}},
\oauthor{\bsnm{Liang}, \binits{P.}}:
On the opportunities and risks of foundation models.
arXiv [cs.LG]
(2021)
{[cs.LG]}
\end{botherref}
\endbibitem

\bibitem[\protect\citeauthoryear{Chithrananda
  et~al.}{2020}]{Chithrananda2020-el}
\begin{botherref}
\oauthor{\bsnm{Chithrananda}, \binits{S.}},
\oauthor{\bsnm{Grand}, \binits{G.}},
\oauthor{\bsnm{Ramsundar}, \binits{B.}}:
{ChemBERTa}: Large-scale self-supervised pretraining for molecular property
  prediction.
arXiv [cs.LG]
(2020)
{[cs.LG]}
\end{botherref}
\endbibitem

\bibitem[\protect\citeauthoryear{Ross et~al.}{2022}]{Ross2022-cm}
\begin{barticle}
\bauthor{\bsnm{Ross}, \binits{J.}},
\bauthor{\bsnm{Belgodere}, \binits{B.}},
\bauthor{\bsnm{Chenthamarakshan}, \binits{V.}},
\bauthor{\bsnm{Padhi}, \binits{I.}},
\bauthor{\bsnm{Mroueh}, \binits{Y.}},
\bauthor{\bsnm{Das}, \binits{P.}}:
\batitle{Large-scale chemical language representations capture molecular
  structure and properties}.
\bjtitle{Nature Machine Intelligence}
\bvolume{4}(\bissue{12}),
\bfpage{1256}--\blpage{1264}
(\byear{2022})
\end{barticle}
\endbibitem

\bibitem[\protect\citeauthoryear{Wang et~al.}{2022}]{Wang2022-hg}
\begin{barticle}
\bauthor{\bsnm{Wang}, \binits{Y.}},
\bauthor{\bsnm{Wang}, \binits{J.}},
\bauthor{\bsnm{Cao}, \binits{Z.}},
\bauthor{\bsnm{Barati~Farimani}, \binits{A.}}:
\batitle{Molecular contrastive learning of representations via graph neural
  networks}.
\bjtitle{Nature Machine Intelligence}
\bvolume{4}(\bissue{3}),
\bfpage{279}--\blpage{287}
(\byear{2022})
\end{barticle}
\endbibitem

\bibitem[\protect\citeauthoryear{Singh et~al.}{2025}]{Singh2025}
\begin{barticle}
\bauthor{\bsnm{Singh}, \binits{R.}},
\bauthor{\bsnm{Barsainyan}, \binits{A.A.}},
\bauthor{\bsnm{Irfan}, \binits{R.}},
\bauthor{\bsnm{Amorin}, \binits{C.J.}},
\bauthor{\bsnm{He}, \binits{S.}},
\bauthor{\bsnm{Davis}, \binits{T.}},
\bauthor{\bsnm{Thiagarajan}, \binits{A.}},
\bauthor{\bsnm{Sankaran}, \binits{S.}},
\bauthor{\bsnm{Chithrananda}, \binits{S.}},
\bauthor{\bsnm{Ahmad}, \binits{W.}},
\bauthor{\bsnm{al.}}:
\batitle{Chemberta-3: An open source training framework for chemical foundation
  models}.
\bjtitle{ChemRxiv}
(\byear{2025})
\doiurl{10.26434/chemrxiv-2025-4glrl}
\end{barticle}
\endbibitem

\bibitem[\protect\citeauthoryear{Allen and Tildesley}{1989}]{Allen1989-nk}
\begin{bbook}
\bauthor{\bsnm{Allen}, \binits{M.P.}},
\bauthor{\bsnm{Tildesley}, \binits{D.J.}}:
\bbtitle{Computer Simulation of Liquids}.
\bpublisher{Clarendon Press},
\blocation{Oxford}
(\byear{1989})
\end{bbook}
\endbibitem

\bibitem[\protect\citeauthoryear{Dror et~al.}{2012}]{Dror2012-di}
\begin{barticle}
\bauthor{\bsnm{Dror}, \binits{R.O.}},
\bauthor{\bsnm{Dirks}, \binits{R.M.}},
\bauthor{\bsnm{Grossman}, \binits{J.P.}},
\bauthor{\bsnm{Xu}, \binits{H.}},
\bauthor{\bsnm{Shaw}, \binits{D.E.}}:
\batitle{Biomolecular simulation: a computational microscope for molecular
  biology}.
\bjtitle{Annu. Rev. Biophys.}
\bvolume{41},
\bfpage{429}--\blpage{452}
(\byear{2012})
\end{barticle}
\endbibitem

\bibitem[\protect\citeauthoryear{Jones et~al.}{2022}]{Jones2022-xt}
\begin{barticle}
\bauthor{\bsnm{Jones}, \binits{D.}},
\bauthor{\bsnm{Allen}, \binits{J.E.}},
\bauthor{\bsnm{Yang}, \binits{Y.}},
\bauthor{\bsnm{Drew~Bennett}, \binits{W.F.}},
\bauthor{\bsnm{Gokhale}, \binits{M.}},
\bauthor{\bsnm{Moshiri}, \binits{N.}},
\bauthor{\bsnm{Rosing}, \binits{T.S.}}:
\batitle{Accelerators for classical molecular dynamics simulations of
  biomolecules}.
\bjtitle{J. Chem. Theory Comput.}
\bvolume{18}(\bissue{7}),
\bfpage{4047}--\blpage{4069}
(\byear{2022})
\end{barticle}
\endbibitem

\bibitem[\protect\citeauthoryear{Berman et~al.}{2000}]{Berman2000-kn}
\begin{barticle}
\bauthor{\bsnm{Berman}, \binits{H.M.}},
\bauthor{\bsnm{Westbrook}, \binits{J.}},
\bauthor{\bsnm{Feng}, \binits{Z.}},
\bauthor{\bsnm{Gilliland}, \binits{G.}},
\bauthor{\bsnm{Bhat}, \binits{T.N.}},
\bauthor{\bsnm{Weissig}, \binits{H.}},
\bauthor{\bsnm{Shindyalov}, \binits{I.N.}},
\bauthor{\bsnm{Bourne}, \binits{P.E.}}:
\batitle{The protein data bank}.
\bjtitle{Nucleic Acids Res.}
\bvolume{28}(\bissue{1}),
\bfpage{235}--\blpage{242}
(\byear{2000})
\end{barticle}
\endbibitem

\bibitem[\protect\citeauthoryear{Wang et~al.}{2019}]{Wang2019-ja}
\begin{barticle}
\bauthor{\bsnm{Wang}, \binits{E.}},
\bauthor{\bsnm{Sun}, \binits{H.}},
\bauthor{\bsnm{Wang}, \binits{J.}},
\bauthor{\bsnm{Wang}, \binits{Z.}},
\bauthor{\bsnm{Liu}, \binits{H.}},
\bauthor{\bsnm{Zhang}, \binits{J.Z.H.}},
\bauthor{\bsnm{Hou}, \binits{T.}}:
\batitle{End-point binding free energy calculation with {MM}/{PBSA} and
  {MM}/{GBSA}: Strategies and applications in drug design}.
\bjtitle{Chem. Rev.}
\bvolume{119}(\bissue{16}),
\bfpage{9478}--\blpage{9508}
(\byear{2019})
\end{barticle}
\endbibitem

\bibitem[\protect\citeauthoryear{Wu et~al.}{2022}]{Wu2022-bw}
\begin{barticle}
\bauthor{\bsnm{Wu}, \binits{F.}},
\bauthor{\bsnm{Jin}, \binits{S.}},
\bauthor{\bsnm{Jiang}, \binits{Y.}},
\bauthor{\bsnm{Jin}, \binits{X.}},
\bauthor{\bsnm{Tang}, \binits{B.}},
\bauthor{\bsnm{Niu}, \binits{Z.}},
\bauthor{\bsnm{Liu}, \binits{X.}},
\bauthor{\bsnm{Zhang}, \binits{Q.}},
\bauthor{\bsnm{Zeng}, \binits{X.}},
\bauthor{\bsnm{Li}, \binits{S.Z.}}:
\batitle{Pre-training of equivariant graph matching networks with conformation
  flexibility for drug binding}.
\bjtitle{Adv. Sci.}
\bvolume{9}(\bissue{33}),
\bfpage{2203796}
(\byear{2022})
\end{barticle}
\endbibitem

\bibitem[\protect\citeauthoryear{Jing et~al.}{2024}]{Jing2024-wj}
\begin{botherref}
\oauthor{\bsnm{Jing}, \binits{B.}},
\oauthor{\bsnm{Stärk}, \binits{H.}},
\oauthor{\bsnm{Jaakkola}, \binits{T.}},
\oauthor{\bsnm{Berger}, \binits{B.}}:
Generative modeling of molecular dynamics trajectories.
arXiv [q-bio.BM]
(2024)
{[q-bio.BM]}
\end{botherref}
\endbibitem

\bibitem[\protect\citeauthoryear{Siebenmorgen
  et~al.}{2024}]{Siebenmorgen2024-pn}
\begin{botherref}
\oauthor{\bsnm{Siebenmorgen}, \binits{T.}},
\oauthor{\bsnm{Menezes}, \binits{F.}},
\oauthor{\bsnm{Benassou}, \binits{S.}},
\oauthor{\bsnm{Merdivan}, \binits{E.}},
\oauthor{\bsnm{Didi}, \binits{K.}},
\oauthor{\bsnm{Mourão}, \binits{A.S.D.}},
\oauthor{\bsnm{Kitel}, \binits{R.}},
\oauthor{\bsnm{Liò}, \binits{P.}},
\oauthor{\bsnm{Kesselheim}, \binits{S.}},
\oauthor{\bsnm{Piraud}, \binits{M.}},
\oauthor{\bsnm{Theis}, \binits{F.J.}},
\oauthor{\bsnm{Sattler}, \binits{M.}},
\oauthor{\bsnm{Popowicz}, \binits{G.M.}}:
{MISATO}: machine learning dataset of protein-ligand complexes for
  structure-based drug discovery.
Nat Comput Sci
(2024)
\end{botherref}
\endbibitem

\bibitem[\protect\citeauthoryear{Prat et~al.}{2024}]{Prat2024-vp}
\begin{botherref}
\oauthor{\bsnm{Prat}, \binits{A.}},
\oauthor{\bsnm{Abdel~Aty}, \binits{H.}},
\oauthor{\bsnm{Pabrinkis}, \binits{A.}},
\oauthor{\bsnm{Bastas}, \binits{O.}},
\oauthor{\bsnm{Paquet}, \binits{T.}},
\oauthor{\bsnm{Kamuntavičius}, \binits{G.}},
\oauthor{\bsnm{Tal}, \binits{R.}}:
{SE}(3) equivariant topologies for structure-based drug discovery.
ChemRxiv
(2024)
\end{botherref}
\endbibitem

\bibitem[\protect\citeauthoryear{Liu et~al.}{2025}]{Liu2025-ef}
\begin{barticle}
\bauthor{\bsnm{Liu}, \binits{M.}},
\bauthor{\bsnm{Jin}, \binits{S.}},
\bauthor{\bsnm{Lai}, \binits{H.}},
\bauthor{\bsnm{Wang}, \binits{L.}},
\bauthor{\bsnm{Wang}, \binits{J.}},
\bauthor{\bsnm{Cheng}, \binits{Z.}},
\bauthor{\bsnm{Zeng}, \binits{X.}}:
\batitle{Molecular dynamics-powered hierarchical geometric deep learning
  framework for protein-ligand interaction}.
\bjtitle{IEEE Trans Comput Biol Bioinform}
\bvolume{22}(\bissue{4}),
\bfpage{1517}--\blpage{1527}
(\byear{2025})
\end{barticle}
\endbibitem

\bibitem[\protect\citeauthoryear{Libouban et~al.}{2025}]{Libouban2025-aw}
\begin{barticle}
\bauthor{\bsnm{Libouban}, \binits{P.-Y.}},
\bauthor{\bsnm{Parisel}, \binits{C.}},
\bauthor{\bsnm{Song}, \binits{M.}},
\bauthor{\bsnm{Aci-Sèche}, \binits{S.}},
\bauthor{\bsnm{Gómez-Tamayo}, \binits{J.C.}},
\bauthor{\bsnm{Tresadern}, \binits{G.}},
\bauthor{\bsnm{Bonnet}, \binits{P.}}:
\batitle{Spatio-temporal learning from molecular dynamics simulations for
  protein-ligand binding affinity prediction}.
\bjtitle{Bioinformatics}
\bvolume{41}(\bissue{8}),
\bfpage{429}
(\byear{2025})
\end{barticle}
\endbibitem

\bibitem[\protect\citeauthoryear{Korlepara et~al.}{2022}]{Korlepara2022-ff}
\begin{barticle}
\bauthor{\bsnm{Korlepara}, \binits{D.B.}},
\bauthor{\bsnm{Vasavi}, \binits{C.S.}},
\bauthor{\bsnm{Jeurkar}, \binits{S.}},
\bauthor{\bsnm{Pal}, \binits{P.K.}},
\bauthor{\bsnm{Roy}, \binits{S.}},
\bauthor{\bsnm{Mehta}, \binits{S.}},
\bauthor{\bsnm{Sharma}, \binits{S.}},
\bauthor{\bsnm{Kumar}, \binits{V.}},
\bauthor{\bsnm{Muvva}, \binits{C.}},
\bauthor{\bsnm{Sridharan}, \binits{B.}},
\bauthor{\bsnm{Garg}, \binits{A.}},
\bauthor{\bsnm{Modee}, \binits{R.}},
\bauthor{\bsnm{Bhati}, \binits{A.P.}},
\bauthor{\bsnm{Nayar}, \binits{D.}},
\bauthor{\bsnm{Priyakumar}, \binits{U.D.}}:
\batitle{{PLAS}-{5k}: Dataset of protein-ligand affinities from molecular
  dynamics for machine learning applications}.
\bjtitle{Sci Data}
\bvolume{9}(\bissue{1}),
\bfpage{548}
(\byear{2022})
\end{barticle}
\endbibitem

\bibitem[\protect\citeauthoryear{Korlepara et~al.}{2024}]{Korlepara2024-wb}
\begin{barticle}
\bauthor{\bsnm{Korlepara}, \binits{D.B.}},
\bauthor{\bsnm{C~S}, \binits{V.}},
\bauthor{\bsnm{Srivastava}, \binits{R.}},
\bauthor{\bsnm{Pal}, \binits{P.K.}},
\bauthor{\bsnm{Raza}, \binits{S.H.}},
\bauthor{\bsnm{Kumar}, \binits{V.}},
\bauthor{\bsnm{Pandit}, \binits{S.}},
\bauthor{\bsnm{Nair}, \binits{A.G.}},
\bauthor{\bsnm{Pandey}, \binits{S.}},
\bauthor{\bsnm{Sharma}, \binits{S.}},
\bauthor{\bsnm{Jeurkar}, \binits{S.}},
\bauthor{\bsnm{Thakran}, \binits{K.}},
\bauthor{\bsnm{Jaglan}, \binits{R.}},
\bauthor{\bsnm{Verma}, \binits{S.}},
\bauthor{\bsnm{Ramachandran}, \binits{I.}},
\bauthor{\bsnm{Chatterjee}, \binits{P.}},
\bauthor{\bsnm{Nayar}, \binits{D.}},
\bauthor{\bsnm{Priyakumar}, \binits{U.D.}}:
\batitle{{PLAS}-{20k}: Extended dataset of protein-ligand affinities from {MD}
  simulations for machine learning applications}.
\bjtitle{Sci. Data}
\bvolume{11}(\bissue{1}),
\bfpage{180}
(\byear{2024})
\end{barticle}
\endbibitem

\bibitem[\protect\citeauthoryear{Su et~al.}{2019}]{Su2019-ni}
\begin{barticle}
\bauthor{\bsnm{Su}, \binits{M.}},
\bauthor{\bsnm{Yang}, \binits{Q.}},
\bauthor{\bsnm{Du}, \binits{Y.}},
\bauthor{\bsnm{Feng}, \binits{G.}},
\bauthor{\bsnm{Liu}, \binits{Z.}},
\bauthor{\bsnm{Li}, \binits{Y.}},
\bauthor{\bsnm{Wang}, \binits{R.}}:
\batitle{Comparative assessment of scoring functions: The {CASF}-2016 update}.
\bjtitle{J. Chem. Inf. Model.}
\bvolume{59}(\bissue{2}),
\bfpage{895}--\blpage{913}
(\byear{2019})
\end{barticle}
\endbibitem

\bibitem[\protect\citeauthoryear{Townshend et~al.}{2020}]{Townshend2020-eh}
\begin{bchapter}
\bauthor{\bsnm{Townshend}, \binits{R.J.L.}},
\bauthor{\bsnm{Vögele}, \binits{M.}},
\bauthor{\bsnm{Suriana}, \binits{P.}},
\bauthor{\bsnm{Derry}, \binits{A.}},
\bauthor{\bsnm{Powers}, \binits{A.}},
\bauthor{\bsnm{Laloudakis}, \binits{Y.}},
\bauthor{\bsnm{Balachandar}, \binits{S.}},
\bauthor{\bsnm{Jing}, \binits{B.}},
\bauthor{\bsnm{Anderson}, \binits{B.}},
\bauthor{\bsnm{Eismann}, \binits{S.}},
\bauthor{\bsnm{Kondor}, \binits{R.}},
\bauthor{\bsnm{Altman}, \binits{R.B.}},
\bauthor{\bsnm{Dror}, \binits{R.O.}}:
\bctitle{{ATOM3D}: Tasks on molecules in three dimensions}.
In: \bbtitle{35th Conference on Neural Information Processing Systems}
(\byear{2020})
\end{bchapter}
\endbibitem

\bibitem[\protect\citeauthoryear{Miller et~al.}{2012}]{Miller2012-tx}
\begin{barticle}
\bauthor{\bsnm{Miller}, \binits{B.R.} \bsuffix{3rd}},
\bauthor{\bsnm{McGee}, \binits{T.D.} \bsuffix{Jr}},
\bauthor{\bsnm{Swails}, \binits{J.M.}},
\bauthor{\bsnm{Homeyer}, \binits{N.}},
\bauthor{\bsnm{Gohlke}, \binits{H.}},
\bauthor{\bsnm{Roitberg}, \binits{A.E.}}:
\batitle{{MMPBSA}.py: An efficient program for end-state free energy
  calculations}.
\bjtitle{J. Chem. Theory Comput.}
\bvolume{8}(\bissue{9}),
\bfpage{3314}--\blpage{3321}
(\byear{2012})
\end{barticle}
\endbibitem

\bibitem[\protect\citeauthoryear{Vander~Meersche
  et~al.}{2024}]{Vander-Meersche2024-vs}
\begin{barticle}
\bauthor{\bsnm{Vander~Meersche}, \binits{Y.}},
\bauthor{\bsnm{Cretin}, \binits{G.}},
\bauthor{\bsnm{Gheeraert}, \binits{A.}},
\bauthor{\bsnm{Gelly}, \binits{J.-C.}},
\bauthor{\bsnm{Galochkina}, \binits{T.}}:
\batitle{{ATLAS}: protein flexibility description from atomistic molecular
  dynamics simulations}.
\bjtitle{Nucleic Acids Res.}
\bvolume{52}(\bissue{D1}),
\bfpage{384}--\blpage{392}
(\byear{2024})
\end{barticle}
\endbibitem

\bibitem[\protect\citeauthoryear{Michaud-Agrawal
  et~al.}{2011}]{Michaud-Agrawal2011-bp}
\begin{barticle}
\bauthor{\bsnm{Michaud-Agrawal}, \binits{N.}},
\bauthor{\bsnm{Denning}, \binits{E.J.}},
\bauthor{\bsnm{Woolf}, \binits{T.B.}},
\bauthor{\bsnm{Beckstein}, \binits{O.}}:
\batitle{{MDAnalysis}: a toolkit for the analysis of molecular dynamics
  simulations}.
\bjtitle{J. Comput. Chem.}
\bvolume{32}(\bissue{10}),
\bfpage{2319}--\blpage{2327}
(\byear{2011})
\end{barticle}
\endbibitem

\bibitem[\protect\citeauthoryear{Gowers et~al.}{2016}]{Gowers2016-nc}
\begin{botherref}
\oauthor{\bsnm{Gowers}, \binits{R.J.}},
\oauthor{\bsnm{Linke}, \binits{M.}},
\oauthor{\bsnm{Barnoud}, \binits{J.}},
\oauthor{\bsnm{Reddy}, \binits{T.J.E.}},
\oauthor{\bsnm{Melo}, \binits{M.N.}},
\oauthor{\bsnm{Seyler}, \binits{S.L.}},
\oauthor{\bsnm{Domański}, \binits{J.}},
\oauthor{\bsnm{Dotson}, \binits{D.L.}},
\oauthor{\bsnm{Buchoux}, \binits{S.}},
\oauthor{\bsnm{Kenney}, \binits{I.M.}},
\oauthor{\bsnm{Beckstein}, \binits{O.}}:
{MDAnalysis}: A python package for the rapid analysis of molecular dynamics
  simulations.
scipy
(2016)
\end{botherref}
\endbibitem

\bibitem[\protect\citeauthoryear{{D.A. Case, H.M. Aktulga, K. Belfon, I.Y.
  Ben-Shalom, J.T. Berryman, S.R. Brozell, D.S. Cerutti, T.E. Cheatham, III,
  G.A. Cisneros, V.W.D. Cruzeiro, T.A. Darden, R.E. Duke, G. Giambasu, M.K.
  Gilson, H. Gohlke, A.W. Goetz, R. Harris, S. Izadi, S.A. Izmailov, K.
  Kasavajhala, M.C. Kaymak, E. King, A. Kovalenko, T. Kurtzman, T.S. Lee, S.
  LeGrand, P. Li, C. Lin, J. Liu, T. Luchko, R. Luo, M. Machado, V. Man, M.
  Manathunga, K.M. Merz, Y. Miao, O. Mikhailovskii, G. Monard, H. Nguyen, K.A.
  O'Hearn, A. Onufriev, F. Pan, S. Pantano, R. Qi, A. Rahnamoun, D.R. Roe, A.
  Roitberg, C. Sagui, S. Schott-Verdugo, A. Shajan, J. Shen, C.L. Simmerling,
  N.R. Skrynnikov, J. Smith, J. Swails, R.C. Walker, J. Wang, J. Wang, H. Wei,
  R.M. Wolf, X. Wu, Y. Xiong, Y. Xue, D.M. York, S. Zhao, and P.A.
  Kollman}}{2022}]{amber2022}
\begin{botherref}
\oauthor{\bsnm{{D.A. Case, H.M. Aktulga, K. Belfon, I.Y. Ben-Shalom, J.T.
  Berryman, S.R. Brozell, D.S. Cerutti, T.E. Cheatham, III, G.A. Cisneros,
  V.W.D. Cruzeiro, T.A. Darden, R.E. Duke, G. Giambasu, M.K. Gilson, H. Gohlke,
  A.W. Goetz, R. Harris, S. Izadi, S.A. Izmailov, K. Kasavajhala, M.C. Kaymak,
  E. King, A. Kovalenko, T. Kurtzman, T.S. Lee, S. LeGrand, P. Li, C. Lin, J.
  Liu, T. Luchko, R. Luo, M. Machado, V. Man, M. Manathunga, K.M. Merz, Y.
  Miao, O. Mikhailovskii, G. Monard, H. Nguyen, K.A. O'Hearn, A. Onufriev, F.
  Pan, S. Pantano, R. Qi, A. Rahnamoun, D.R. Roe, A. Roitberg, C. Sagui, S.
  Schott-Verdugo, A. Shajan, J. Shen, C.L. Simmerling, N.R. Skrynnikov, J.
  Smith, J. Swails, R.C. Walker, J. Wang, J. Wang, H. Wei, R.M. Wolf, X. Wu, Y.
  Xiong, Y. Xue, D.M. York, S. Zhao, and P.A. Kollman}}}:
Amber 2022.
University of California, San Francisco,
(2022).
University of California, San Francisco
\end{botherref}
\endbibitem

\bibitem[\protect\citeauthoryear{Zhang}{}]{Zhang2022-jx}
\begin{botherref}
\oauthor{\bsnm{Zhang}, \binits{X.}}:
conveyorlc: A pipeline to do virtual screening.
\url{https://github.com/XiaohuaZhangLLNL/conveyorlc}.
Accessed: 2022-5-16
\end{botherref}
\endbibitem

\bibitem[\protect\citeauthoryear{Gilmer et~al.}{2017}]{Gilmer2017-ul}
\begin{botherref}
\oauthor{\bsnm{Gilmer}, \binits{J.}},
\oauthor{\bsnm{Schoenholz}, \binits{S.S.}},
\oauthor{\bsnm{Riley}, \binits{P.F.}},
\oauthor{\bsnm{Vinyals}, \binits{O.}},
\oauthor{\bsnm{Dahl}, \binits{G.E.}}:
Neural message passing for quantum chemistry.
arXiv [cs.LG]
(2017)
{[cs.LG]}
\end{botherref}
\endbibitem

\bibitem[\protect\citeauthoryear{Bruna et~al.}{2013}]{Bruna2013-kf}
\begin{botherref}
\oauthor{\bsnm{Bruna}, \binits{J.}},
\oauthor{\bsnm{Zaremba}, \binits{W.}},
\oauthor{\bsnm{Szlam}, \binits{A.}},
\oauthor{\bsnm{LeCun}, \binits{Y.}}:
Spectral networks and locally connected networks on graphs.
arXiv [cs.LG]
(2013)
{[cs.LG]}
\end{botherref}
\endbibitem

\bibitem[\protect\citeauthoryear{Kipf and Welling}{2016}]{Kipf2016-sp}
\begin{botherref}
\oauthor{\bsnm{Kipf}, \binits{T.N.}},
\oauthor{\bsnm{Welling}, \binits{M.}}:
Semi-supervised classification with graph convolutional networks.
arXiv [cs.LG]
(2016)
{[cs.LG]}
\end{botherref}
\endbibitem

\bibitem[\protect\citeauthoryear{Duvenaud et~al.}{2015}]{Duvenaud2015-kl}
\begin{bchapter}
\bauthor{\bsnm{Duvenaud}, \binits{D.K.}},
\bauthor{\bsnm{Maclaurin}, \binits{D.}},
\bauthor{\bsnm{Iparraguirre}, \binits{J.}},
\bauthor{\bsnm{Bombarell}, \binits{R.}},
\bauthor{\bsnm{Hirzel}, \binits{T.}},
\bauthor{\bsnm{Aspuru-Guzik}, \binits{A.}},
\bauthor{\bsnm{Adams}, \binits{R.P.}}:
\bctitle{Convolutional networks on graphs for learning molecular fingerprints}.
In: \bbtitle{Advances in Neural Information Processing Systems},
pp. \bfpage{2224}--\blpage{2232}
(\byear{2015})
\end{bchapter}
\endbibitem

\bibitem[\protect\citeauthoryear{Ganea et~al.}{2021}]{Ganea2021-tj}
\begin{botherref}
\oauthor{\bsnm{Ganea}, \binits{O.-E.}},
\oauthor{\bsnm{Huang}, \binits{X.}},
\oauthor{\bsnm{Bunne}, \binits{C.}},
\oauthor{\bsnm{Bian}, \binits{Y.}},
\oauthor{\bsnm{Barzilay}, \binits{R.}},
\oauthor{\bsnm{Jaakkola}, \binits{T.}},
\oauthor{\bsnm{Krause}, \binits{A.}}:
Independent {SE}(3)-equivariant models for end-to-end rigid protein docking.
arXiv [cs.AI]
(2021)
{[cs.AI]}
\end{botherref}
\endbibitem

\bibitem[\protect\citeauthoryear{Paszke et~al.}{2019}]{Paszke2019-us}
\begin{botherref}
\oauthor{\bsnm{Paszke}, \binits{A.}},
\oauthor{\bsnm{Gross}, \binits{S.}},
\oauthor{\bsnm{Massa}, \binits{F.}},
\oauthor{\bsnm{Lerer}, \binits{A.}},
\oauthor{\bsnm{Bradbury}, \binits{J.}},
\oauthor{\bsnm{Chanan}, \binits{G.}},
\oauthor{\bsnm{Killeen}, \binits{T.}},
\oauthor{\bsnm{Lin}, \binits{Z.}},
\oauthor{\bsnm{Gimelshein}, \binits{N.}},
\oauthor{\bsnm{Antiga}, \binits{L.}},
\oauthor{\bsnm{Desmaison}, \binits{A.}},
\oauthor{\bsnm{Köpf}, \binits{A.}},
\oauthor{\bsnm{Yang}, \binits{E.}},
\oauthor{\bsnm{DeVito}, \binits{Z.}},
\oauthor{\bsnm{Raison}, \binits{M.}},
\oauthor{\bsnm{Tejani}, \binits{A.}},
\oauthor{\bsnm{Chilamkurthy}, \binits{S.}},
\oauthor{\bsnm{Steiner}, \binits{B.}},
\oauthor{\bsnm{Fang}, \binits{L.}},
\oauthor{\bsnm{Bai}, \binits{J.}},
\oauthor{\bsnm{Chintala}, \binits{S.}}:
{PyTorch}: An imperative style, high-performance deep learning library.
arXiv [cs.LG]
(2019)
{[cs.LG]}
\end{botherref}
\endbibitem

\bibitem[\protect\citeauthoryear{}{}]{amd_mi300a}
\begin{botherref}
{AMD} {INSTINCT™} {MI300A} {APU}.
\url{https://www.amd.com/content/dam/amd/en/documents/instinct-tech-docs/data-sheets/amd-instinct-mi300a-data-sheet.pdf}.
Accessed: 2025-8-21
\end{botherref}
\endbibitem

\bibitem[\protect\citeauthoryear{}{}]{RDKit}
\begin{botherref}
{RDKit}: Open-source cheminformatics.
\url{https://www.rdkit.org}.
Accessed: [Date of Access, e.g., 2025-08-14]
\end{botherref}
\endbibitem

\end{thebibliography}

\backmatter

\bmhead{Acknowledgments}



This research was supported by funds from the UC National Laboratory Fees Research Program of the University of California, Grant Number L23GF6259. This work was supported in part by CRISP and PRISM, centers in  JUMP 1.0 and 2.0, SRC programs sponsored by DARPA, SRC \#236160. Computing support for this work came  from the Lawrence Livermore National Laboratory (LLNL) Institutional Computing Grand Challenge program. This work was funded in part by the Defense Threat Reduction Agency (DTRA) HDTRA1036045. Part of this research was also supported by the American Heart Association under CRADA TC02274-4. All work performed at Lawrence Livermore National Laboratory is  performed  under  the  auspices  of  the  U.S.  Department  of  Energy  under  Contract  DE-AC52-07NA27344,  LLNL-JRNL-2010833-DRAFT.

\bmhead{Author contributions}
D.J., J.E.A, and T.S.R designed the research study. Y.Y. performed the molecular docking, molecular dynamics, and MMGBSA calculations. D.J. developed the method, wrote the code, and performed the analysis. D.J. and J.E.A wrote the manuscript. D.J., J.E.A, N.M., and T.S.R. edited the manuscript. 
\bmhead{Competing interests}
The authors declare no competing interests.

\bmhead{Supplementary information}




\clearpage
\section{Dataset Splits}\label{sec:dataset_splits}

\begin{table}[!h]
    \centering
    \begin{tabular}{c|c|c|c}
     \textbf{Fold} & \textbf{Train}    & \textbf{Val} & \textbf{Test} \\ \hline
    0 & 172 & 20 & 49 \\ 
        1 & 173 & 20 & 48 \\
        2 & 173 & 20 & 48 \\
        3 & 173 & 20 & 48 \\
        4 & 173 & 20 & 48 \\
    \end{tabular}
    \caption{PDB structure count for each split of the dataset.}
    \label{tab:dataset_split_pdb}
\end{table}

\vspace{1cm}
\section{Model Architecture Details}\label{sec:encoder_parameters}

\begin{table}[!h]
    \centering
    \begin{tabular}{c|c|c|c}
     \textbf{Encoder} & \textbf{\# Params.}    & \textbf{dim} & \textbf{$\mathcal{L}$} \\ \hline
        GNN & 445,312 & 128 & 4 \\
        EGNN & 495,872 & 128 & 4 \\
        EGMN & 5,237,445 & 128 & 6\\
        MLP & 8,321 & - & - \\
    \end{tabular}
    \caption{Trainable parameter counts for each neural network considered in our work.}
    \label{tab:model_params}
\end{table}

\clearpage
\section{Sequence similarity of each training set versus the corresponding test set by fold}\label{secA1}

\clearpage
\section{Ligand Similarity Clustermaps}\label{sec:ligand_sim_clustermap}

\clearpage
\section{Sequence similarity of the ProtMD training set versus each test set}\label{sec:seq_sim_protmd}

\clearpage
\section{PCA Projections of Model Representations}\label{sec:model_pca}

\end{document}